\documentclass[twocolumn,pre,floatfix,superscriptaddress, nofootinbib]{revtex4}
\usepackage[T1]{fontenc}
\usepackage[latin9]{inputenc}
\usepackage{mathrsfs}
\usepackage{amsmath}
\usepackage{amsthm}
\usepackage{amssymb}
\usepackage{graphicx}
\usepackage{braket}
\usepackage{color}
\usepackage[svgnames]{xcolor}
\usepackage{comment}
\usepackage{multirow}
\usepackage{natbib}
\usepackage{pifont} 
\usepackage{dsfont}

\makeatletter

\usepackage{listings}

\usepackage{hyperref}

\hypersetup{
  colorlinks=true,
  linkcolor=blue,
  citecolor=ForestGreen,
  urlcolor=DarkOrchid
}

\newcommand{\HP}{\mathbb{H}}

\begin{document}

\title{Quantum Compilation Toolkit for Rydberg Atom Arrays with Implications for Problem Hardness and Quantum Speedups}

\author{Martin J. A. Schuetz}
\thanks{These authors contributed equally.}
\affiliation{Amazon Advanced Solutions Lab, Seattle, Washington 98170, USA}
\affiliation{AWS Center for Quantum Computing, Pasadena, CA 91125, USA}

\author{Ruben S. Andrist}
\thanks{These authors contributed equally.}
\affiliation{Amazon Advanced Solutions Lab, Seattle, Washington 98170, USA}

\author{Grant Salton}
\affiliation{Amazon Advanced Solutions Lab, Seattle, Washington 98170, USA}
\affiliation{AWS Center for Quantum Computing, Pasadena, CA 91125, USA}

\author{Romina Yalovetzky}
\affiliation{Global Technology Applied Research, JPMorgan Chase, New York, NY 10017 USA}

\author{Rudy Raymond}
\affiliation{Global Technology Applied Research, JPMorgan Chase, New York, NY 10017 USA}

\author{Yue Sun}
\affiliation{Global Technology Applied Research, JPMorgan Chase, New York, NY 10017 USA}

\author{Atithi Acharya}
\affiliation{Global Technology Applied Research, JPMorgan Chase, New York, NY 10017 USA}

\author{Shouvanik Chakrabarti}
\affiliation{Global Technology Applied Research, JPMorgan Chase, New York, NY 10017 USA}

\author{Marco Pistoia}
\altaffiliation[]{These authors acted as Co-PIs.}
\affiliation{Global Technology Applied Research, JPMorgan Chase, New York, NY 10017 USA}

\author{Helmut G. Katzgraber}
\altaffiliation[]{These authors acted as Co-PIs.}
\affiliation{Amazon Advanced Solutions Lab, Seattle, Washington 98170, USA}

\date{\today}
\begin{abstract}

We propose and implement a comprehensive quantum compilation toolkit for solving the maximum independent set (MIS) problem on quantum hardware based on Rydberg atom arrays. 
Our end-to-end pipeline involves three core components to efficiently map generic MIS instances onto Rydberg arrays with unit-disk connectivity, 
with modules for graph reduction, hardware compatibility checks, and graph embedding.  
The first module (reducer) provides hardware-agnostic and deterministic reduction logic that iteratively reduces the problem size via lazy clique removals. 
We find that real-world networks can typically be reduced by orders of magnitude on sub-second time scales, thus significantly cutting down the eventual load for quantum devices. 
Moreover, we show that reduction techniques may be an important tool in the ongoing search for potential quantum speedups, given their ability to identify hard problem instances. In particular, for Rydberg-native MIS instances, we observe signatures of an easy-hard-easy transition and quantify a critical degree indicating the onset of a hard problem regime.  
The second module (compatibility checker) implements a hardware compatibility checker that quickly determines whether or not a given input graph may be compatible with the restrictions imposed by Rydberg quantum hardware.  
The third module (embedder) describes hardware-efficient graph embedding routines to generate (approximate) encodings with controllable overhead and optimized ancilla placements. 
We exemplify our pipeline with experiments run on the QuEra Aquila device available on Amazon Braket.
In aggregate, our work provides a set of tools that extends the class of problems that can be tackled with near-term Rydberg atom arrays. 
\end{abstract}

\maketitle

\section{Introduction}
\label{introduction}

The field of combinatorial optimization (CO) involves the search for the extremum of an objective function (such as a cost or profit value) within a finite (but usually large) set of candidate solutions. 
Despite often being simple to conceptualize, many CO problems are hard to solve and even NP-hard, representing some of the most exquisite, yet challenging computational problems known \citep{papadimitriou:98, korte:12}.  
Given its close ties to the maximum clique, minimum vertex cover, and set packing problems \citep{wurtz:22}, the maximum independent set (MIS) problem represents an important paradigmatic CO problem, with practical applications in virtually any industry sector. These include network design \citep{hale:80}, vehicle routing \citep{dong:22}, and quantitative finance \citep{boginski:05, kalra:08, herman:13}, to name a few.

\textbf{Quantum optimization.} Over the last few decades, quantum optimization (QO) has emerged as a novel paradigm for solving such discrete optimization problems \citep{abbas:23}, 
in pursuit of quantum speedups over the best-known classical algorithms for practically relevant problems.  
Prominent QO algorithms include quantum annealing algorithms (QAA) \citep{kadowaki:98, farhi:00, farhi:01, das:08, hauke:20} as well as hybrid (quantum-classical) algorithms such as the quantum approximate optimization algorithm (QAOA) \citep{farhi:14, zhou:20}, among others.
In these approaches, typically the optimal solution to the classical optimization problem at hand is encoded in the ground state of quantum many-body systems that can be implemented and controlled with (special-purpose) quantum hardware \citep{lucas:14, glover:18}.

\textbf{Optimization with Rydberg arrays.} Over the last few years, analog neutral-atom quantum machines in the form of Rydberg atom arrays have emerged as a novel class of programmable and scalable special-purpose quantum devices geared towards optimization workloads \citep{pichler:18, pichler:18computational, zhou:20, serret:20, ebadi:22, cain:23, schiffer:23, finzgar:23, finzgar:23b, kim:23}. 
In particular, recent experiments with up to 289 Rydberg atoms have reported a potential super-linear quantum speedup over classical simulated annealing for the MIS problem, with pioneering implementations of both QAA and QAOA within the same experimental setup \citep{ebadi:22, andrist:23}. 
Because of the isotropic nature of the Rydberg blockade mechanism \citep{lukin:01, levine:19, saffman:10}, these first-generation experiments were inherently limited to maximum independent set (MIS) problem instances on a restricted class of geometric graphs known as unit-disk (UD) graphs \citep{clark:90, pichler:18, pichler:18computational}. 
Ultimately, however, the usefulness of these devices will depend on the scope of problems they can tackle beyond the hardware-native MIS-UD problem set.  

\textbf{Embedding schemes.} Embedding techniques are designed to expand the scope of problems supported by a given quantum device by mapping \textit{logical} input problems to \textit{physical} representations compatible with connectivity-constrained quantum hardware. 
Given the local topology of the underlying physical hardware and limitations in natively accommodating long-range couplings, embedding is usually achieved via the introduction of ancilla qubits to effectively distribute logical information across distant locations, thereby boosting the system's connectivity at the expense of an (often quadratic) overhead in the number of qubits~\citep{koenz:21}. 
Specifically, for superconducting (SC) annealing devices, so-called \textit{minor embedding techniques} are well established \citep{bunyk:14, vinci:15, koenz:21, sugie:21}, providing a (heuristic) mapping from a logical graph 
to a sub-graph of a target hardware graph (with sparse connectivity), with chains of several physical qubits making up single (compound) logical variables.
For Rydberg atom arrays, embedding schemes have recently been put forward that are akin to the embedding schemes for SC annealers. These schemes were developed within a larger effort to expand the class of problems that can be tackled with near-term, analog Rydberg devices, beyond the hardware-native MIS problem on UD graphs (dubbed MIS-UD hereafter). 
Existing proposals, however, are either restricted to a specific class of problem instances \citep{byun:22, byun:23}, or come with (potentially demanding) experimental requirements, such as the need for three-dimensional arrays \citep{kim:22}, four-body interactions \citep{dlaska:22}, or local detuning \citep{nguyen:23}, with some schemes incurring an (\textit{a priori} undefined) overhead in the number of ancilla qubits \citep{byun:23}. 
Recently, first experiments have successfully verified the ability to embed and solve maximum weighted independent set problems on non-UD graphs, albeit for relatively simple instances at small scales with five logical and nine physical nodes, respectively \citep{deoliveira:24}.  

\textbf{Research objectives.} Our main goal is to contribute to this line of recent research by providing a set of tools that help expand the scope and size of problems compatible with near-term Rydberg atom arrays.  
We note that the ability to coherently shuttle around individual atoms---arguably one of the most distinctive features of modern Rydberg arrays \citep{bluvstein:21, bluvstein:24}---with its potential to generate effectively non-local interactions in real time, brings exciting possibilities for future Rydberg embedding technology. 
For the sake of relatively low experimental complexity, however, here we focus on developing tools that are either hardware-agnostic or can already be implemented with static Rydberg array positioning.  

\section{Overview of main results}
\label{overview}

\begin{figure*}
  \leftskip=0pt
  \includegraphics[width=2.075\columnwidth]{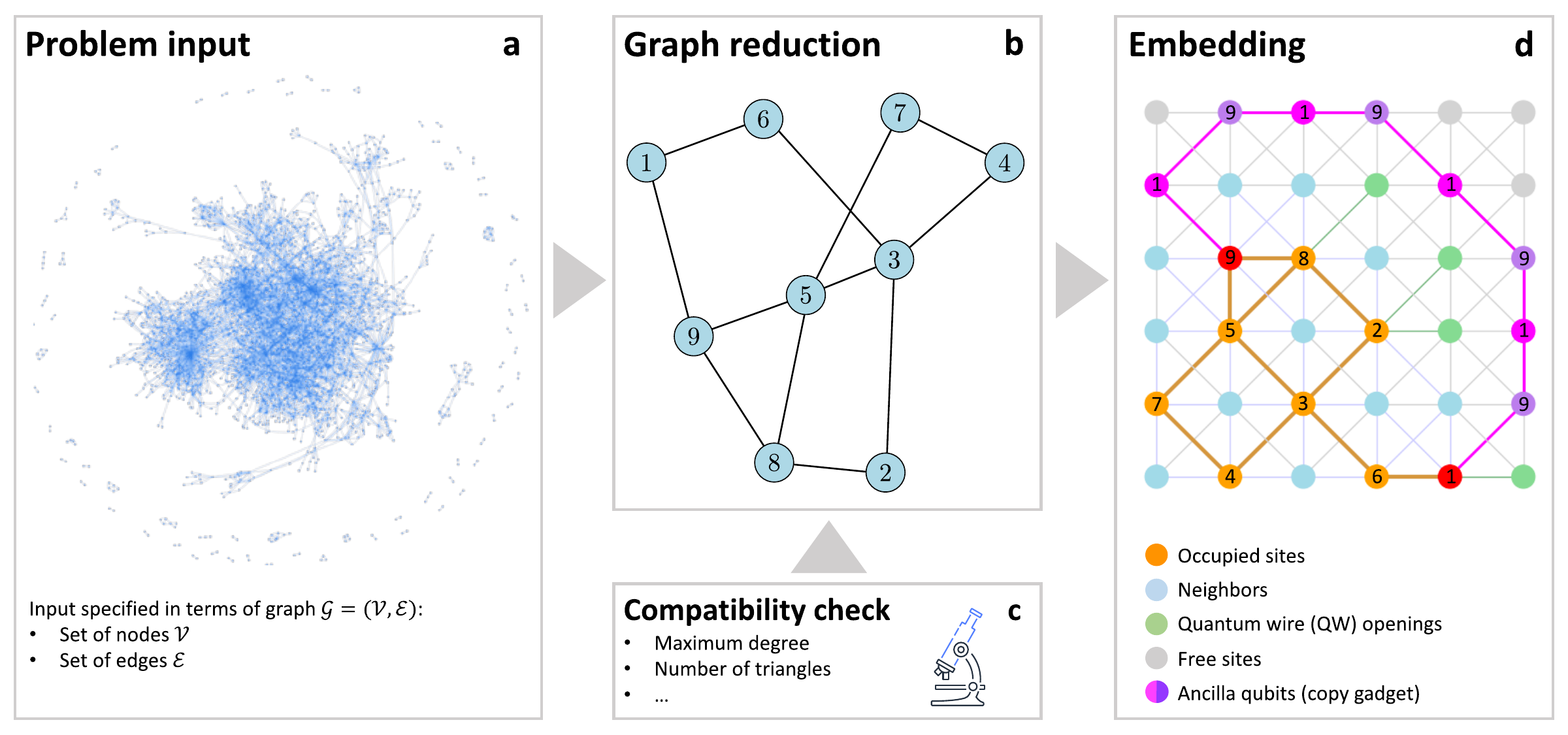}
  \caption{
    Schematic illustration of the proposed compilation pipeline, shown here for the Cora citation graph with $\sim 2700$ nodes and $\sim 5300$ edges. 
      \textbf{(a) Problem input}: The computational problem is defined in terms of an input graph $\mathcal{G}=(\mathcal{V}, \mathcal{E})$. 
      The ultimate goal is to solve the MIS problem on this graph, using Rydberg atom arrays. The scale and connectivity of the graph $\mathcal{G}$ is arbitrary, and, in general, incompatible with near-term Rydberg arrays. 
      \textbf{(b) Graph reduction}: Using an iterative and deterministic reduction algorithm, subgraphs of the original graph can be removed, while maintaining optimality, leaving a graph kernel with one or more smaller components that can be solved individually. 
      After reduction, the largest component for the Cora graph consists of the graph shown with 9 nodes and 12 edges.  
      \textbf{(c) Compatibility check}: Using a fixed set of criteria (based on graph properties such as the maximum degree, number of triangles, etc.), the graph compatibility checker quickly assesses whether or not the reduced graph is (in principle) compatible with the restrictions imposed by the Rydberg quantum hardware.  
      \textbf{(d) Embedding}: Heuristic, random-key algorithms are used to find a hardware-efficient mapping from the logical graph in (b) to the Rydberg quantum hardware with unit-disk connectivity (with Rydberg atoms corresponding to the nodes of the graph). 
      Here we assume a simple, two-dimensional square lattice with Union-Jack (UJ)-like connectivity to nearest and next-nearest neighbors only, but generalizations are straightforward. 
      Our (bottom-up) embedding logic first optimizes the similarity between the logical input graph and the physical (embedded) graph, given the limited hardware connectivity. 
      For the example graph in (b) an optimized atom placement (shown as occupied sites in orange and red) can be found with an edit distance of one (only missing the edge between nodes 1 and 9), without incurring any overhead. 
      This overhead-free embedding can be further refined using additional ancilla atoms arranged into a quantum wire (shown in pink and purple), the placement of which is optimized with a breadth-first search algorithm solving a constrained shortest-path problem on the underlying square lattice between the start and end nodes marked in red. 
      Overall, the MIS problem on the original Cora graph has been mapped to the MIS problem on a hardware-native graph with 9 nodes, 11 UJ edges, and one quantum wire with 8 ancilla nodes.
      \label{fig:scheme}
  }
\end{figure*}

Here we complement the existing Rydberg embedding toolbox by providing a comprehensive, modular compilation pipeline that maps generic (potentially large-scale) instances of the MIS problem to (potentially much smaller) MIS problems on UD graphs, which are natively encodable with Rydberg atom arrays (see Fig.~\ref{fig:scheme} for a schematic illustration).  
The pipeline consists of three core modules, with the goal of providing approximate, low-overhead embeddings for Rydberg atom arrays that keep the number of required qubits small and controllable.
We now provide a brief description of these modules. 

\textbf{Reducer.} Our first module, the \textit{reducer}, takes a generic graph $\mathcal{G}=(\mathcal{V}, \mathcal{E})$, with node set $\mathcal{V}$ and edge set $\mathcal{E}$, as problem input and provides a smaller (\textit{core}, or \textit{kernel}) graph (with one or more components) as output. These output graph components can then be treated independently, thereby distilling the most important logical variables from the original problem definition, while minimizing the eventual workload for the quantum device. 
Specifically, we implement a procedure known as isolated clique removal \citep{butenko:02, butenko:07, strash:16, hespe:19, lamm:19}. 
Just like peeling an onion, this scheme proceeds by iteratively removing sub-graphs of the input graph (such as isolated nodes, dangling bonds, and larger isolated cliques) while maintaining optimality, until no further reduction can be achieved. 
Optimal solutions for the reduced (core) instance can then be easily expanded to an optimal solution for the original (potentially much larger) instance by reversing previously applied reductions.

For illustration, consider a simple dangling bond (i.e., a clique of size two) involving an exposed (degree-1) node $u$ connected to the remainder of the larger graph via an edge $(u,v)$ to node $v$ only. 
It is easy to see that node $u$ has to be included in at least a subset of all (degenerate) MIS solutions; leaving both node $u$ as well as node $v$ out of the set is clearly suboptimal, and at least one of them should be marked.   
By deterministically selecting the exposed node $u$ and removing the entire clique from the graph (in accordance with the independence constraint), our reduction logic is minimally invasive to the remainder of the graph, 
ensuring that at least one optimal MIS solution can be found, while systematically shrinking the problem size. 
The underlying clique checks are performed \textit{lazily}, because the sub-graph removals may require a do-over. 
The scheme then starts over, applying the same simplification logic to the reduced graph where nodes $u$ and $v$ have now been removed, with the exposed node $u$ tracked as member of the set. 

We provide systematic numerical experiments, testing the performance of this reduction technique for both synthetic and real-world graphs, with up to tens of thousands of nodes. 
In particular, we find that structured, real-world graphs can often be reduced by orders of magnitude, in some cases even down to an empty null graph (in which case the reducer amounts to an MIS solver), typically in linear time (as demonstrated below for both synthetic and real-world graphs). 
For example, we find that the well-known Cora \citep{cora:00} and Pubmed \citep{pubmed:12} citation graphs can be reduced from $\sim 2700$ and $\sim 2\times 10^4$ nodes to small core graphs with just nine and eleven nodes, respectively, on sub-second timescales on a laptop. 
We also discuss and analyze reduction in the context of the on-going search for potential quantum speedups, based on the reducer's ability to efficiently single-out hard problem instances. 
Specifically, for Rydberg-native MIS instances, we observe signatures of an easy-hard-easy transition and quantify the corresponding transition points.

\textbf{Checker.} Our second module, the \textit{compatibility checker}, takes a graph as an input and checks if the graph is, in principle, compatible with an overhead-free embedding on a Rydberg array. The checker can determine if a graph \textit{cannot} be natively embedded, and it outputs a Boolean flag to this effect. The flag constitutes a necessary but not sufficient condition for native encoding; a graph may pass the compatibility check but may not be natively encoded in hardware, but a graph that fails the check \textit{cannot} be encoded without overhead. Thus, the checker provides information that can be utilized in any down-stream embedding scheme. 
For concreteness, we will focus on UD graphs compatible with Rydberg atoms arranged on a two-dimensional square lattice with nearest and next-nearest (diagonal) couplings, generating a Union-Jack (UJ)-like connectivity pattern [cf. Fig.~\ref{fig:scheme}(d)], as realized experimentally in, e.g.,  Ref.~\cite{ebadi:22} and currently available on Amazon Braket \citep{wurtz:23}. Generalizations to alternative geometries, however, should be straightforward.
For a simple illustration, consider the maximum degree of a given graph. 
As evident from Fig.~\ref{fig:scheme}(d), the UJ Rydberg hardware can only accommodate graphs with maximum degree $d_{\mathrm{max}}\leq8$. 
Therefore, for any input graph with $d_{\mathrm{max}}>8$, the compatibility checker will raise a flag, classifying the input graph as a non-native graph that will require ancilla qubits for any perfect embedding. 
Beyond the maximum degree, additional graph properties (such as the number of triangles) are evaluated in the compatibility assessment, providing a simple yet fine-grained (binary) classification for any given input graph.

\textbf{Embedder.} Our third module, the \textit{embedder}, takes a given (reduced) graph as input and provides an efficient embedding, with optimized overhead and ancilla wiring. 
Specifically, we design two complementary embedding strategies, including a (bottom-up) approach that \textit{learns} approximate embeddings with minimal resource requirements, as well as a (top-down) scheme that constructively generates exact embeddings with MIS optimality guarantees, while minimizing the required qubit overhead. 
See Fig.~\ref{fig:scheme}(d) for an illustration of our bottom-up embedder.       
Here, we design efficient assignment heuristics that \textit{learn} hardware-efficient embeddings by optimizing a given similarity metric (such as the edit-distance) between the (logical) input graph and the (physical) embedded graph, while adhering to the connectivity constraints imposed by the hardware. 
This hardware-efficient embedding can then be further improved by surgically placing vertex wires \citep{kim:22, nguyen:23} to generate long-distance edges. 
In our scheme, the placement of the latter is optimized with a breadth-first search algorithm solving a constrained shortest-path problem to efficiently connect distant nodes.  
Our second (top-down) embedding scheme builds on the generic (exact) embedding scheme outlined in Ref.~\citep{nguyen:23}, and systematically reduces the required overhead using a custom simulated annealing solver that identifies optimized qubit arrangements.

After reduction and embedding, quantum algorithms (e.g., those shown in Ref.~\citep{ebadi:22}) can be used to approximately solve the MIS on the embedded (physical) graph, with site-resolved projective measurements for readout of the final quantum many-body state involving both system and ancilla atoms. Simple post-processing can then distill the solution for the original (logical) input graph. 

\textbf{Modular integration.} The modular nature of our scheme allows for plug-and-play solutions, combining a subset of the building blocks detailed here with some of the existing schemes discussed above. 
For example, it should be straightforward to integrate the reduction logic proposed here with the embedding scheme described in Ref.~\citep{nguyen:23}. 
Finally, for an experimental realization, our scheme is tailored towards neutral atom experiments supporting two-dimensional atom trapping and the Rydberg blockade mechanism, as successfully demonstrated with hundreds of atoms in, e.g., Ref.~\citep{ebadi:22} and available in the cloud on Amazon Braket via QuEra's Aquila device \citep{wurtz:23}. These architectures have the potential to scale to systems with thousands of atoms in the near future \citep{gyger:24, manetsch:24}. 

\textbf{Structure}. The remainder of this paper is organized as follows. 
In Sec.~\ref{preliminaries}, we first formalize the problem we consider and provide some more background. 
Next, in Sec.~\ref{framework}, we describe in detail our larger pipeline with an in-depth discussion for all core modules. 
In Sec.~\ref{numerics}, we then assess the performance of our pipeline with a series of numerical experiments for both synthetic and real-world graphs. 
Finally, in Sec.~\ref{conclusion}, we draw conclusions and give an outlook on future directions of research.

\section{Preliminaries and problem specification}
\label{preliminaries}

Our work is motivated by recent advances in experiments involving neutral-atom quantum machines \citep{adams:20, henriet:20}, in particular in the field of quantum optimization \citep{abbas:23, ebadi:22}, and 
our main goal is to extend both the size and scope of problems compatible with these devices.   

\textbf{Rydberg atom arrays.} In Rydberg atom arrays, atomic qubits are (typically) defined by an internal ground state $|0\rangle_{i}$ and a highly excited, long-lived Rydberg state $|1\rangle_{i}$, where the label $i=1, \dots, n$ runs over the $n$ qubits. Within these two-dimensional subspaces, we define the usual Pauli matrices $\hat{\sigma}_{i}^{\alpha}$, where $\alpha=x,y,z$, and the corresponding lowering operators $\hat{\sigma}_{i}^{-} = |0\rangle_{i} \langle 1|$. The quantum dynamics of Rydberg atom arrays is then described by the Hamiltonian $\hat{H}=\hat{H}_{\mathrm{drive}}+\hat{H}_{\mathrm{cost}}$, with ($\hbar=1$)
\begin{equation}
\label{eq:rydberg-hamiltonian}
\begin{aligned}
    \hat{H}_{\mathrm{drive}} &= \frac{\Omega(t)}{2} \sum_i \left(e^{i\phi(t)} \hat{\sigma}_{i}^{-} + \mathrm{h.c.}\right),
    \\
    \hat{H}_{\mathrm{cost}} &=
    - \sum_i \Delta_{i}(t) \hat{n}_i
    +
    \sum_{i<j} V_{i j} \hat{n}_i \hat{n}_j,
\end{aligned}
\end{equation}
where $\hat{n}_i:=|1\rangle_{i} \langle 1|=(\mathds{1}_2+\sigma_{i}^{z})/2$ is the number operator counting the number of Rydberg excitations for atom $i$, $\Omega(t)$ is the (time-dependent) global Rabi frequency, $\phi(t)$ is the laser phase of the Rabi drive, $\Delta_{j}(t)$ is the laser detuning for atom $j$, and $V_{ij}$ is the static pairwise interaction potential between atoms $i$ and $j$. The interaction term is given by $V_{ij}=C_6/\|\mathbf{x}_{i}-\mathbf{x}_{j}\|^{6}$
for atoms positioned at $\mathbf{x}_{i}$ and $\mathbf{x}_{j}$, respectively, 
where $C_6$ is the (atom-species-dependent) van der Waals coefficient.   
Schedules (or drives) specified in terms of the  laser parameters $\{\Omega(t), \phi(t), \Delta_{i}(t)\}$ and atom arrangement $\{\mathbf{x}_{i}\}$ define the complete problem input for Rydberg-based analog quantum simulators available today with 256 qubits \cite{wurtz:23}.
Specifically, for a given (programmable) atom arrangement, one can then explore the quantum many-body dynamics generated by the Hamiltonian $\hat{H}$ by tuning (for example) quantum-annealing and/or bang-bang-type schedules \citep{pichler:18, ebadi:22, finzgar:23}. 

\textbf{Rydberg blockade.} The (induced) dipole-dipole interactions $V_{ij}$ in Eq.~(\ref{eq:rydberg-hamiltonian}) give rise to the so-called \textit{Rydberg blockade} mechanism \citep{lukin:01, saffman:10, levine:19}, 
whereby simultaneous excitation of nearby atoms is strongly suppressed when their spatial separation is smaller than the Rydberg blockade radius $R_b \equiv (C_{6}/\Omega_{\mathrm{max}})^{1/6}$, with $\Omega_{\mathrm{max}} = \mathrm{max}_{t} \Omega(t)$. Typically, $R_{b} \sim 1-10\mu \mathrm{m}$ \citep{adams:20}.
With the Rydberg blockade effectively preventing two neighboring atoms from being simultaneously in the excited Rydberg state (provided they are within the Rydberg blockade radius $R_b$), 
the dynamics of Rydberg atom arrays are natively restricted to coherent superpositions of independent sets \citep{pichler:18}, i.e., subsets of nodes in a graph
for which no two nodes within the set are adjacent, as pertinent to the MIS problem.  

\textbf{Rydberg-native MIS-UD problem.} The Rydberg blockade mechanism allows for a hardware-efficient encoding for the (hardware-native) MIS-UD problem (and the generalized maximum-weight independent set problem on UD graphs), with Rydberg atoms placed at the nodes of the problem (target) graph and Rydberg interactions inducing edges within the (tunable) unit-disk radius $R_d=R_b$ \citep{pichler:18, ebadi:22}.  
Specifically, setting $\Omega=0$ and carefully choosing $\Delta_{i}=\Delta>0$ in Eq.~(\ref{eq:rydberg-hamiltonian}), optimal solutions to the MIS-UD problem are encoded in the ground states of the Hamiltonian $\hat{H}$. 
In this classical limit, up to corrections from the Rydberg interaction tails, the Rydberg Hamiltonian effectively reduces to the classical MIS cost function 
\begin{equation}
H = -\sum_{i \in V} n_{i} + U \sum_{(i,j) \in E} n_{i}n_{j},
\label{eq:hamiltonian-mis}
\end{equation}
with $n_{i}=1$ if node $i=1, \dots, N$ belongs to the independent set, and $n_{i}=0$ otherwise.
By virtue of the Rydberg blockade mechanism, the problem in Eq.~(\ref{eq:hamiltonian-mis}) is defined on a \textit{physical} unit-disk graph $G=(V,E)$, with set of nodes $V$ and edges $E$, respectively. 
Energetically, this (classical) Hamiltonian favors having each variable in the state $n_{i}=1$ unless a pair of nodes is connected by an edge; for $U>1$, the ground state is guaranteed to be an optimal MIS solution, because it is strictly more favorable to have at most one vertex per edge in the set as opposed to both vertices being marked. 
As such, the Hamiltonian $H$ captures the MIS problem with membership in the MIS described by nodes marked as $n_{i}=1$.
As demonstrated in Refs.~\citep{ebadi:22} and \citep{finzgar:23}, one can then search for the MIS (encoded in the ground state of the Hamiltonian $H$) via, for
example, coherent quantum-annealing-type approaches using quantum tunneling induced by the quantum mixer $\hat{H}_{\mathrm{drive}}$. 
For atoms arranged on a two-dimensional square lattice with lattice spacing $a$ chosen such that $\sqrt{2}a \leq R_{d} <2a$ the UD connectivity results in a Union-Jack-like pattern (cf.~Fig.~\ref{fig:scheme} d), as experimentally studied (for example) in Refs.~\citep{ebadi:22, finzgar:23, kim:23}. If not stated otherwise, we will focus on this family of hardware-native instances, but generalizations should be straightforward. 

\textbf{Generic MIS problem.} Generically, graphs of interest for real-world applications will not necessarily be UD graphs and may feature system sizes (specified in terms of the number of \textit{logical} nodes) much larger than the number of qubits provided by near-term Rydberg devices. 
Here, we design and implement a set of tools that help expand the scope and size of problems compatible with near-term Rydberg atom arrays, allowing to solve (potentially large-scale) MIS instances on \textit{logical} graphs $\mathcal{G}=(\mathcal{V}, \mathcal{E})$ not restricted to the family of UD graphs.  

\textbf{Embedding overhead.} A comprehensive, exact embedding scheme for Rydberg atom arrays has recently been put forward in Ref.~\citep{nguyen:23}.
In particular, this work introduces a set of gadgets whereby logical variables are redundantly encoded in one-dimensional chains of ancilla atoms to effectively distribute logical information across distant locations and couple pairs of distant qubits, while adhering to the underlying unit-disk connectivity. 
Generically, however, this scheme comes with a quadratic overhead that requires $N \lesssim 4n(n-1)$ physical qubits (with $N=|V|$) for the faithful embedding of a logical problem with $n=|\mathcal{V}|$ variables. 
Accordingly, relatively small (logical) problems with just $n \sim 10$ variables (requiring $N\sim 360$ physical qubits) can already exceed the capabilities of today's devices with $\sim 250$ qubits \citep{wurtz:23}.  
To address this challenge, we propose and implement tools that effectively identify and remove redundant variables in the original problem, and we provide efficient embedding schemes with small and optimized overhead, thereby expanding the scope and size of MIS problems that can be tackled with near-term Rydberg atom arrays. 

\section{Theoretical Framework}
\label{framework}

In this section, we describe our larger quantum compilation pipeline for Rydberg atom arrays, as illustrated in Fig.~\ref{fig:scheme}. 
While our three core modules are detailed in the context of the full quantum compilation pipeline, we emphasize that they can be utilized independently, given their stand-alone, modular design.  

\subsection{Graph Reduction}

\begin{figure*}
  \hspace{-0.5mm}\includegraphics[width=2.07 \columnwidth]{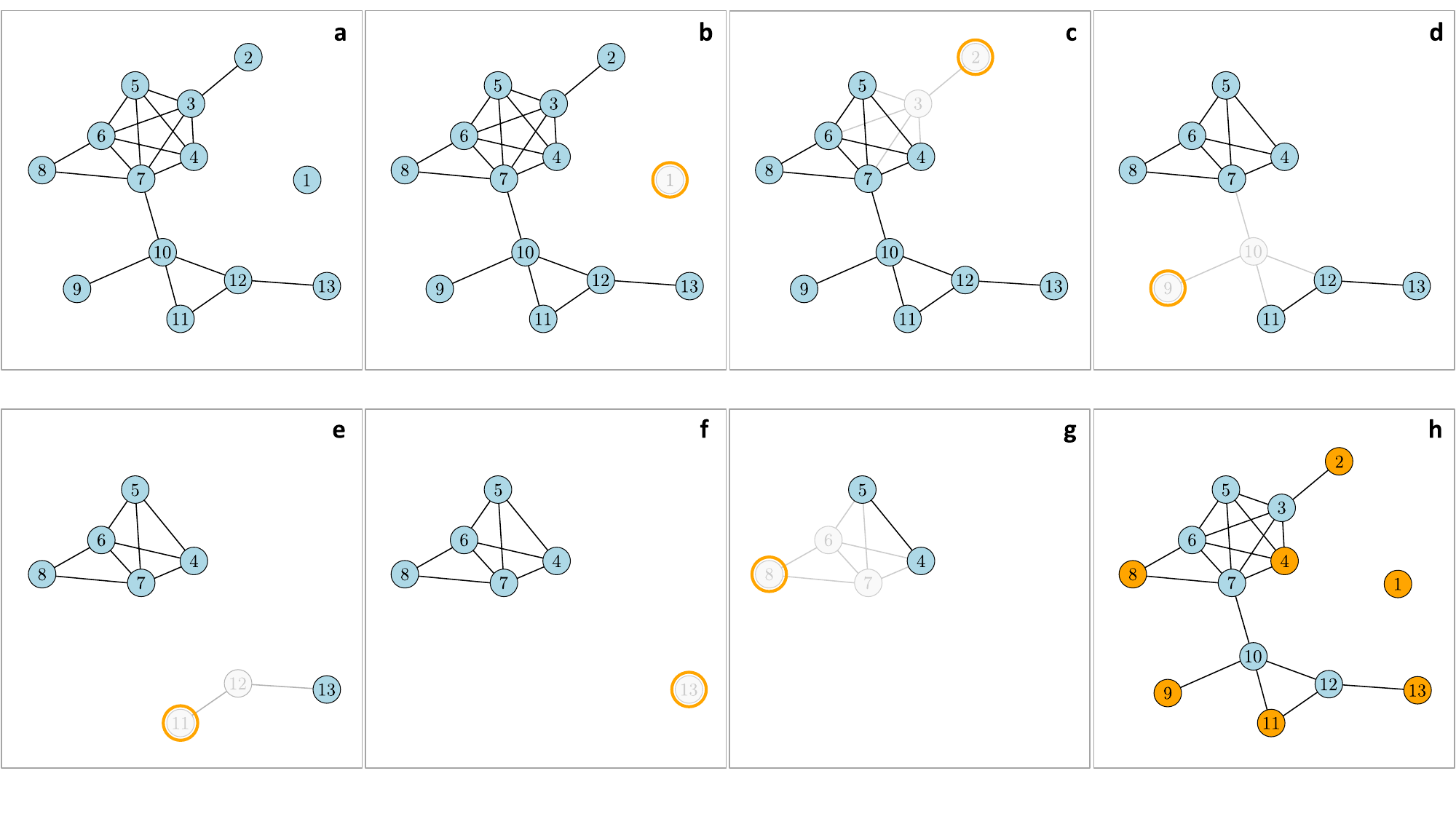}
  \vspace{-10mm}
  \caption{
    \textbf{(a)} Schematic illustration of our clique-based graph reduction scheme for
    an example input graph with $13$ nodes. 
    The set of \textit{exposed} nodes is given by $\{1, 2, 4, 5, 8, 9, 11, 13\}$. 
    Our reduction algorithm proceeds deterministically by identifying exposed nodes that can be added to the independent set and removed from the graph along with their neighborhood, in ascending order of node degree and node label, while preserving the ability to compute an optimal solution. 
    \textbf{(b)} First, any isolated nodes (i.e., cliques of size one, such as node 1 here) are added to the set and removed from the graph. 
    Next, as shown in steps \textbf{(c)} and \textbf{(d)}, dangling bonds (i.e., cliques of size two) are removed, adding nodes 2 and 9 to the set, respectively, while removing adjacent nodes 3 and 10 from the graph without selection, in accordance with the independence constraint. 
    \textbf{(e)} Node 11 is selected as the lowest degree node with smallest label, thereby removing adjacent node 12. 
    \textbf{(f)} Node 13, now isolated, has been added to the set. 
    \textbf{(g)} After selection of exposed node 8, we are left with a simple bond from which node 4 is selected. 
    \textbf{(h)} Because this graph can be reduced completely (amounting to a reduction factor of $\xi = 1$), the MIS with size $|\mathrm{MIS}|=7$ is fully determined by the set of nodes previously selected in the reduction routine (given here by the set $\{1, 2, 4, 8, 9, 11, 13\}$ highlighted in orange). 
    The degenerate MIS solution $\{1, 2, 5, 8, 9, 11, 13\}$ where node 4 is swapped for node 5 within the 5-clique subgraph can be found with the same logic after random re-labeling of the nodes, or (alternatively) via simple local search.  
      \label{fig:reduction}
  }
\end{figure*}
 
Reduction (or kernelization) techniques are known to play an integral part in state-of-the-art (SOTA) MIS solvers, for both efficient exact algorithms and heuristics~\citep{butenko:02, butenko:07, butenko:09, strash:16, chang:17, lamm:17, hespe:19, lamm:19}.
By selecting vertices that are provably part of some maximum(-weight) independent set and removing well-defined subgraphs according to relatively simple reduction rules, kernelization techniques are able to efficiently shrink the input graph to an irreducible kernel.
If an MIS has been found for the kernel, each reduction step can be undone, eventually providing a provably optimal MIS for the original graph, thanks to the optimality guarantees built into the reduction rules.  
For exact MIS solvers, leading approaches implement some form of branch-and-reduce search \citep{lamm:19}, where a suite of reductions intermixes branching and reduction to shrink the input size. 
For those (hard) instances that cannot be solved exactly, ReduMIS has emerged as one of the leading heuristic approaches \citep{lamm:17}. In ReduMIS, reduction techniques are integrated with a heuristic, evolutionary algorithm, with recent work suggesting potential further improvements via memetic algorithms combining genetic algorithms with local search \citep{grossmann:23}. 

To the best of our knowledge, classical reduction techniques have so far not been adopted for quantum computing purposes in the Rydberg community. 
Here, we provide a self-contained implementation of a simple yet efficient, lazy reduction algorithm known as isolated clique removal \citep{butenko:02, butenko:07, strash:16, hespe:19}.
Our contributions are the following: 
(i) We position reduction techniques as a new tool for Rydberg atom arrays, and for the larger QO community, with applications in quantum compilation, hybrid (quantum-classical) algorithms integrating classical reduction with quantum algorithms, and the search for quantum speed-ups.  
(ii) Beyond the MIS problem, we propose a simple generalization of our reduction logic towards other (NP-hard) optimization problems, such as maximum cut or graph coloring. 
(iii) We show that our lazy implementation of clique-based reduction has linear run-time scaling across a large set of instances, without any limit in clique size.   

\textbf{Reduction scheme.} Our classical graph reduction algorithm is illustrated in Fig.~\ref{fig:reduction}. The algorithm takes a generic graph $\mathcal{G}$ as input and, after reduction, outputs a kernel $\mathcal{K}=\mathrm{red}(\mathcal{G})$ of smaller or equal size (in polynomial time), 
thereby helping to identify the hard core of the MIS problem on graph $\mathcal{G}$, while stripping off the simple parts, 
yet preserving all information required to find an MIS for the input graph $\mathcal{G}$.

\textbf{Reduction factor.} The overall amount of reduction is measured in terms of the normalized reduction factor $\xi$ as
\begin{equation}
\xi = \frac{n(\mathcal{G})-n(\mathcal{K})}{n(\mathcal{G})},
\label{eq:reduction-factor}
\end{equation}
where $n=n(\mathcal{G})$ and $n(\mathcal{K})$ denotes the number of nodes of the original graph $\mathcal{G}$ and graph kernel $\mathcal{K}$, respectively. 
For $\xi=0$, no reduction is possible, and the reduced graph is identical to the input graph. For $\xi=1$, complete reduction has been achieved from the input graph down to the null graph, and a MIS solution has been found by mere reduction; cf.~Fig.~\ref{fig:reduction}.
For $0 < \xi <1$, we obtain partial reduction, as is the case for the reduced Cora graph displayed in Fig.~\ref{fig:scheme}(b).
Generically, the larger (smaller) the reduction factor $\xi$ is, the easier (harder) the given problem instance is. 

\textbf{Exposed corner nodes.} Our clique-based reduction scheme is based on the simple observation that many real-world graphs feature some community-like structure in the form of cliques (i.e., complete subgraphs where every two distinct nodes in the clique are adjacent) that, by definition, can feature at most one member in the MIS. For example, see the 2-clique, 3-clique and 5-clique spanned by nodes $\{2, 3\}$,  $\{10, 11, 12\}$, and $\{3, 4, 5, 6, 7\}$, respectively, in Fig.~\ref{fig:reduction}.
Our algorithm recursively identifies and removes certain cliques, importantly without having to solve the NP-hard maximum clique problem. 
To this end we introduce the notion of \textit{exposed} nodes (or, equivalently, \textit{corner} nodes) to identify nodes that \textit{only} have connections within their host clique.
In the literature, those nodes are also known as simplicial vertices or isolated vertices \citep{butenko:02, butenko:07, strash:16, hespe:19,  lamm:19}, and
corner nodes of degree one are also referred to as pendant vertices \citep{lamm:17}.
For example, for the graph shown in Fig.~\ref{fig:reduction} all nodes in $\{1, 2, 4, 5, 8, 9, 11, 13\}$ qualify as exposed nodes. 
Conversely, node $7$ is not a corner node; given its degree, it is a candidate for a 7-clique but its neighborhood lacks the required complete connectivity. 
Viewed differently, node $7$ is part of the 5-clique $\{3, 4, 5, 6, 7\}$, but, given its outside connections $(7,8)$ and $(7, 10)$ it does not qualify as a corner node. 
Recursive selection of corner nodes only then maximizes the number of marked vertices within a given clique without blocking the selection of nodes outside of this clique. 
For example, selection of node $11$ in Fig.~\ref{fig:reduction} maximizes the number of selected nodes in the underlying clique without blocking node selections outside of this clique, 
whereas a (sub-optimal) selection of non-corner node $7$ would not only block the underlying 5-clique, but would also block nodes $8$ and $10$ from further addition to the independent set.  
 
\begin{figure}
  \includegraphics[width=1.0 \columnwidth]{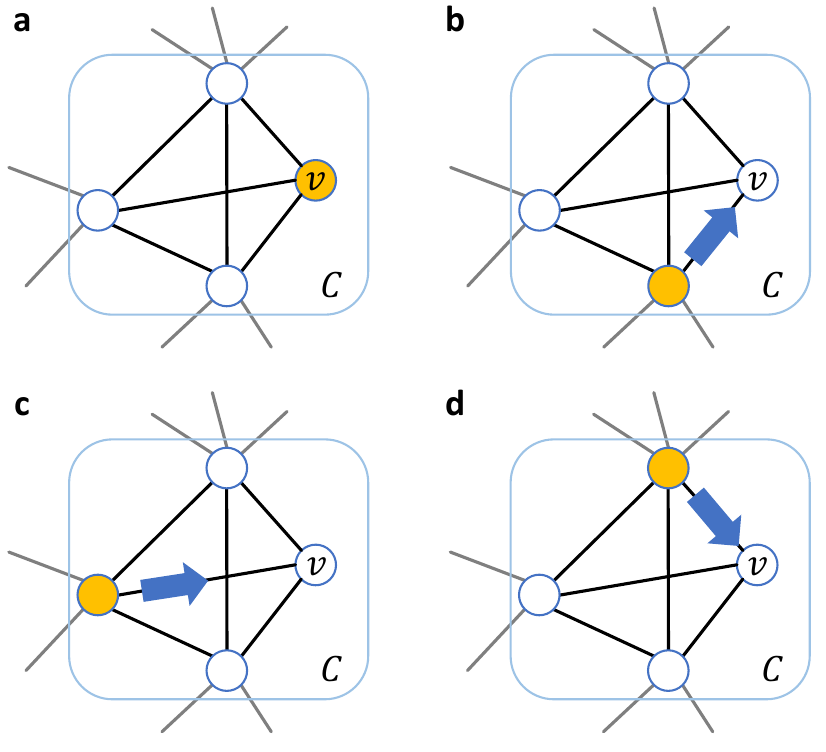}
  \caption{Schematic illustration of the cut-and-paste argument underlying the provably optimal selection of corner nodes.
  Node $v$ is part of a 4-vertex clique $C$. Because $v$ has no neighbors outside of the clique $C$, it is an (exposed) corner node that must be part of some MIS. 
  At most one node $u \in C$ can be selected. 
  If $u \neq v$, one can always swap (i.e., cut and paste) the selection of $u$ for the selection of $v$, without causing any constraint violations while preserving the independent set size.  
  Therefore, per reduction logic we opt to add $v$ to the MIS, and remove $v$ and its neighborhood $\mathcal{N}(v)=\{u | (v,u)\in\mathcal{E}\}$ from the graph.
      \label{fig:reduction-cut-and-paste}
  }
\end{figure}

\textbf{Optimality.} This repeated selection of corner nodes is \textit{optimal} because those nodes are provably always part of \textit{some} MIS, as follows from a simple cut-and-paste argument \citep{butenko:09, lamm:17, hespe:19, lamm:19}; cf.~Fig.~\ref{fig:reduction-cut-and-paste} for a schematic illustration. To see this, we note that at most one node from a given clique can be part of any MIS. Either it is the corner node $v$, or, if a neighbor of $v$ is in an MIS, then simple reduction opts to select $v$ instead (while preserving the independent set size).  

\textbf{Reduction algorithm.} Formally, our implementation of isolated clique removal proceeds by considering each node as a potential \textit{corner} node, for which there are only $n$ candidates in total, by starting with low-degree nodes and working in order of increasing node degree. The algorithm proceeds in this way because smaller neighborhoods are cheaper to check and remove, with the added benefit of potentially shrinking more expensive, larger cliques prior to the neighborhood completeness check. 
For example, in Fig.~\ref{fig:reduction}, the largest 5-clique was never evaluated, simply because smaller adjacent cliques had previously been removed (by selecting the exposed nodes $2$ and $8$, respectively, while removing their neighborhoods). 
In the case of tie-breaks between exposed nodes of the same degree, we choose to remove by node label in ascending order, noting that node removals can expose more corners, but cannot make any corner node irremovable, independent of the removal ordering.

The run time for clique-based reduction is polynomial, rather than exponential as in the case of the maximum clique problem. The reason for this scaling is that we are only looking for exposed cliques (i.e., those where at least one node has minimal degree of $k-1$ within a $k$-clique for $k=1,2,\ldots$), rather than searching over all cliques. 
For every node, we only need to check if all of its neighbors form a $k-1$ clique. As such, (potentially maximum) cliques without exposure are not detected.  
Given that there are $n$ corner candidates and that the cost of clique checks are upper-bounded by $d_{\mathrm{max}}^2$ (where $d_{\mathrm{max}}$ refers to the maximum degree) we obtain a rough upper bound on the algorithmic run-time $T_{\mathrm{red}}$ given by $T_{\mathrm{red}} \lesssim d_\mathrm{max}^2 n$, in the worst case (when $d_{\mathrm{max}} \sim n$) resulting in a run time given by $\mathcal{O}(n^3)$. 
However, for \textit{sparse} instances with bounded maximum degree (i.e., $d_{\mathrm{max}} = \mathrm{const.}$) we obtain a \textit{linear} run-time scaling. 
In particular, we find that the average run time $\bar{T}_{\mathrm{red}}$ can be approximated well by a linear scaling as 
\begin{equation}
\bar{T}_{\mathrm{red}} \approx \gamma(\bar{d})n,
\label{eq:reduction-runtime}
\end{equation}
with a pre-factor $\gamma(\bar{d})$ that roughly captures the structure of the instance via the average degree $\bar{d}$. 
As demonstrated in Sec.~\ref{reduction-hardness} and Sec.~\ref{numerics}, we have empirically verified Eq.~(\ref{eq:reduction-runtime}) for both synthetic and real-world graphs. 

By only selecting corner nodes, our reduction algorithm is able to preserve the MIS size (also referred to as independence number) but cannot generate all possible (degenerate) MIS solutions.   
However, additional MIS solutions not found by clique-based reduction can potentially be generated via simple local search starting from clique-based solutions. 
This is done by swapping selected corner nodes for neighbors when possible; for example, in Fig.~\ref{fig:reduction}(h) one can swap selection of node $4$ for the selection of node $5$ without introducing any constraint violation, thereby generating the MIS solution $\mathrm{MIS} = \{1, 2, 5, 8, 9, 11, 13\}$.
Alternatively, one can use randomization techniques as described below. 

\textbf{Node elimination via exact splitting.} Our clique-based reduction algorithm terminates once no further corner nodes can be found;
see, for example, the reduced core graph shown in Fig.~\ref{fig:scheme}(b), which does not feature any corner nodes, and, as such, cannot be reduced further via clique removals. 
In these situations, further simplification may be achieved via complementary reduction schemes.
For example, one can perform exact node elimination of $K$ nodes via splitting (branching) into $2^K$ branches to track and evaluate a bounded number of possibilities.
Heuristically, one may choose to split on high-degree nodes, as those splits provide the largest footprint, by eliminating a large number of nodes and edges. 
We showcase this scheme with a concrete example below in Section \ref{reduction-hardness}, whereby one node is eliminated from an irreducible core in order to unblock the clique-based reduction scheme, which can then continue to iterate with its reduction logic in another round of clique-based reduction via removal of corner nodes. 
In our numerical experiments, we study and quantify the reduction already possible with only the simple clique-based scheme, leaving potential extensions (such as node splitting) and integration with alternative, existing graph reduction or sparsification schemes (as discussed for example in Refs.~\citep{liu:18, hashemi:24, lewis:17, glover:18b, narimani:17, djidjev:20}) to future research.

\textbf{Randomization.} Our scheme is designed to be deterministic by recursively identifying and removing exposed nodes and their neighborhoods, in ascending order of node degree and (arbitrary) node label. 
To generate a larger pool of solutions (if those exist), alternative solutions can potentially be found as needed by simply randomly re-shuffling the node labels. 
Similarly, randomization of our scheme could be implemented by randomly picking a node in the case of tie-breaks between nodes with the same degree. 

\textbf{Reduction of weighted problems.} Our reduction scheme can also be generalized towards the maximum-weight independent set (MWIS) problem with node weights $\{ w_{i}\}$, e.g., in the form of simple reduction heuristics. 
Specifically, for instances where the distribution of node weights $\{ w_{i}\}$ is relatively narrow, the graph reduction logic presented here should provide good (heuristic) results that may be analyzed statistically with randomization schemes as discussed above. 
Conversely, if the distribution of node weights $\{ w_{i}\}$ is broad, simple (hybrid) greedy heuristics that first add the largest-weight nodes are expected to provide good reduction results. 
For details on SOTA reduction techniques for the MWIS problem we refer to Refs.~\citep{lamm:19} and \citep{grossmann:23}.

\textbf{Problem hardness.} The clique-based reduction scheme presented here is conceptually simple and fast, as demonstrated numerically in Sec.~\ref{numerics}.  
Overall, we think that classical reduction schemes are of interest to near-term quantum devices for at least two reasons: 
First, reduction techniques help minimize the eventual load for quantum devices, thereby effectively giving quantum devices access to problem instances otherwise outside of their hardware capabilities.  
In particular, this is important in the near term where the number of available qubits is still in the range of hundreds to thousands. 
As such, integration of reduction techniques is poised to play a pivotal role in the design of new hybrid (quantum-classical) algorithms. 
Second, with their ability to efficiently distill a problem down to its core, reduction techniques can be an important tool in the on-going search for classically hard problem instances, for which a potential quantum speed-up is more likely to occur \citep{ebadi:22, schiffer:23, andrist:23, kim:23}, thereby furthering the quantum community's ongoing efforts towards quantum advantage. 
To further qualify this hypothesis, we apply our reduction logic to a family of random (Rydberg-native) UJ instances in Sec.~\ref{reduction-hardness}.

\textbf{Generalization.} The reduction scheme introduced here has been applied to the MIS problem. 
However, similar considerations can be applied to other (NP-hard) combinatorial optimization problems often discussed in the quantum optimization community as benchmark problems, such as maximum cut (MaxCut) or the graph coloring problem (GCP).  
For example, just as done above for both MaxCut and the GCP, dangling bonds (exposed corner nodes with degree one) can be removed recursively and labeled optimally in simple post-processing routines by mere neighborhood association, given the assignment for the reduced core graph.
In particular, such reduction schemes could find applications when dealing with (structured) real-world networks that often feature (approximately) power-law degree distributions, with the majority of nodes having a low degree (e.g., one, two, or three) and a few highly connected hubs (e.g., large hub airports in transportation networks) \citep{newman:10}. 
 
\subsection{Hardware Compatibility Checks}

\begin{figure}
  \includegraphics[width=1.0 \columnwidth]{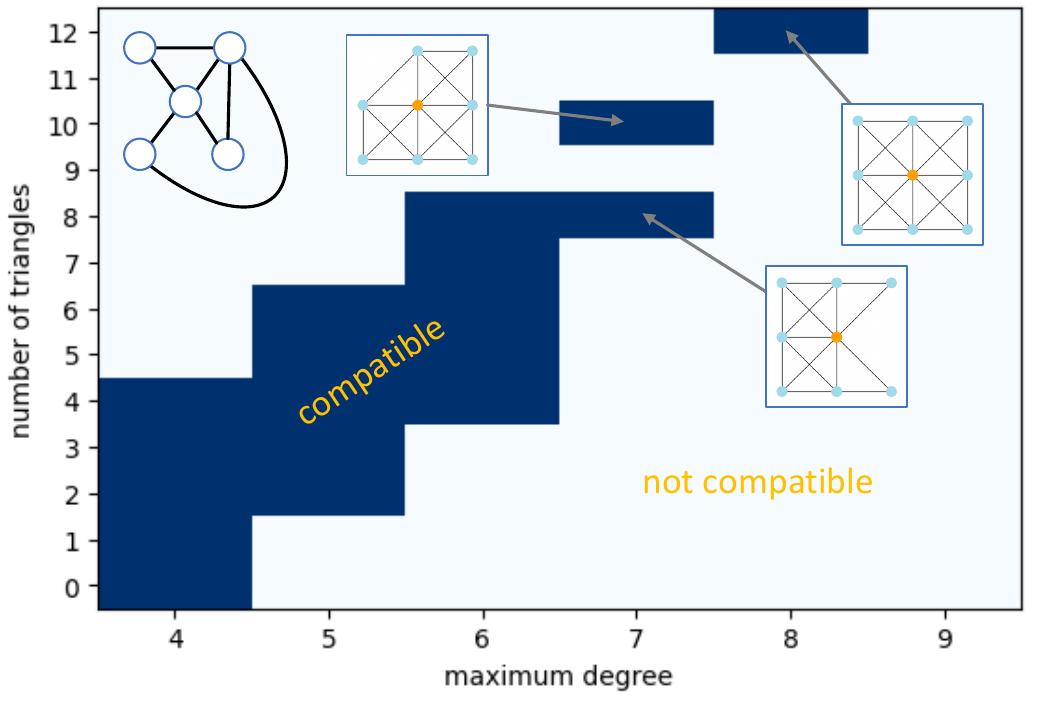}
  \caption{Hardware compatibility diagram. 
  Graphs are classified according to their maximum degree $d_{\mathrm{max}}$ and the number of triangles adjacent to maximum degree nodes $n_{\triangle}$. 
  Regimes not compatible with the Rydberg hardware are shaded light, while candidates for a native embedding are shaded dark. 
  Example UJ-native graphs are shown for $d_{\mathrm{max}}=7$, $8$, together with corresponding number of triangles.  The compatibility check provides a necessary, but not sufficient condition, as illustrated by the simple non-native graph with $d_{\mathrm{max}}=4$ and commensurate $n_{\triangle}$ shown in the upper left-hand corner. 
      \label{fig:compatibility}
  }
\end{figure}

We now describe in detail the Rydberg hardware compatibility checker (HCC). 
The HCC logic performs a series of $l \geq 1$ simple checks to determine if a given graph can be encoded natively with Rydberg hardware or not. 
For concreteness, following experimental demonstrations in Refs.~\citep{ebadi:22}, \citep{finzgar:23}, and \citep{kim:23}, we focus on setups where Rydberg atoms are placed on a two-dimensional square lattice with nearest and next-nearest interactions, resulting in a Union-Jack-like (UJ) connectivity with vacancies, as illustrated in Fig.~\ref{fig:scheme}(c).  

\begin{figure}
  \includegraphics[width=1.0 \columnwidth]{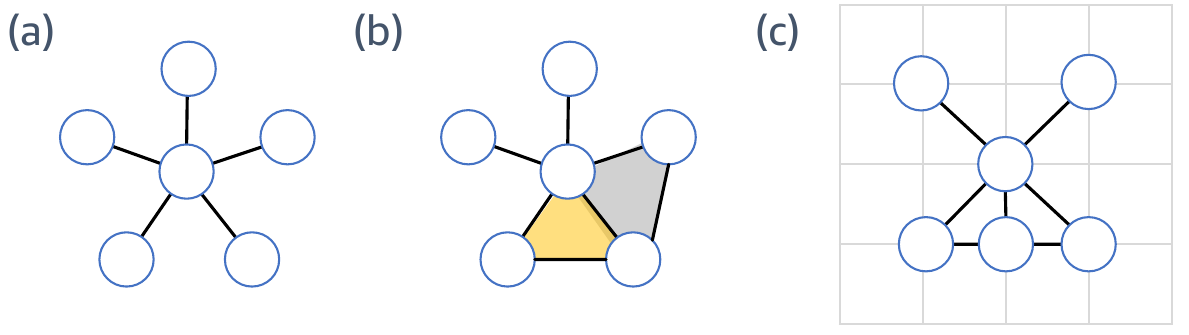}
  \caption{Example for simple hardware compatibility checks.
  \textbf{(a)} Simple star graphs with $d_{\mathrm{max}}\geq5$ are not UJ-compatible, because triangle-free graphs are supported natively only up to $d_{\mathrm{max}}=4$. 
  \textbf{(b)} Similar, but denser graphs with $d_{\mathrm{max}}=5$ and $n_{\triangle}=2$ are UJ-compatible, as shown in panel \textbf{(c)}.
      \label{fig:cc_example}
  }
\end{figure}

Simple ($l=2$) compatibility checks can be performed based on the diagram shown in Fig.~\ref{fig:compatibility}.
Here, a given graph is classified according to its maximum degree $d_{\mathrm{max}}$ and the number of \textit{triangles} (consisting of three vertices and three edges, in the form of a triangle) adjacent to the maximum degree nodes $n_{\triangle}$. 
Accordingly, graphs with maximum degree $d_{\mathrm{max}}>8$ \textit{cannot} be supported natively by Rydberg arrays with UJ connectivity, 
while graphs with $d_{\mathrm{max}} \leq 8$ \textit{may} be compatible only if the number of adjacent of triangles $n_{\triangle}$ is commensurate with the restricted UJ connectivity. 
For example, as shown in Fig.~\ref{fig:compatibility} graphs with $d_{\mathrm{max}}=8$ can only be UJ-compatible if $n_{\triangle}=12$, while 
graphs with $d_{\mathrm{max}}=7$ can only be UJ-compatible if $n_{\triangle}=8$ or $n_{\triangle}=10$ (depending on whether a boundary or corner node is missing with respect to the unit-filling case). 
Accordingly, a graph with $d_{\mathrm{max}}=7$ and $n_{\triangle}=9$ is not compatible with the UJ connectivity and will require ancilla qubits for a faithful embedding.  
We emphasize that the maximum degree (or, equivalently, \textit{local} density) is not a good indicator for native embeddability per se. 
For illustration, see Fig.~\ref{fig:cc_example} showing that a simple star graph with $d_{\mathrm{max}}=5$ is not UJ-compatible, 
while a similar graph with higher density and $n_{\triangle}=2$ can be embedded natively with UJ connectivity.   

The simple HCC outlined above provides a necessary, but not sufficient criterion for native embeddability. 
For example, the $d_{\mathrm{max}}=4$ graph shown in the inset of Fig.~\ref{fig:compatibility} passes the check but cannot be embedded natively. 
Such false positives could be caught with extended, more fine-grained checks involving, for example, other graph properties (such as the number of chordless cycles) in a deeper ($l>2$) catalog of checks.  

Finally, we note that the classification outlined here does not have to be aggregated on the graph level, 
but may be performed locally on the node level.  
Specifically, given its properties (such as degree and number of adjacent triangles) every single node may be classified as (potentially) hardware compatible or not, thereby providing local information that can be utilized in down-stream embedding logic. 
For example, with the central ($d_{\mathrm{max}}=5$) node in Fig.~\ref{fig:cc_example}(a) classified as non-compatible, one could ``roll-out'' this node using a simple copy gadget \citep{nguyen:23} to effectively support the required connectivity with the help of ancilla qubits. 

\subsection{Graph Embedding}

We now outline two complementary embedding strategies for Rydberg atom arrays. 
Our first (bottom-up) approach targets \textit{approximate} embeddings with minimal resource requirements, 
while our second (top-down) scheme provides \textit{exact} embeddings with optimality guarantees, at the expense of a larger, yet optimized overhead. 

\subsubsection{Bottom-Up Embedding Scheme}

We first describe an approximate, hardware-efficient (bottom-up) embedding heuristic we refer to as \textit{GAGE} (Generative Approximate Graph Embedder), in which optimized atomic (vertex) positions and the corresponding physical graph are learned via iterative training. 
Our scheme makes use of the random-key formalism to efficiently identify a physical (hardware-native) graph $G$ that is optimized to be as similar as possible to the target (input) graph $\mathcal{G}$, according to some desired metric.
In the following, we first briefly review the random-key formalism, and then we outline its application to the embedding problem at hand. 

\textbf{Random key optimizer (RKO).} A \textit{random key} is a real number in the continuous interval $[0,1)$, and 
a vector $\chi$ of random keys is an array $\chi \in [0, 1)^D$.
The RKO formalism is based on the idea that solutions to optimization problems can be encoded as vectors of random keys, providing an abstract (problem-independent) embedding for solution candidates in the latent (hypercube) space $[0, 1)^D$, with the encoding dimension $D$ chosen by the user. 
Such a vector $\chi$ is mapped to a feasible solution of the optimization problem with the help of a deterministic (problem-dependent) \textit{decoder} that takes as input a vector of random keys and returns a feasible solution $\mathcal{D}(\chi)$ to the optimization problem, as well as the cost of the solution. 
Based on a clear separation of problem-independent and problem-dependent modules, given by the encoding space $\mathcal{H}_{D}=[0, 1)^D$ and the decoder $\mathcal{D}(\cdot)$, respectively, RKO provides standardized plug-ins for a plethora of optimization paradigms \citep{chaves:24rko}, such as genetic algorithms \citep{londe:24a, londe:24b}, annealing-type algorithms \citep{schuetz:22}, or greedy randomized adaptive search procedures \citep{chaves:24}, among others, all within one unified framework. 
For more details on the random key formalism and RKO we refer to Refs.~\citep{londe:24a}, \citep{londe:24b}, and references therein. 

\begin{figure}
  \includegraphics[width=1.0 \columnwidth]{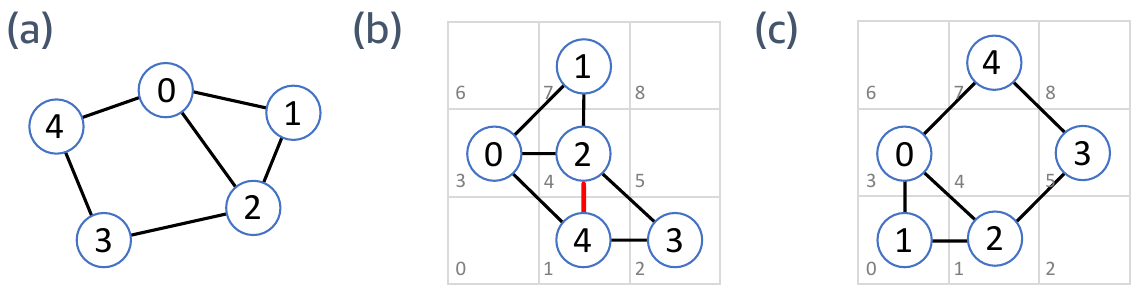}
  \caption{Example for the RKO-based generative approximate graph embedder (GAGE), for an underlying square lattice with $N_{\mathrm{sites}}=9$ sites labeled as $i= 0, 1, \dots, 8$.
  \textbf{(a)} Input graph $\mathcal{G}$ with five nodes.  
  \textbf{(b)} Approximate (hardware-native) embedding $G$ with edit distance $d_{\mathrm{edit}}=1$. 
  This node placement corresponds to the example random-key vector given in Eq.~(\ref{eq:random-key}).
  Edges are drawn for nearest and next-nearest neighbors corresponding to native UJ connectivity, here resulting in one undesired edge $(2, 4)$ highlighted in bold red. 
  Note that this embedding, even though approximate, supports MIS solutions such as $\{0, 3\}$ or $\{1, 4\}$ that are optimal solutions for the original input graph $\mathcal{G}$. 
  \textbf{(c)} Perfect (hardware-native) embedding $G$ with edit distance $d_{\mathrm{edit}}=0$, 
  as encoded by the example random-key vector $\chi =(0.16, 0.23, 0.71, 0.05, 0.62, 0.29, 0.79, 0.47, 0.98)$. 
      \label{fig:rko_example}
  }
\end{figure}

\textbf{RKO embedder.} Here we utilize RKO to generate a physical graph $G$ with adjacency matrix $A$ that is similar to the logical graph $\mathcal{G}$ with adjacency matrix $\mathcal{A}$, yet easy to implement on Rydberg atom arrays with native connectivity constraints. 
We focus on embeddings within two-dimensional square lattices with lattice spacing $a$, as realized experimentally in Ref.~\citep{ebadi:22}; generalizations to other settings (such as those with continuous atomic positions) should be straightforward.  
Our decoder takes a vector $\chi$ of $N_{\mathrm{sites}}=L_{x}\cdot L_{y}$ random keys as input (where $N_{\mathrm{sites}}$ is the number of available atomic positions in the underlying lattice), and sorts the keys in increasing order. 
Similar to canonical optimization problems, such as the traveling salesperson problem or vehicle routing \citep{schuetz:22}, the indices of this sorted vector make up the solution, where $n<N_{\mathrm{sites}}$ atoms are placed according to the first $n$ indices associated with the sorted vector. 
This design ensures placement of at most one atom per available site, as desired, while also adhering to placements within the constrained space with width $L_x$ and height $L_y$. 
For a given placement, one can then easily extract the corresponding graph $G$ for a given blockade radius $r=R_{b}/a$. 
The cost (or, equivalently, fitness) associated with such a placement can be calculated via graph similarity measures, such as the standard graph edit distance, e.g., 
\begin{equation}
d_{\mathrm{edit}} = \frac{1}{2} \sum_{ij} (\mathcal{A}_{ij} - A_{ij})^2,
\label{eq:rko-cost}
\end{equation}
which is then fed back to the optimizer as a feedback signal to be optimized over. 
RKO then learns a hardware-native embedding, guided by this similarity measure as a feedback signal, by traversing the random-key space in the form of, e.g., genetic \citep{londe:24a, londe:24b} or annealing-based \citep{schuetz:22} updates. 
Upon completion, after a series of training steps, RKO outputs an approximate graph $G$ of low cost that is as similar as possible to the input graph $\mathcal{G}$, given the hardware connectivity constraints. 

Let us illustrate the GAGE embedding scheme as described above with a simple example; cf.~Fig.~\ref{fig:rko_example} for illustration. 
Consider a simple input graph $\mathcal{G}$ with five nodes as illustrated in Fig.~\ref{fig:rko_example}(a). 
We search for a faithful embedding of $\mathcal{G}$ on a square lattice with $L_x=L_y=3$ and Union-Jack connectivity ($\sqrt{2} \leq r < 2$). 
Let us assume, for example, that the current solution candidate is given by the random-key vector 
\begin{equation}
\chi = (0.84, 0.34, 0.27, 0.07, 0.18, 0.42, 0.71, 0.13, 0.54).
\label{eq:random-key}
\end{equation}
The decoder takes this vector $\chi$ and applies simple sorting logic (in ascending order), resulting in $\mathrm{sort}(\chi)=(0.07, 0.13, \dots, 0.84)$. 
The corresponding vector of indices is given by $\mathrm{argsort}(\chi)=(3, 7, 4, 2, 1, 5, 8, 6, 0)$.
Accordingly, the first node is placed at position 3, the second one at position 7, and so on until all five nodes are placed, leaving sites $(5, 8, 6, 0)$ unoccupied. 
Assuming hardware-native UJ connectivity, we then readily obtain the corresponding candidate graph shown in Fig.~\ref{fig:rko_example}(b). 
This is an example of a low-cost (high-quality) solution candidate $G$ with Union-Jack connectivity, with edit distance $d_{\mathrm{edit}}=1$. 
Further (heuristic) optimization in additional training steps and re-starts will search for even better solutions, with the potential to ultimately identify the best possible Union-Jack embedding, as shown in Fig.~\ref{fig:rko_example}(c) for a perfect embedding with $d_{\mathrm{edit}}=0$.   

\textbf{Embedding refinement.} In the simple decoder design outlined above, we systematically search for the best possible hardware-native embedding with zero qubit overhead.  
The quality of the embedding (measured, e.g., in terms of graph similarity) may, however, be improved with the help of additional ancilla qubits, using (for example) the gadgets described in Ref.~\citep{nguyen:23}. 
Specifically, within an extended decoder design, one might also be able to learn the optimized placement of ancilla atoms, as encoded in additional ancilla keys.  
Alternatively, we can take the (approximate) overhead-free embedding solution as generated by GAGE and try to refine this embedding via optimized placement of ancilla quantum wires (QWs); see Fig.~\ref{fig:scheme}(d) for an example illustration.  
To this end, we introduce the notion of \textit{QW openings}, which refer to vacant sites whose neighborhood is empty except for a connection to the original node. In Fig.~\ref{fig:scheme}(d), we show an example in which potential QW openings have been highlighted in green, and the QW openings for nodes $1$ and $9$ have been used to effectively couple those through an even numbered ancilla chain at the expense of eight additional qubits.  
This ancilla chain effectively acts as an edge that is missing in the zero-overhead embedding, thereby improving the embedding quality from $d_{\mathrm{edit}}=1$ down to $d_{\mathrm{edit}}=0$.
To systematically find the placement of QW wires, we frame the ancilla placement problem as a constrained shortest path problem (SPP) within free (i.e., available) sites connecting two desired QW openings.
To solve this SPP, we use a breadth first search (BFS) algorithm that preserves cardinality (by restricting ancilla wires to be even numbered for proper redundant encoding) and avoids sharp turns via direction tracking(so as to not introduce undesired edges within the wire).  
 
\textbf{Characteristics and limitations.} As exemplified in our experiment underlying Fig.~\ref{fig:scheme}(c), our RKO-based embedder, in conjunction with potential refinement through quantum wires, can learn and identify (typically approximate) low-overhead embeddings. 
In particular, the zero-overhead embedding found via RKO can be used as a another tool to assess the hardware \textit{friendlessness} of a given input graph, as quantified by the edit distance $d_{\mathrm{edit}}$. 
While the example graph displayed in Fig.~\ref{fig:scheme}(b) is found to be hardware-friendly, further numerical experiments have shown that generic input graphs are typically not hardware-friendly (as expressed by a large edit distance), i.e., not even approximately compatible with the limited UJ hardware connectivity. 
To address this challenge, we design and implement a complementary top-down embedding scheme described below. 

\subsubsection{Top-Down Embedding Scheme}

\begin{figure}
    \includegraphics[width=1.0 \columnwidth]{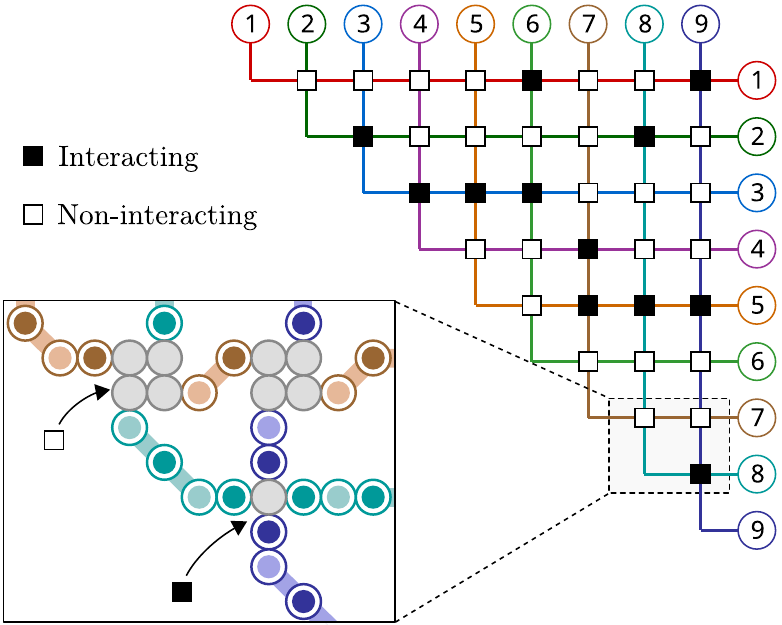}
    \caption{
        Representation of the generic embedding scheme from Ref.~\citep{nguyen:23} as a wire
        representation (main panel), with individual qubits forming intersection
        gadgets (inset). Filled squares represent interacting crossings, while
        empty squares are non-interacting gadgets. The graph shown here corresponds
        to the logical graph in Fig.~\ref{fig:scheme}(b).
    \label{fig:top-down-full}
    }
    \vspace{-1mm}
\end{figure}

Our second embedding approach builds on the generic embedding scheme outlined in Ref.~\citep{nguyen:23}.
This scheme is designed to provide a (physical, hardware-native) embedding for any potential (logical) input graph, regardless of its size,
edge density or specific interactions, with an exact one-to-one correspondence between the ground states of the logical and embedded graphs.
The recipe calls for the representation of all logical nodes in the form of extended qubit chains (i.e., copy gadgets),
arranged such that all pairs intersect at exactly one crossing. Edges between logical nodes are implemented by selecting an ``interacting'' intersection gadget for the crossing, while crossings of node-chains without an edge in the logical graph use the ``non-interacting'' crossing gadget (see Fig.~\ref{fig:top-down-full}).
We employ a version with odd-length chains and detunings adjusted accordingly.
For a graph with $n$ nodes and $m$ edges, this scheme then results in an embedding with the following qubit count:
\begin{equation}
    \label{eq:topdown-overhead}
    N_{\mathrm{Qubits}} \approx 8 \cdot \frac{n(n-1)}{2} - m + 5n\,,
\end{equation}
with $8$ qubits per crossing (reduced by $m$ because interacting gadgets use
$7$ vs $8$ qubits) and $5$ extra qubits per chain for the ends and corners. Accordingly, the
overhead in the number of qubits scales roughly as $\sim 4n^2$, assembled within an embedding area of
$(4n+4)\times(4n+3)$, which scales as $\sim 16n^2$.

\begin{figure}
    \includegraphics[width=1.0 \columnwidth]{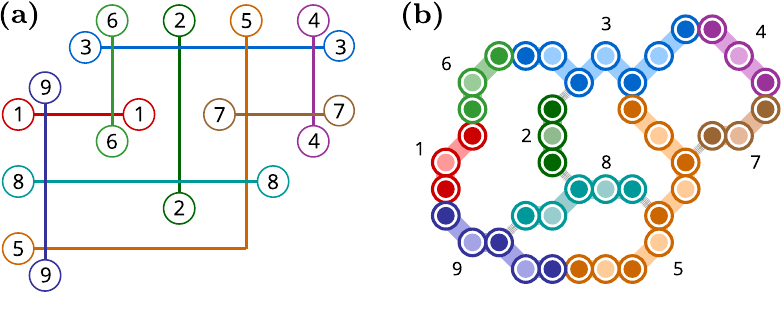}
    \caption{
        \textbf{(a)} Rearranged and shortened chains for the graph
        from Fig.~\ref{fig:top-down-full}. Here, a geometry has been identified
        where no interacting gadgets are necessary: All interactions happen
        at the end of at least one of the chains involved. As a result, usage of costly crossing gadgets can be completely avoided.
        \textbf{(b)} Visualization of the optimized qubit placement for chains with
        the interactions dictated by the simplified chain geometry. For
        each terminal crossing in Fig.~\ref{fig:top-down-opt}(a), the
        placed chains are touching at odd-numbered offsets (indicated
        by darker-filled circles), while ancilla qubits (lighter
        filling) only touch odd-numbered qubits of the same chain.
    \label{fig:top-down-opt}
    }
\end{figure}

\textbf{Overhead reduction.} We aim to systematically reduce this overhead while preserving the ability to accurately represent a given graph --
in particular for lower densities and graphs typically found after reduction. To this end,
we make use of the following four observations (cf.~Fig.~\ref{fig:top-down-opt}):

First, the suggested arrangement of chains with all pairs intersecting allows for an embedding scheme with the same
placement for any graph. However, for a specific graph, we are free to choose a different
chain geometry. Starting from the upper-right triangle arrangement in Fig.~\ref{fig:top-down-full},
we can change the ordering of node labels on the right/top such that some pairs never intersect.
As long as we only drop non-interacting intersections with this reordering, we preserve the
ability for an exact representation, while reducing the overhead through removal of redundant non-interacting gadgets.

Second, while chains extending to the top/right of the arrangement are convenient for readout,
this is not a necessity. With odd-length chains, we can instead choose to shorten each chain from both ends up to their first/last \emph{interacting} crossing.

Third, interactions at the end of a chain can be implemented without a gadget, by placing
the terminal qubit next to an odd-numbered qubit of the chain it needs to interact
with, thereby allowing us to avoid additional interacting intersections.

Fourth, the resulting chain representation can be
translated into an actual qubit arrangement with straights represented by copy
gadgets. We can further reduce the qubit overhead by relaxing gadget positioning
and rewriting chain paths (provided that parity and spacing
requirements are observed).

\textbf{Embedding algorithm.} In practice we use simulated annealing \citep{kirkpatrick:83} with restarts (from random initial order) to
identify an optimized end-point ordering. This search uses the number of
gadgets needed after optimization steps 1 -- 3 as its merit function. After
translation to a qubit arrangement, we further reduce the overhead by
adjusting gadget positions in a secondary Markov Chain process: At each step,
a gadget and all the chains connecting to direct neighbors are removed, followed
by a proposed location adjustment and rewiring. The new connections are
constructed using a biased breadth-first search that takes qubit interactions
into account (both within the chain itself and with existing chains/gadgets), while
ensuring that the resulting length has the correct parity. If successful, the
new placement is accepted based on a standard Metropolis criterion on the change in qubits
used.

\section{Implications for Problem Hardness and Quantum Speedups}
\label{reduction-hardness}

To illustrate the standalone value of the reduction module outlined above, we now first apply our reducer to the Rydberg-native family of random UJ instances, as previously studied in, e.g.,  Refs.~\citep{ebadi:22}, \citep{andrist:23}, \citep{finzgar:23}, and \citep{kim:23}. 
We find that a large fraction of these instances can be reduced substantially or even solved to optimality in polynomial time by reduction alone, thereby providing us with an efficient tool to classify instances as easy or (potentially) hard. 
We also pinpoint signatures of an ``easy-hard-easy'' transition that allows us to tune problem hardness. 
Based on this evidence, we conclude that classical reduction routines (and generalizations thereof) provide a useful tool in the on-going search for hard problem instances and quantum speedups, and as such should be integrated into future hybrid (quantum-classical) end-to-end MIS and MWIS solvers.  

\begin{figure*}
  \includegraphics[width=2.0 \columnwidth]{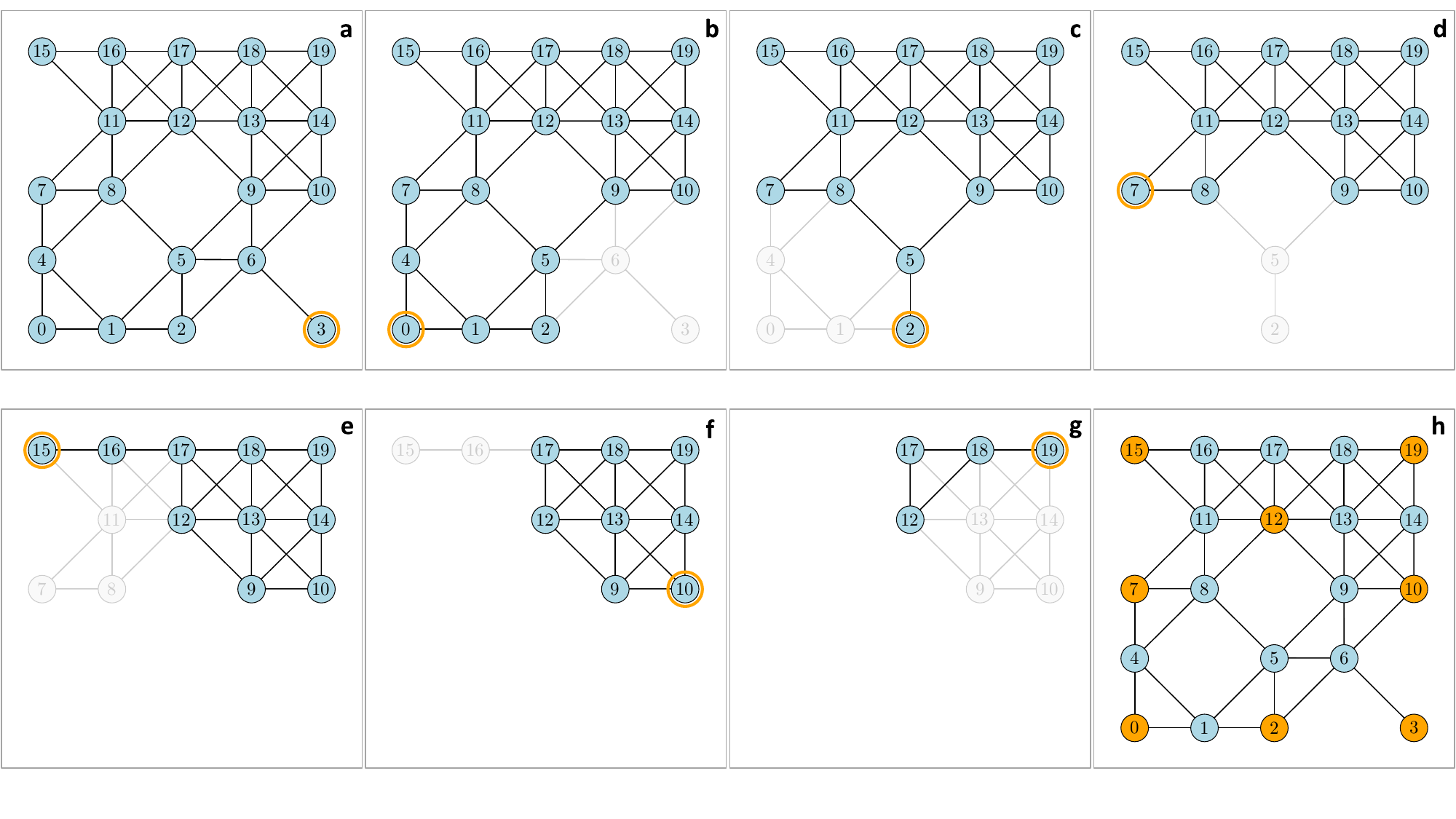}
  \caption{
  Example application of our general-purpose reducer to a hardware-native random UJ instance with $n=20$ nodes on a square lattice with $L=5$ (i.e., $25$ sites) and filling $\varrho=0.8$. 
  The logic follows the one detailed in Fig.~\ref{fig:reduction}.
  By iteratively removing exposed corner nodes, the input graph in panel \textbf{(a)} can be fully reduced [panels \textbf{(b)} -- \textbf{(g)}], deterministically providing an optimal solution with independence number $|\mathrm{MIS}|=8$, as highlighted in panel \textbf{(h)}. Additional MIS solutions can be generated via local search (e.g., swapping node $2$ for node $5$ or node $12$ for node $17$). 
  Given that this example instance can be solved to optimality by reduction alone, it is classified as \textit{easy}.  
      \label{fig:reduction_uj_example}
  }
\end{figure*} 

\textbf{Easy UJ example instance.} For illustration, first consider the small random UJ instance shown in Fig.~\ref{fig:reduction_uj_example} (taken from Ref.~\citep{cain:23}), with $n=20$ nodes on a square lattice with $L=5$ and filling fraction $\varrho=0.8$. 
By iteratively removing exposed corner nodes, we straightforwardly find an optimal solution with independence number $|\mathrm{MIS}|=8$ via reduction. 
As such, this example instance can be classified as \textit{easy}, representing a candidate instance that is unlikely to showcase potential future quantum speedups. 
Quantum speedups are more likely to be found for \textit{hard} instances that feature a large kernel after reduction \citep{hespe:19}.
We note that such a definition of \textit{hard instances} should involve a full suite of reduction algorithms, beyond the clique-based reducer implemented and tested here. 
Specifically, it is easy to construct UJ instances that are immune to clique-based reduction, by simply removing all four exposed corner nodes. 
However, those instances could be unblocked with additional reduction logic, e.g., in the form of the splitting techniques discussed above. 
Nonetheless, here we focus on clique-based reduction only, because it provides a simple tool with which to identify easy instances, and we will assess its reduction performance in detail.

\begin{figure*}
  \includegraphics[width=2.0 \columnwidth]{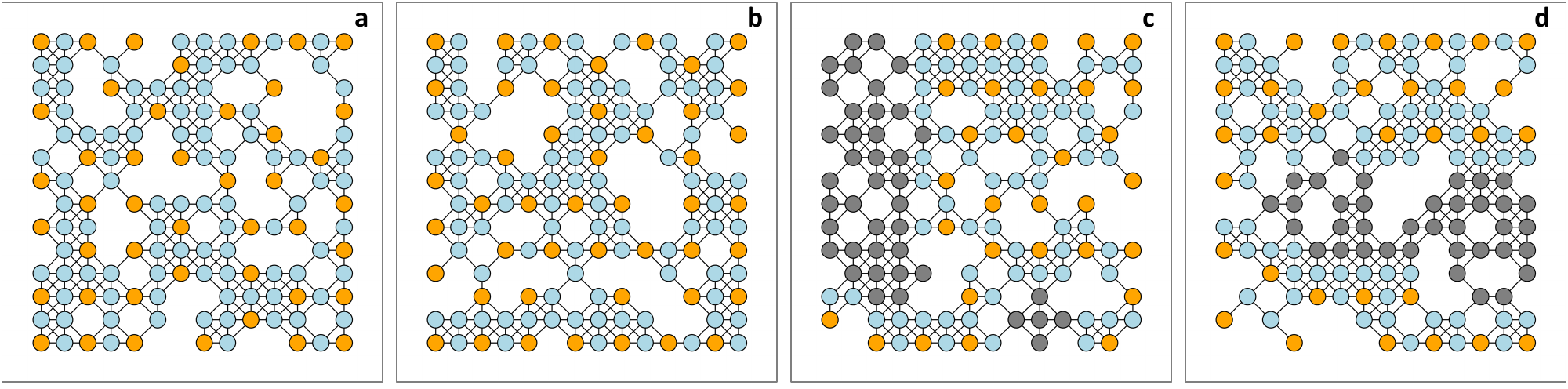}
  \caption{
  Reduction results for larger, hardware-native, random UJ instance with $n=137$ nodes on a square lattice with $L=14$ and filling fraction $\varrho \sim 0.7$.
  The hardness parameter of these four instances ranges across three orders of magnitude, with $\mathbb{H} \sim 1.478$ in panel \textbf{(a)}, $\mathbb{H} \sim 14.12$ in panel \textbf{(b)}, $\mathbb{H} \sim 125.5$ in panel \textbf{(c)}, and $\mathbb{H} \sim 1435$ in panel \textbf{(d)}. 
  Orange nodes have been selected $(n_{i}=1)$, while blue nodes have not been selected $(n_{i}=0)$. Kernel nodes are shown in dark gray. 
  The instances in panels \textbf{(a)} and \textbf{(b)} can be solved to optimality by simple reduction, with $|\mathrm{MIS}|=45$ and $|\mathrm{MIS}|=47$, respectively. 
  The instances in panels \textbf{(c)} and \textbf{(d)} can be reduced by $\xi \sim 70\%$ and $\xi \sim 73\%$, with $31$ and $32$ nodes selected outside of the kernel, leaving smaller kernel instances of size $41$ and $37$, respectively. 
  It is straightforward to see that the resulting kernel graphs do not feature any exposed (corner) nodes. 
      \label{fig:reduction_uj_4hp}
  }
\end{figure*}

\begin{figure}
  \includegraphics[width=1.0 \columnwidth]{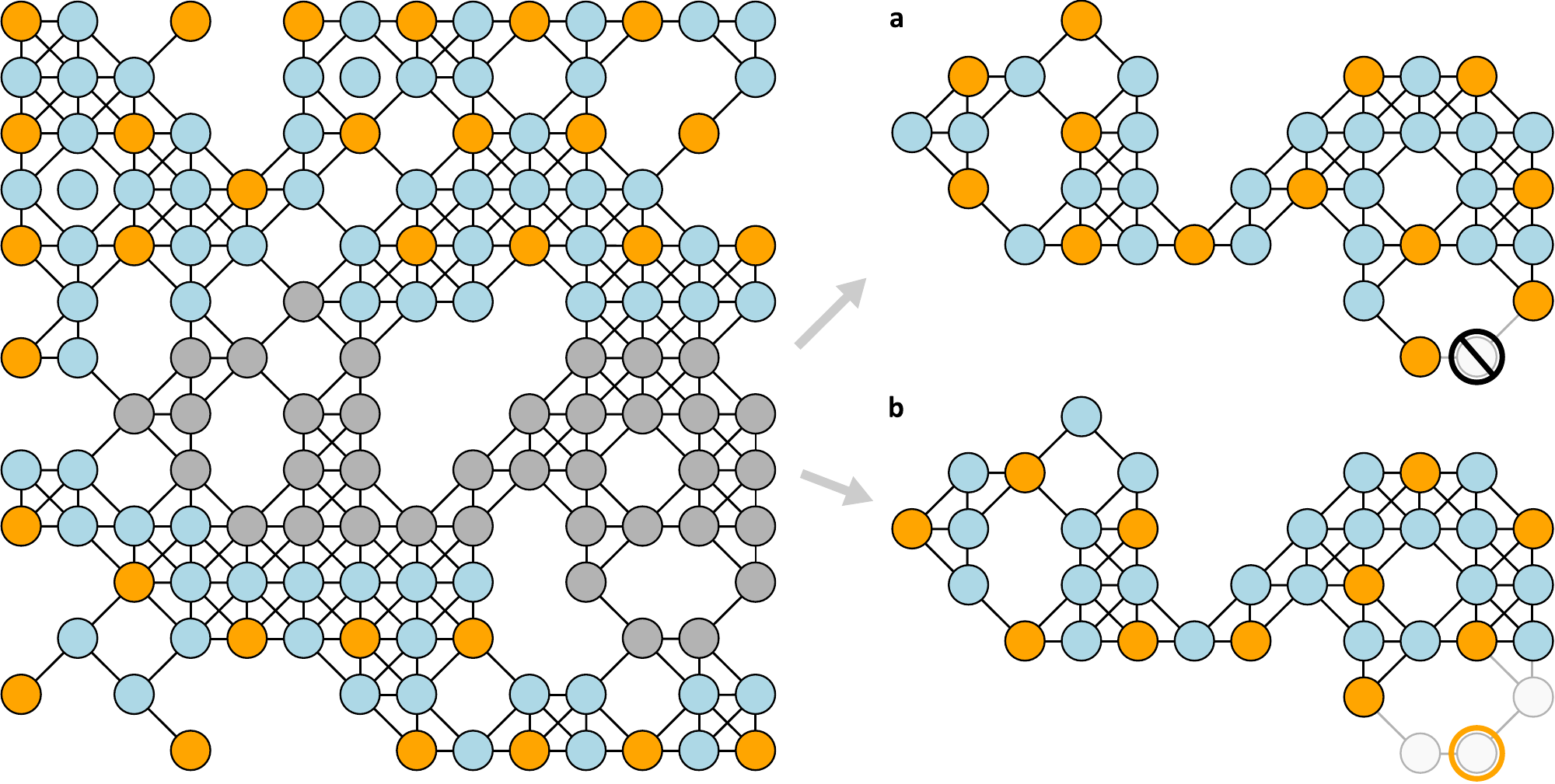}
  \caption{
  Optimal MIS solution through repeated reduction post elimination via exact splitting. 
  The irreducible kernel shown in Fig.~\ref{fig:reduction_uj_4hp}(d) has been split based on the node in the lower right hand corner, given that this node (like any other node) can only be selected [panel \textbf{(b)}] or not [panel \textbf{(a)}]. 
  Removal of the selected node (and its neighborhood if selected) unblocks clique-based reduction. 
  Post splitting, the remaining kernels can be solved to optimality via mere reduction, with 13 (11) nodes selected in scenario \textbf{(a)} and \textbf{(b)}, respectively.   
      \label{fig:reduction_uj_4hp_splitting}
  }
\end{figure}

\textbf{Larger UJ example instances.} Next, we consider four larger example instances previously studied in Refs.~\citep{finzgar:23} and \citep{perseguers:24}. 
While those four instances are all random UJ instances of the same size (with $n=137$ on a square lattice with side length $L=14$), their \textit{hardness parameters} $\mathbb{H}$ span three orders of magnitude~\citep{ebadi:22, cain:23, andrist:23}, with $\mathbb{H}$ defined as $\HP =D_{\mathrm{|MIS|-1}}/(|\mathrm{MIS}|\cdot D_{|\mathrm{MIS}|})$,
where $D_{\alpha}$ denotes the degeneracy of the independent sets of size $\alpha$.
For both classical and quantum Markov chain Monte Carlo (MCMC) algorithms, the conductance-like hardness parameter $\HP$ captures the likelihood to get stuck in a local minimum of size $|\mathrm{MIS}|-1$, as opposed to reaching a ground state of size $|\mathrm{MIS}|$ \citep{ebadi:22}. 
The corresponding success probability to find the MIS in a single algorithmic run (shot), denoted as $P_{\mathrm{MIS}}$, shows an exponential dependence on $\HP$. Specifically, $P_{\mathrm{MIS}} \approx 1 - \exp(-C \HP^{-\alpha})$, 
where $\alpha$ depends on the given MCMC algorithm, and $C$ refers to a positive fitted constant that could have polynomial dependence on the system size in general.
In particular, $P_{\mathrm{MIS}} \approx 0.478$, $P_{\mathrm{MIS}} \approx 0.004$, $P_{\mathrm{MIS}} \approx 0.019$, and $P_{\mathrm{MIS}}=0$ has been reported for the four instances shown in Fig.~\ref{fig:reduction_uj_4hp},
notably with $P_{\mathrm{MIS}}=0$ for the hardest instance ($\HP \sim 1435$), even for an optimized quantum algorithm \citep{perseguers:24}. 
Conversely, our reduction algorithm is insensitive to the hardness parameter $\HP$ tailored towards MCMC-based algorithms, by design. 
As demonstrated in Fig.~\ref{fig:reduction_uj_4hp}, all four instances can be either be solved to optimality, or reduced substantially using our simple clique-based reduction logic alone. In particular, for two instances we find $\xi \sim 70\%$ and $\xi \sim 73\%$, respectively, while the other two instances are even solved to optimality by reduction alone.  
As such, given this efficiency of the inexpensive reduction logic, usage of relatively expensive quantum algorithms (both in terms of wall clock run time or actual monetary cost) can either be made more efficient (with a reduced workload for the QPU and higher success rate given the known size dependence of the hardness parameter \citep{ebadi:22, andrist:23}) or even become unnecessary. 

This statement can be generalized through repeated (optimal) reduction, as exemplified in Fig.~\ref{fig:reduction_uj_4hp_splitting}. 
Here, it is shown that the irreducible kernel taken from Fig.~\ref{fig:reduction_uj_4hp}(d) can be unblocked via node elimination (that exposes new corner nodes) and subsequent repeated reduction. 
Overall, with just two rounds of reduction and the combination of selected nodes over all reduction steps, we then find the optimal solution for this problem of $|\mathrm{MIS}|=32+13=45$, with $P_{\mathrm{MIS}} = 1$, thereby outperforming the optimized quantum algorithm from Ref.~\citep{perseguers:24}.
Further extensions of this idea (with potential plug-ins for quantum computers as co-processors) will be discussed below in Sec.~\ref{conclusion}.

\begin{figure*}
  \includegraphics[width=1.0 \columnwidth]{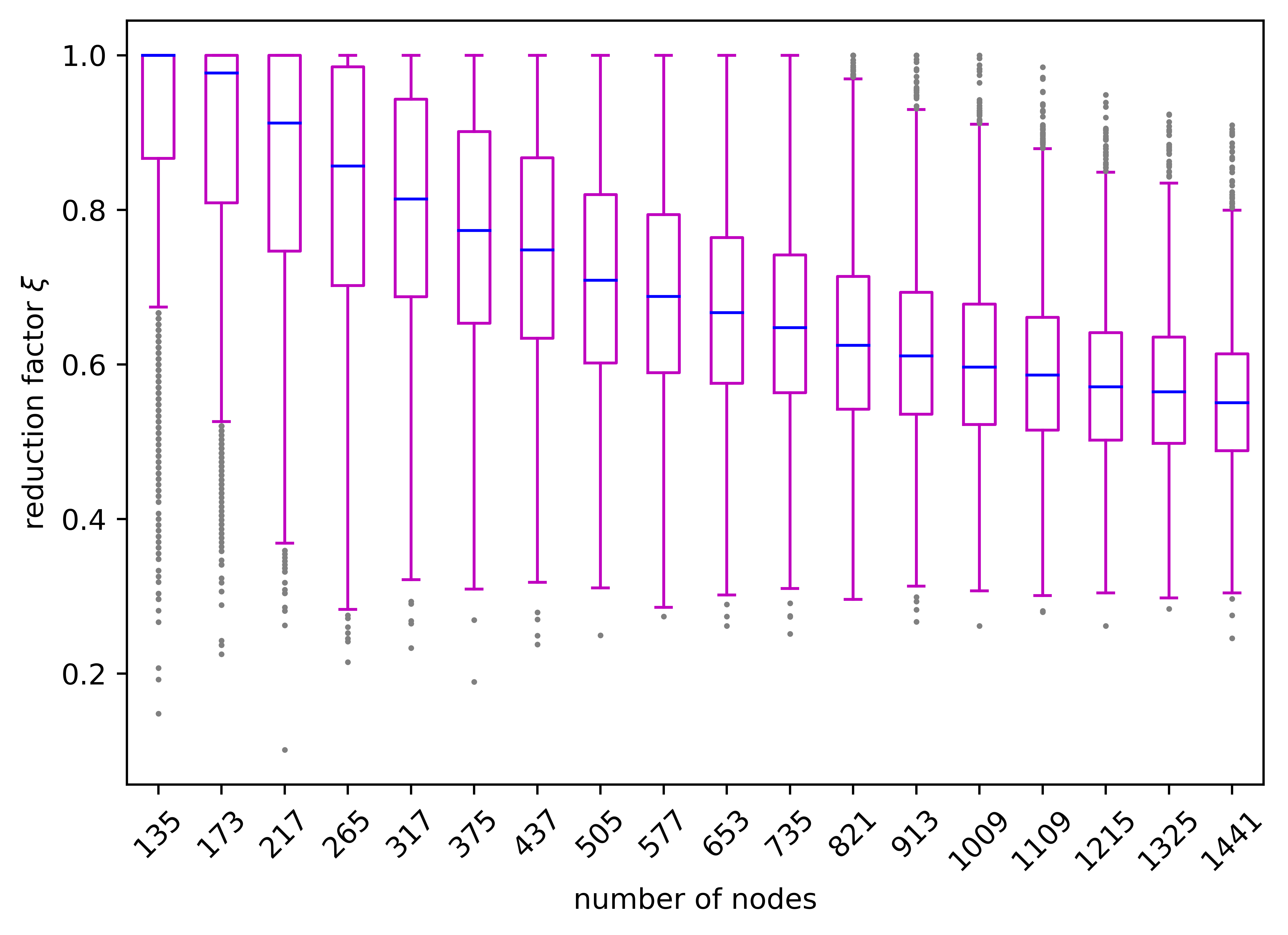}
  \includegraphics[width=1.0 \columnwidth]{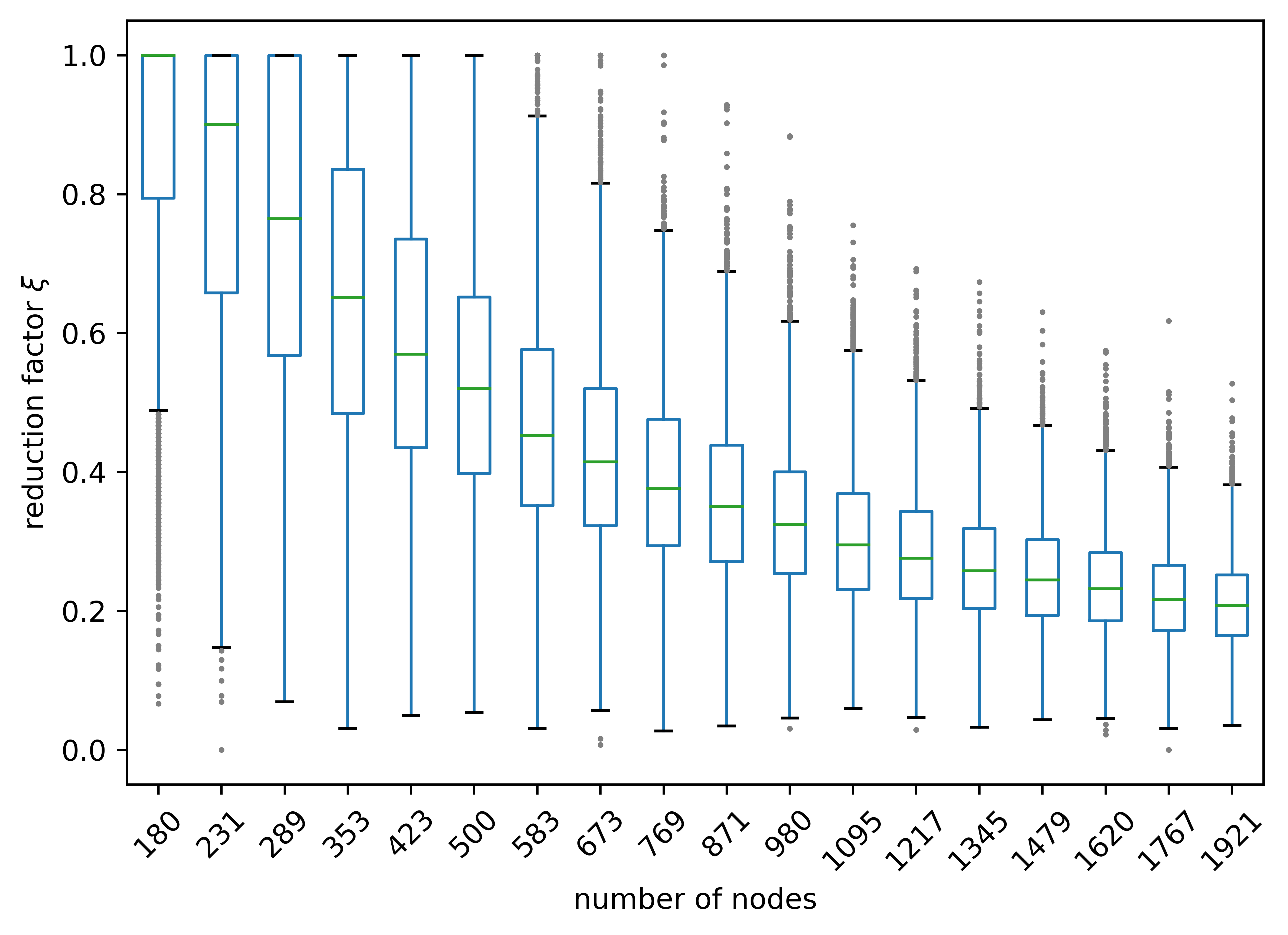}
  \caption{
  Box plot for the reduction factor $\xi$ as a function of the number of nodes for random UJ instances with filling fractions $\varrho=0.6$ (left) and $\varrho=0.8$ (right), respectively.
  The underlying sizes of the square lattices range from $L=15$ up to $L=49$ (with $n=\varrho L^2$), with $5000$ random seeds per size. 
  Boxes extend from the first quartile (25th percentile) to the third quartile (75th percentile), with a horizontal colored line at the median. 
  Whiskers extend from the box to the farthest data point lying within 1.5 times the inter-quartile range (IQR) from the box. 
  Outliers outside of the whiskers are shown individually as dots. 
      \label{fig:reduction-size}
  }
\end{figure*}

\textbf{Systematic experiments.} Next, we perform systematic experiments to quantify typical reduction values across a large set of random UJ instances. 
Specifically, we analyze the reduction factor $\xi$ as a function of system size (given in terms of the number of nodes $n$) for different values of the lattice filling fraction $\varrho$, 
with large reduction pinpointing easy instances with a small kernel of size $(1-\xi)n$ (i.e., the larger $\xi$, the easier the instance).  
Our results are displayed in Fig.~\ref{fig:reduction-size}, for UJ instances with $L=15$, $17$, \ldots, $49$, resulting in graphs with up to $n=1921$ nodes (still beyond the scale of today's quantum hardware with $L \approx 16$ \citep{wurtz:23}). 
All reduction experiments completed on sub-second timescales per instance on a laptop (taking on average only $\sim 0.02\mathrm{s}$ for instances with $\sim 1000$ nodes), following an approximately linear run-time scaling. For more details on algorithmic run time, we refer to Appendix \ref{additional-information}. 
Within the range of instances probed, we find that typical instances can be reduced substantially, with strong instance-to-instance variations even for fixed system size, 
somewhat reminiscent of the strong spread in hardness reported in Refs.~\citep{ebadi:22} and \citep{andrist:23}. 
For small instances compatible with today's quantum hardware (with $L \leq 15$), the median reduction is found to amount to $\xi = 1$, meaning that a given random (small) UJ instance can be solved to optimality via mere reduction with at least 50\% probability (akin to the example shown in Fig.~\ref{fig:reduction_uj_example}). 
For larger instances with $L>15$, we observe a slow decrease in the average reduction, suggesting a potential plateau with a non-zero (density-dependent) fixed-point in the asymptotic limit.
This density dependence is already evident from Fig.~\ref{fig:reduction-size}, with denser instances with $\varrho = 0.8$ typically showing smaller reduction values than sparser instances with $\varrho = 0.6$ for graphs of comparable size $n$.  
 
 \begin{figure}
  \includegraphics[width=1.0 \columnwidth]{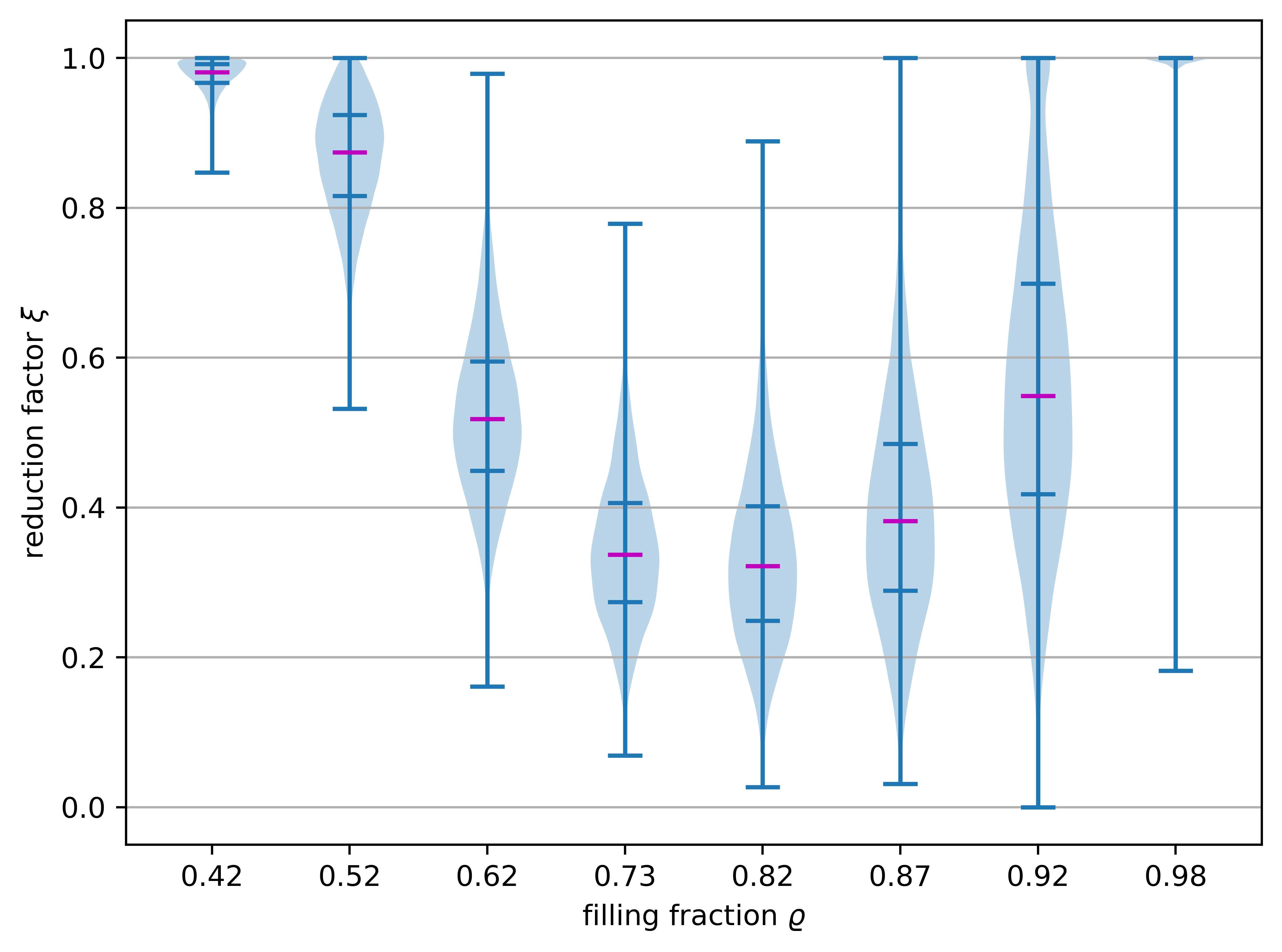}
  \caption{
  Violin plot for the reduction factor $\xi$ as a function of the filling fraction $\varrho$ for random UJ instances with $n=1000$ nodes.
  The underlying sizes of the square lattices range from $L=32$ for dense instances with $\varrho \approx 0.98$ up to $L=49$ for sparser instances with $\varrho \approx 0.42$, with $10^4$ random seeds per size. 
  Extrema, first quartile (25th percentile), median (50th percentile) and third quartile (75th percentile) are marked with horizontal lines. The data suggest an easy-hard-easy crossover as a function of the filling fraction.
  \label{fig:reduction-density}
  }
\end{figure}
 
\textbf{Easy-hard-easy transition.} To probe this density dependence further, we analyze the reduction factor $\xi$ as a function of the filling fraction $\varrho$ for a fixed number of nodes $n$.
Our results for $n=1000$ are shown in Fig.~\ref{fig:reduction-density}. 
We observe indications of an easy-hard-easy crossover, as expected, because both sparse and complete UJ instances are prone to complete reduction (with $\xi =1$). 
The smallest average reduction (corresponding to the hardest instances for reduction logic) is observed for filling fractions around $\sim 80\%$, resulting in an average degree of $\bar{d} \sim 6$, which is consistent with the results in Ref.~\citep{andrist:23} reporting the hardest random UJ instances at similar filling values. 
For filling fractions in the range $\sim 0.8$ -- $0.92$, we observe a large instance-to-instance spread in reduction values, featuring UJ instances effectively immune to reduction (with $\xi \approx 0$) as well as easy instances (with $\xi \approx 1$), even for a fixed filling fraction $\varrho$. This is reminiscent of the phase transition and empirical hardness spike observed for random propositional satisfiability (SAT) problems \citep{leyton-brown:14}, however we did not determine in detail if this was an actual phase transition or a crossover.  A more detailed analysis thereof will be left for future research.  

\textbf{Implications for problem hardness and quantum speedups.} While we have focused on random UJ instances, similar conclusions should hold for other instances previously studied with Rydberg atom arrays (e.g., Cayley tree instances \citep{song:21}), simply because those instances feature exposed corner nodes making them prone to reduction. 
Given that reduction algorithms effectively reduce the problem size as $n \rightarrow (1-\xi)n$ with polynominal run times, they can reduce the generically exponential run time of exact downstream solvers from $\sim 2^{\alpha n}$  to $\sim 2^{\tilde{\alpha} n}$ with $\tilde{\alpha}=(1-\xi)\alpha$, meaning that a reduction of, for example, $\sim 50\%$ can effectively result in a quadratic speedup.
As such, reduction algorithms can have a strong impact on problem hardness as measured by, e.g.,  the overall time to solution for generic pipelines combining reduction techniques with down-stream solvers. 
Therefore, adopting the notion put forward in the classical literature according to which only those MIS instances with a large kernel size are considered \textit{hard} \citep{hespe:19}, our results should help the Rydberg and larger quantum community zero in on truly hard instances, thereby helping to direct efforts in the on-going search for potential quantum advantage.   

\section{Numerical Experiments}
\label{numerics}

We now provide additional numerical results illustrating the performance of our larger compilation pipeline. 
First, we discuss an end-to-end example for the well-known Cora citation graph, featuring results obtained from the QuEra Aquila device available on Amazon Braket.
Next, we report on systematic experiments providing a comprehensive assessment of the achievable reduction factor for both real-world networks and families of synthetic graphs. Finally, we provide numerical experiments to quantify the typical embedding overhead for a family of graph instances.  

\subsection{End-to-End Example}

\begin{figure*}
  \includegraphics[width=2.0 \columnwidth]{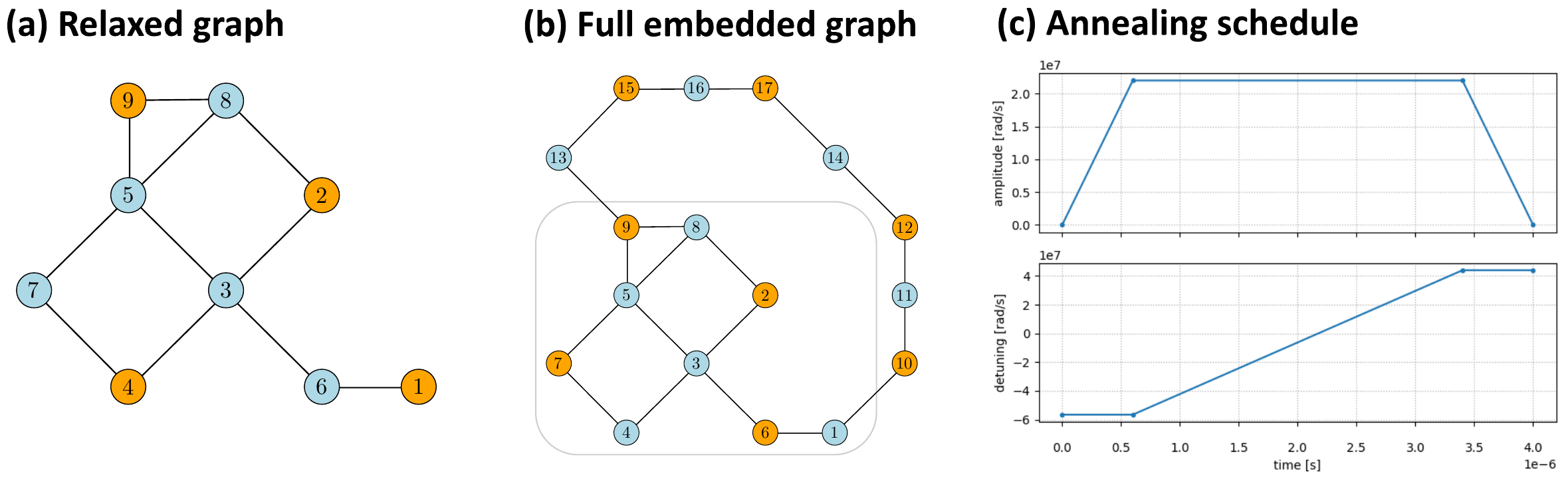}
  \caption{
  MIS solutions obtained for the largest component of the Cora core graph via analog Hamiltonian simulation (AHS) on Amazon Braket. 
  \textbf{(a) Relaxed graph}: For the overhead-free, relaxed graph, we predominantly find the MIS solution $\{ 1, 2, 4, 9\}$ with a probability of $p \approx 61\%$. 
  This solution is infeasible for the original problem graph, because the edge $(1, 9)$ is missing in the relaxed graph, but is present in the original graph. This issue can be easily fixed via local search (e.g., by swapping selection of node $1$ with selection of node $6$ and/or swapping node $9$ for node $5$).
  \textbf{(b) Full embedded graph}: For the full embedded graph involving 17 qubits, we predominantly find the feasible MIS solution $\{ 2, 6, 7, 9\}$ (with probability of $p \approx 25\%$), with the 8-qubit quantum wire (outside of the gray box) effectively suppressing the simultaneous selection of nodes $1$ and $9$. 
  \textbf{(c) Annealing schedule}: All results were obtained using a piecewise linear schedule and $1000$ shots. 
  Numerical parameters: Maximum Rabi frequency $\Omega_{\mathrm{max}}/2\pi = 3.5 \mathrm{MHz}$, minimal detuning $\Delta_{\mathrm{min}}/2\pi = -9 \mathrm{MHz}$, maximum detuning $\Delta_{\mathrm{max}}/2\pi = 7 \mathrm{MHz}$, laser phase $\phi=0$, total annealing time $\tau = 4 \mu \mathrm{s}$, and lattice spacing $a=4.5\mu\mathrm{m}$.
      \label{fig:quantum-demo}
  }
\end{figure*}

For illustration, we first consider the well-known Cora citation graph with $\sim 2700$ nodes and $\sim 5300$ edges \citep{cora:00}; see Fig.~\ref{fig:scheme}.
The size of this network goes well beyond the scale supported by today's quantum hardware (which is limited to hundreds of qubits with restricted connectivity \citep{wurtz:23}). 

In an effort to make instances of this scale potentially compatible with near-term quantum hardware, we first apply our clique-based reduction scheme, and find that the Cora graph can be reduced by $\sim 97\%$ to a core with 79 nodes and 94 edges, with maximum degree $d_{\mathrm{max}}=4$ (compared to $d_{\mathrm{max}}=168$ for the original graph), within sub-second run time on a laptop.
Moreover, this core is found to decompose into 16 connected components that can each be solved individually in parallel. 
The majority of these components (9 out of 16) are simple cycle graphs that are immune to clique-based reduction, but are trivial to solve (i.e., every other node can be added to the independent set).  
The largest component of the core graph is displayed in Fig.~\ref{fig:scheme}(b) and has nine nodes, with the second largest having just six nodes, overall resulting in a large reduction in complexity and load for any down-stream MIS solver. 

After reduction, the core graph can be solved with a variety of MIS solvers. 
Here, we utilize our embedding logic to make the remaining core graph compatible with Rydberg atom arrays that support the implementation of \textit{quantum} algorithms to approximately solve the MIS problem on the embedded (physical) graph \citep{ebadi:22}. 
The individual components of the core can be embedded and solved individually, with the potential to fit more than one component on an annealing device, provided both enough space and qubits are available.  
For illustration, we focus on the largest component. 
As displayed in Fig.~\ref{fig:scheme}(c), we find that this core graph (with nine nodes and twelve edges) can be embedded onto a hardware-native graph with nine nodes, eleven UJ edges, and one quantum wire with eight ancilla nodes. 
We then solve the MIS problem on this embedded graph via quantum-annealing techniques that leverage neutral atoms \citep{pichler:18, finzgar:23}. 
To this end, nodes are mapped to qubits subject to analog quantum dynamics under the many-body Hamiltonian given in Eq.~(\ref{eq:rydberg-hamiltonian}). 

Our results are shown in Fig.~\ref{fig:quantum-demo}, with the underlying simple, piecewise linear annealing schedule detailed in Fig.~\ref{fig:quantum-demo}(c).
In particular, we set $a=4.5\mu \mathrm{m}$ and drive parameters were chosen such that $\sqrt{2} \leq R_b/a < 2$ as required for UJ connectivity.  
Here we outline two complementary experimental strategies: 
(i) \textit{Relaxed graph}: To keep the embedding overhead minimal, we first solve the MIS problem on a \textit{relaxed}, hardware-native graph that is optimized to approximate the original input graph up to one missing edge (resulting in an edit distance of one) while adhering to UJ connectivity constraints. 
Exact (noise-free) simulations of this 9-qubit system's quantum dynamics show that the Rydberg solver acts as a \textit{biased} MIS solver, favoring those MIS solutions that minimize the Rydberg interaction tails. 
Specifically, we find that out of eight possible MIS solutions for this instance, the Rydberg solver predominantly finds three MIS solutions with probabilities of $\sim 61\%$ [shown in Fig.~\ref{fig:quantum-demo}(a)], $\sim 37\%$, and $\sim 1\%$, whereas an unbiased MIS sampler should generate these with uniform probabilities of $1/8=12.5\%$. 
It is possible, however, to potentially generate additional MIS solutions through simple post-processing routines, taking the solutions generated by the Rydberg solver as initial seed for local search heuristics.  
Specifically, additional MIS solutions may found by swapping nodes in the set with available neighbors. 
For example, starting from the MIS solution $\{ 1, 2, 4, 9\}$ shown in Fig.~\ref{fig:quantum-demo}(a) it is easy to identify the swap candidates $1 \leftrightarrow 6$, $4 \leftrightarrow 7$, and $5 \leftrightarrow 9$, generating in total $\binom{3}{1}+\binom{3}{2}+\binom{3}{3}=7$ potential MIS candidates that can be accepted if feasible, or dismissed otherwise. 
Similar logic can be applied in an effort to achieve feasibility on the original input graph for solutions found on the companion relaxed graph, as is the case for the solution shown in Fig.~\ref{fig:quantum-demo}(a) when mapped to the original graph shown in Fig.~\ref{fig:scheme}(b) with an additional edge $(1, 9)$.  
(ii) \textit{Full embedded graph}: Such post-processing routines can be avoided via an \textit{exact} embedding, albeit at the expense of an increased overhead, as exemplified in our second experiment; see Fig.~\ref{fig:quantum-demo}(b). 
Here we solve for the MIS on the embedded graph, using an analog Hamiltonian simulation (AHS) program for 17 qubits representing 9 logical nodes and 8 additional ancillas. 
We find that the Rydberg solver favors the MIS solution $\{2, 6, 7, 9\}$ shown in Fig.~\ref{fig:quantum-demo}(b) with a probability of $p \approx 25\%$, with the ancilla quantum wire copying the selection of node 9 across every other node along the wire as desired. 

Finally, we have run our AHS program on real quantum hardware provided by QuEra available on Amazon Braket. 
Given the limited laser power budget for driving the ground-Rydberg transition \citep{wurtz:23}, here we used slightly different parameters, with reduced Rabi frequency $\Omega_{\mathrm{max}}/2\pi = 2.5\mathrm{MHz}$, while adjusting the lattice spacing to $a=4.8\mu\mathrm{m}$ so as to keep $r=R_{b}/a \sim 1.74$, comparable to $r\sim 1.7$ as used in Ref.~\citep{ebadi:22}.
All other parameters are listed in the caption of Fig.~\ref{fig:quantum-demo}.
The most likely outcome is found to be identical to our result shown in Fig.~\ref{fig:quantum-demo}(b) based on noise-free simulations, with a reduced success probability of $p\approx 9.80\%$, a drop that we attribute to noise present on the quantum hardware.  

\subsection{Reduction Experiments}

We now complement the reduction results presented above with additional experiments on both real-world networks and synthetic (random) graphs. 
We observe that substantial reduction can typically be achieved up to some critical average degree, with structured graphs being more prone to reduction than structure-less, random graphs. 

\begin{table*}
\centering
 \begin{tabular}{| l | c c c | c c c c | c | c |} 
 \hline
 \hline
\multirow{2}{8em}{\hspace*{2.0em}network} & \multicolumn{3}{|c|}{input graph $\mathcal{G}$} & \multicolumn{4}{|c|}{kernel $\mathcal{K}$ of input graph $\mathcal{G}$} & \multicolumn{1}{|c|}{run time} & \multicolumn{1}{|c|}{reduction}\\
 & nodes & edges & average degree & nodes & edges & components & largest component & [ms] & $\xi$ \\ [0.5ex] 
 \hline
 Florentine & 15 & 20 & 2.67 & 0 & 0 & --- & --- & 0.38 & 100\%  \\ 
 Zachary Karate & 34 & 78 & 4.59  & 4 & 4 & 1 & 4 & 0.97 & 88.2\%  \\ 
 Dolphins & 62 & 159 & 5.13 & 20 & 30 & 2 & 12 & 1.51 & 67.7\%  \\ 
 Les Miserables & 77 & 254 & 6.60 & 0 & 0 & --- & --- & 2.21 & 100\%  \\ 
 Jazz & 198 & 2742 & 27.70 & 83 & 580 & 1 & 83 & 17.74 & 58.1\%  \\ 
 C.~Elegans & 438 & 1519 & 6.94 & 19 & 26 & 1 & 19 & 14.32 & 95.7\%  \\ 
 Email & 1133 & 5451 & 9.62 & 315 & 818 & 1 & 315 & 44.05 &  72.2\% \\  
 Cora & 2708 & 5278 & 3.90 & 79 & 94 & 16 & 9 & 116.93 & 97.1\%  \\ 
 Citeseer & 3264 & 4536 & 2.78 & 217 & 341 & 26 & 83 & 67.30 & 93.4\% \\
 PubMed & 19714 & 44281 & 4.49 & 16 & 23 & 2 & 11 & 665.06 & 99.9\% \\ [1ex] 
 \hline
 \hline
 \end{tabular}
\caption{
Numerical reduction results for real-world graphs. 
The first block describes the input graph $\mathcal{G}$ in terms of number of nodes, number of edges, and average degree. 
The second block details the kernel of the graph $\mathcal{K}$ after clique-based reduction in terms of nodes and edges (post reduction), the number of components, and the size of the largest component. 
The reduction result is summarized in terms of the reduction factor $\xi$, and the run time of the reduction logic (on a laptop). 
Note that nodes with self loops have been removed.
Further details are provided in the main text. 
\label{tab:social-networks}} 
\end{table*}

\textbf{Real-world networks.} We have applied our reduction logic to ten well-known benchmark networks, involving small social networks with $\sim 10$ nodes up to relatively large citation networks with $\sim 10^4$ nodes (including the Cora citation network featured above as our guiding example).
Specifically, we consider the following publicly available benchmark instances. 
The Florentine graph describes a social network of families in 15th century Florence related by marriage.
The Zachary graph refers to another social network of a university karate club.
The Dolphins graph represents a social network of frequent associations between 62 dolphins in a community living off New Zealand.  
The Les Miserables graph is a coappearance network of characters in the novel Les Miserables. 
The Jazz graph is a social collaboration network of Jazz musicians. 
The C.~Elegans graph is a metabolic network of the nematode (roundworm) c.~elegans. 
The Email graph describes a university email network. 
The Cora and Citeseer graphs describe networks of computer science publications (with nodes representing publications and edges referring to citations), and 
the PubMed graph is another citation network based on articles related to diabetes from the PubMed database. 
All graphs can be sourced directly from the NetworkX library or downloaded from the network repository website \citep{rossi:15}. 

Our numerical results are summarized in Tab.~\ref{tab:social-networks}. 
Across our set of ten network instances, we consistently observe significant reduction, with $0.58 \leq \xi \leq 1$, an average reduction of $\sim 87\%$, and a relatively strong Pearson correlation coefficient of $\sim -0.73$ between the reduction factor $\xi$ and the average degree of the input graph. These results  
suggest that graphs with a larger average degree may be more resilient to clique-based reduction. 
In particular, we find that the largest graph considered (PubMed) can be reduced by $\sim 99.9\%$, from $\sim 2\times 10^4$ nodes to a core with 16 nodes in sub-second run time on a laptop. 
For some graphs (such as Cora, Citeseer, and PubMed) the core is found to decompose into several, smaller components which can then be solved down-stream individually in parallel. 
In line with our experiments for random UJ graphs, all reduction experiments completed on sub-second timescales on a laptop, with the run time following an approximately linear run-time scaling $\sim n$; see Appendix \ref{reduction-uj} for more details. 

\textbf{Synthetic graphs.} We have performed additional, systematic benchmark experiments across three families of synthetic (random) graphs, including random geometric (RG) graphs, Erd\H{o}s-R\'enyi (ER) graphs, and Barab\'asi-Albert (BA) graphs. 
Here we focus on our results for Rydberg-native RG graphs. Qualitatively identical results have been obtained for ER and BA graphs, and quantitative differences across these families of random graphs will be discussed below. 

RG graphs belong to the larger class of spatial networks, as found in infrastructure networks in transportation and telecommunication systems, where the probability of a connection between two nodes is a decreasing function of the distance between those nodes \citep{barthelemy:11}.
In particular, $d$-dimensional RG graphs are graphs where each node is assigned a random coordinate in the unit box $[0, 1]^{d}$, and only nodes within the connection distance $R$ are connected by an edge. 
Here we focus on two-dimensional RG graphs as native to Rydberg atom arrays with arbitrary positioning in two space dimensions and $R=R_{b}$. 

\begin{figure*}
  \includegraphics[width=2.05 \columnwidth]{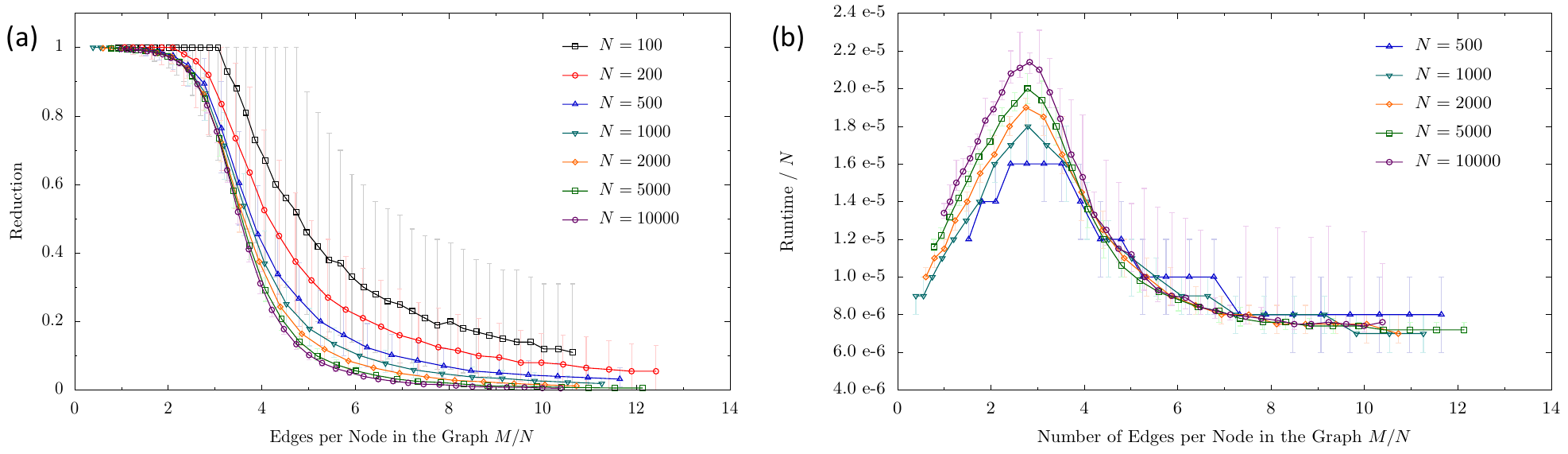}
  \caption{
  Reduction applied to random geometric graphs in two space dimensions. 
  \textbf{(a)} Reduction factor $\xi$ as a function of the number of edges $M$ per node, as can be tuned via the radius $R$. 
  \textbf{(b)} Reduction run time (in seconds, scaled in terms of the number of nodes) as a function of the number of edges per node.
  We observe a peak in run time at the critical average degree of $\bar{d}^{*}=2M/N\sim 6$.
  Substantial reduction is possible for sparse instances with average degree smaller than $\bar{d}^{*}$, while denser instances with average degree larger than $\bar{d}^{*}$ are immune to clique-based reduction. The average run time is well approximated by $\bar{T}_{\mathrm{red}}/n \approx \gamma(\bar{d})$, in agreement with Eq.~(\ref{eq:reduction-runtime}).
      \label{fig:reduction-rgg}
  }
\end{figure*}

Our reduction results for RG graphs in two dimensions (with up to $10^4$ nodes) are shown in Fig.~\ref{fig:reduction-rgg}.
By tuning the parameter $R$ we can effectively tune the number of edges $M$ and, as such, the average degree $\bar{d}=2M/N$.  
We observe large reduction up to a certain (critical) average degree $\bar{d}^{*} \sim 6$, with a clear  transition from macroscopic reduction to minimal reduction in the asymptotic limit of large system sizes. Moreover, this reduction is accompanied by a notable run-time peak around the critical degree, akin to phase transitions in uniform-random 3-SAT instances around the critical clause-to-variable ratio \citep{leyton-brown:14}. However, additional studies would have to be performed to determine if this is an actual phase transition or an easy-hard-easy crossover.
Consistent with our results for random UJ instances and real-world networks, reduction run times are sub-second for sparse instances as large as $\sim 10^{4}$ nodes, with a scaling that is approximately linear in the number of nodes. 

Similar to our findings for random UJ instances, our results suggest that Rydberg-native RG instances with an average degree below the critical degree $\bar{d}^{*}$ are easy, and, as such, unlikely to display potential future quantum speedups. 
Conversely, RG instances with sufficient connectivity are immune to clique-based reduction and have a large kernel, making them interesting candidates for future quantum experiments. 
Beyond Rydberg-native UJ and RG graphs, we have observed a similar critical degree of $\bar{d}^{*} \sim 6$ for scale-free BA graphs that feature hubs and power-law degree distribution \citep{barabasi-albert:99}, approximately similar to many real-world networks. 
For sufficiently sparse graphs, such structured networks are likely prone to reduction, e.g., via simple removal of simple dangling bonds (with degree one).  
Conversely, given their inherent lack of structure, ER graphs with a Poissonian degree distribution appear to be the most robust against reduction with a smoother transition towards irreducibility appearing for average degree values as low as $\sim 3$ to  $4$; see Appendix \ref{reduction-synthetic} for more details. 

\subsection{Embedding Experiments}

\begin{figure}
    \includegraphics[width=1.0 \columnwidth]{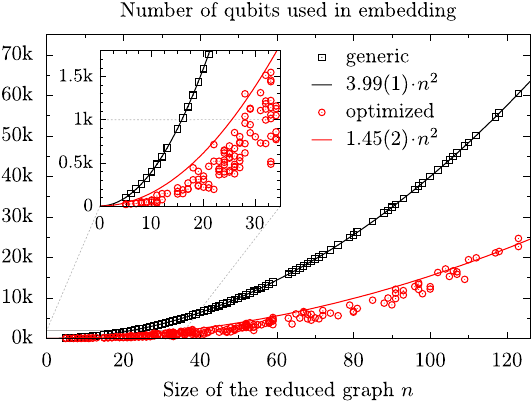}
    \caption{
        Number of qubits required for hardware-native embedding for the generic scheme
        (black squares) and optimized approach (red circles). Each
        point indicates the overhead for a \emph{residual graph kernel} with
        $n$ nodes, as obtained by applying
        reduction to random Erd\H{o}s--R\'enyi graphs with $1.7$ edges per node.
        The fitted functions indicate a reduction in the pre-factor
        from $3.99$ to $1.45$. Inset: Number of qubits required
        for (small) logical graphs $n<35$, with the horizontal line
        indicating hyptothetical quantum hardware with $10^3$
        physical qubits.
        \label{fig:top-down-overhead}
    }
\end{figure}

In our proposed pipeline, embedding techniques are used to make
the residual kernel graph compatible with Rydberg quantum hardware (after reduction has been applied), ultimately allowing us to solve for the MIS with quantum algorithms. 
Here, using our top-down embedding scheme, we perform
numerical experiments to assess the qubit overhead for
a set of graphs involving
250 \emph{reduced} graphs. These are generated by applying reduction
to large Erd\H{o}s--R\'enyi (ER) graphs with a density that corresponds to $1.7$
edges per node. The resulting residual (kernel) graphs are smaller than the original ER graphs, and used as a
representation of typical irreducible cores for sparse graphs.
Our results are shown in Fig.~\ref{fig:top-down-overhead}, demonstrating
that the required number of qubits can be reduced from $\sim 3.99(1) n^2$
for the generic embedding strategy per Ref.~\citep{nguyen:23} down to $\sim 1.45(2) n^2$ when adopting our optimized top-down embedding scheme.
Accordingly, the overhead remains quadratic in the number of nodes $n$,
yet the smaller pre-factor allows to embed graphs with $\sim 1.66$ as many nodes. 
For example, for the set of graphs studied here and hypothetical hardware with 1000
physical qubits, the optimized top-dowm method is able to embed all instances with up to $n=27$ nodes
(and several larger instances), while the full generic scheme reaches this
limit already at $n=15$.
While the specific pre-factors are likely dependent on the class of graphs studied,
this indicates that optimized graph-specific top-down embedding can be
leveraged efficiently for larger sparse instances.

\section{Conclusion and Outlook}
\label{conclusion}

In summary, we have proposed and analyzed a comprehensive suite of tools to support the implementation of quantum optimization algorithms with near-term Rydberg atom arrays, for generic (non-native) MIS instances, potentially at problem scales previously inaccessible. 
Our toolbox features three main modules. 

The reducer module provides an efficient tool to shrink the size of generic MIS problem instances while preserving optimality, by iteratively removing sub-graphs via selection of exposed nodes that are provably part of some optimal MIS solution. 
Optimal solutions for the reduced (core) instance can then be easily expanded to an optimal solution for the original instance by unrolling previously applied reductions. 
We have shown that this simple reduction technique can be a powerful asset in the on-going search for potential quantum speedups, because it allows to identify hard problem instances (where quantum speedups are more likely to be found) as those instances with a large kernel. 
In particular, for Rydberg-native MIS instances we have observed signatures of an easy-hard-easy transition, showing that problem hardness could be tuned as a function of atomic density.

The checker module quickly assesses the hardware ``friendliness'' of a given input graph via simple checks on graph properties, such as the maximum degree and the number of triangles. We have used this tool to generate a hardware compatibility diagram that can inform down-stream embedding logic.

Lastly, we have described in detail two complementary embedding strategies for Rydberg atom arrays, including a (bottom-up) approach that \textit{learns} approximate embeddings with minimal resource requirements, as well as a (top-down) scheme that constructively generates exact embeddings with MIS optimality guarantees, while minimizing the required qubit overhead. 
Using this module, we have systematically analyzed the typical overhead for a class of generic input instances and quantified problem scales accessible with near-term devices. 

In aggregate, these three modules can be integrated to set up an end-to-end pipeline for solving generic (i.e., non-native) instances of the MIS problem on quantum hardware based on Rydberg atom arrays. 
We have exemplified this pipeline with a demo run on quantum hardware provided by QuEra on Amazon Braket for the Cora citation network with $\sim 2700$ nodes, after optimality-preserving reduction to a kernel with a largest component of just nine nodes. 

Finally, we now highlight possible extensions of research going beyond our present work.
First, in an effort to streamline on-going work towards potential quantum advantage, it would be interesting to use state-of-the-art reduction procedures (involving, for example, generalized dominance reduction instead of isolated clique removal \citep{chang:17}) to establish a standard testbed featuring provably hard MIS instances as a go-to resource for benchmarking new quantum algorithms.  
Second, one could study extended embedding schemes in which atomic positions are not necessarily restricted to sites on an underlying square lattice, but are allowed to take on arbitrary (continuous) coordinates. 
Third, given the wide-ranging applications of the MIS problem, e.g., in network design \citep{hale:80}, vehicle routing \citep{dong:22}, or quantitative finance \citep{kalra:08}, it would be interesting to apply and systematically test the performance of our pipeline for such real-world problems.  
Finally, one could expand the set of accessible optimization problems through a series of extensions, including, for example,  
(i) the generalization to the maximum weight independent set problem (MWIS) based on reductions for the MWIS problem \citep{lamm:19}, with real-world applications in map-labeling, 
(ii) generalized reduction logic for other graph-based optimization problems (such as maximum cut or graph coloring) where similar sub-graph removal techniques should apply, and 
(iii) the integration of state-of-the-art reduction logic with non-native hybrid quantum solvers as presented in Ref.~\citep{wurtz:24}. 
In particular, inspired by the results presented in Figs.~\ref{fig:reduction_uj_4hp} and \ref{fig:reduction_uj_4hp_splitting}, it would be interesting to consider \textit{quantum} versions of state-of-the-art classical MIS heuristics such as ReduMIS \citep{lamm:17} in which exact reduction is applied repeatedly in tandem with a quantum algorithm (rather than a classical evolutionary heuristic) to select a subset of independent kernel vertices to include, thereby opening up the reduction space after removal of the selected vertices and their neighbors.   
Similar to classical state-of-the-art MIS heuristics \citep{lamm:17, grossmann:23}, the quantum computer would then be used as a co-processor to inform the classical wrapper about vertices that are likely to be in large independent sets.  

\begin{figure}
  \includegraphics[width=1.0 \columnwidth]{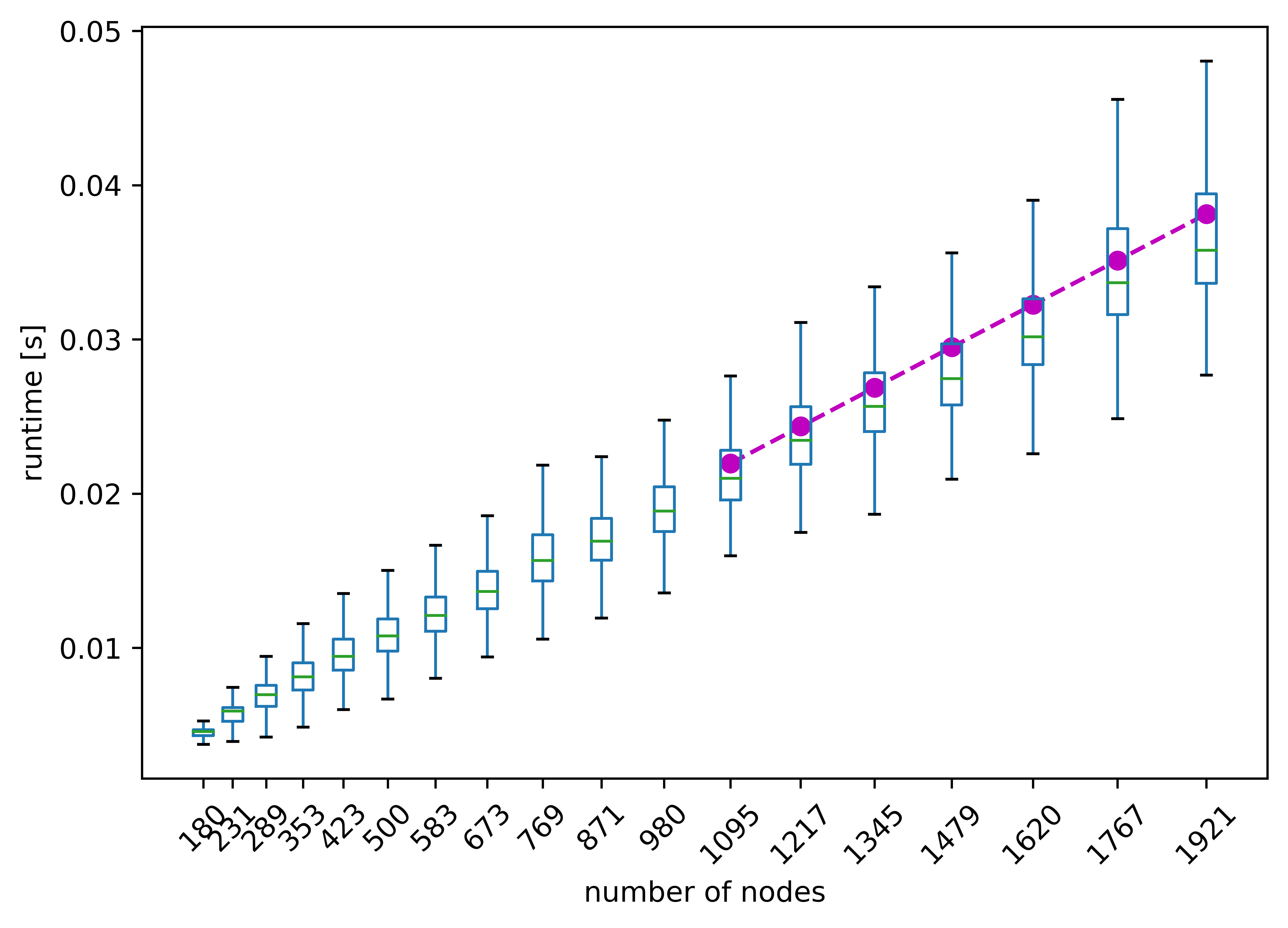}
  \caption{Box plot for algorithmic run time (in seconds) for our reduction logic as a function of the number of nodes for random UJ instances.
  Data and the definition of boxes and whiskers are identical to those shown in Fig.~\ref{fig:reduction-size} for filling fraction $\varrho=0.8$.
  Outliers are omitted for clarity. 
  The best power-law fit $\sim n^{\alpha}$ for all data for larger instances with $n>1000$ (including outliers) is shown as a dashed line (magenta), resulting in an approximately linear scaling with $\alpha \sim 0.981(5)$.     
      \label{fig:runtime-reduction-uj}
  }
\end{figure}

\begin{figure}
  \includegraphics[width=1.0 \columnwidth]{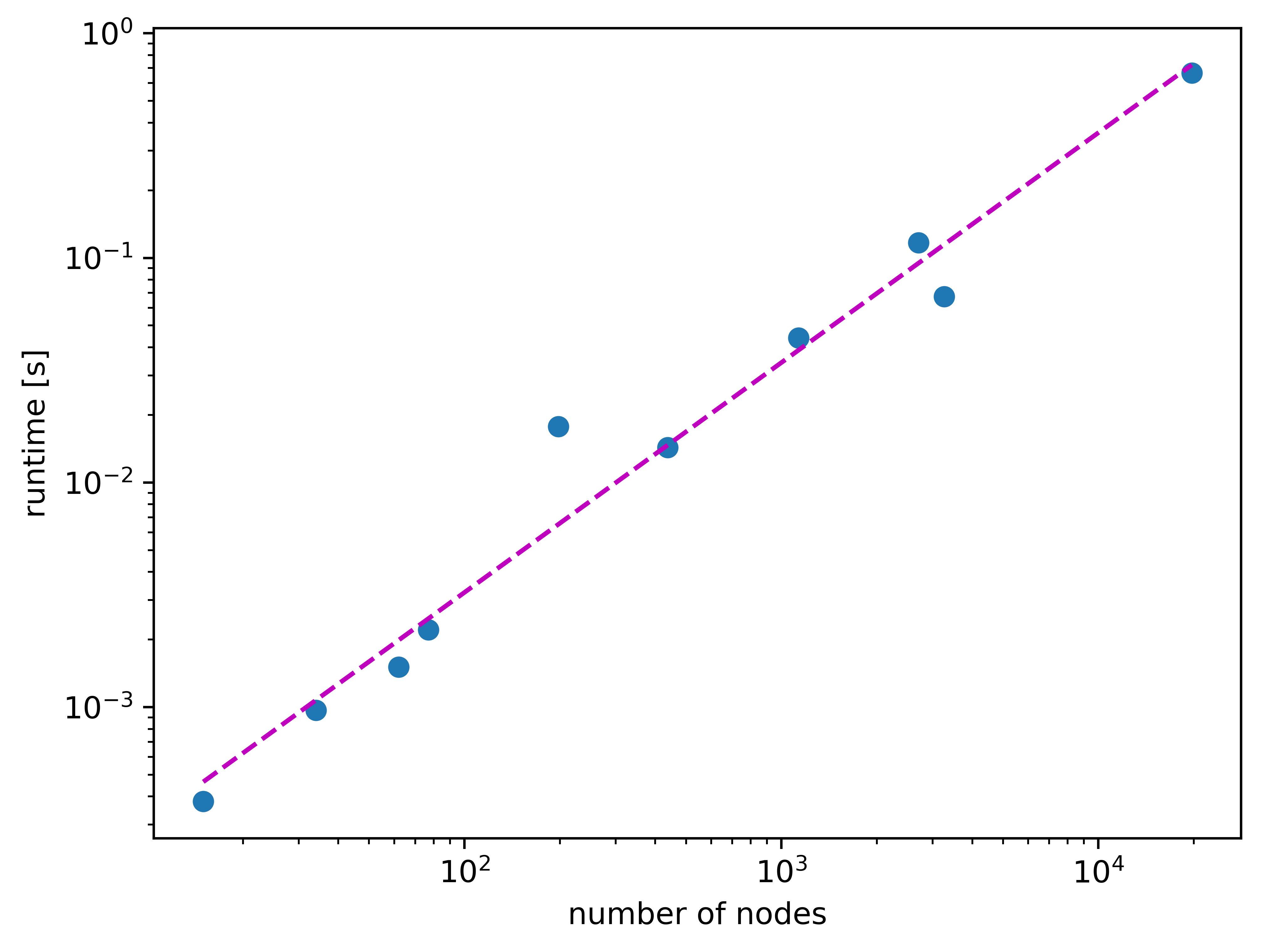}
  \caption{Algorithmic run time (in seconds) for our reduction logic as a function of the number of nodes for the ten networks considered in Tab.~\ref{tab:social-networks}.
  The best power-law fit $\sim n^{\alpha}$ is shown as a dashed line (magenta), resulting in an approximately linear scaling with $\alpha \sim 1.02(6)$.     
      \label{fig:runtime-reduction-networks}
  }
\end{figure}


\begin{acknowledgments}

We thank Victor Bocking, Alexander Buts, Peter Komar, Lou Romano, Peter Sceusa, and Rajagopal Ganesan for their support.
We thank Darren Strash for useful comments on the manuscript.

\end{acknowledgments}

\begin{figure*}
  \includegraphics[width=2.0 \columnwidth]{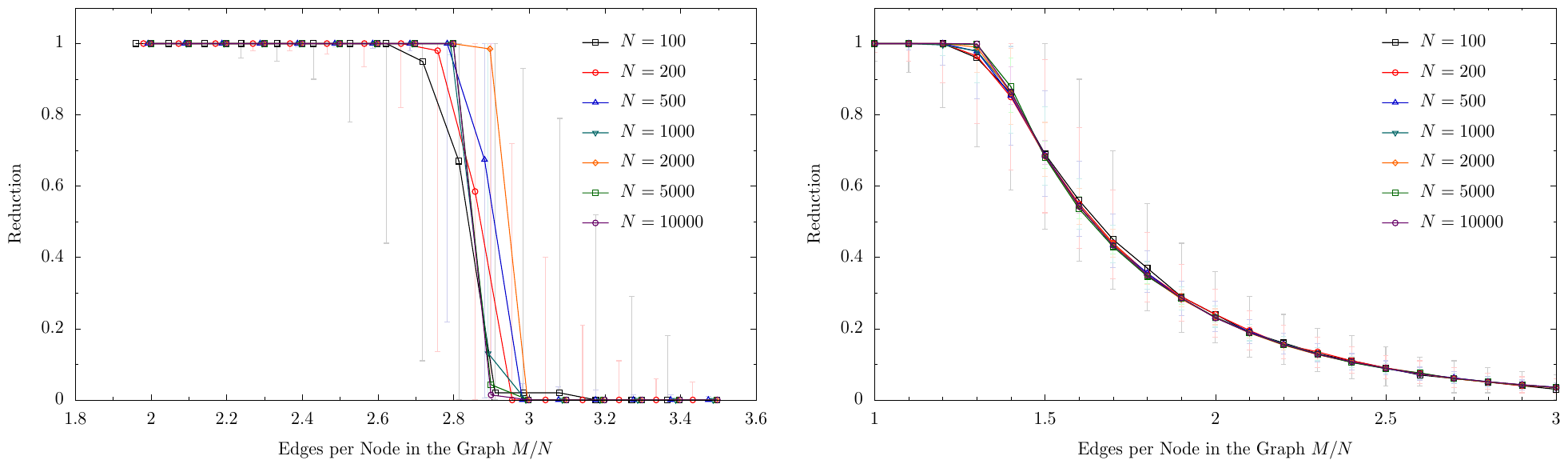}
  \caption{
  Reduction factor $\xi$ as a function of the number of edges $M$ per node, for random BA (left) and ER (right) graphs with up to $10^4$ nodes. 
  For scale-free BA graphs, we observe a sharp transition at a critical average degree of $\bar{d}^{*} \sim 6$, similar to our results for random RG graphs.
  For structureless ER graphs, we observe a smoother transition at relatively small average degree values. 
      \label{fig:reduction-ba-er}
  }
\end{figure*}

\appendix

\section*{Disclaimer}
This paper was prepared for informational purposes with contributions from the Global Technology Applied Research center of JPMorgan Chase \& Co. This paper is not a product of the Research Department of JPMorgan Chase \& Co. or its affiliates. Neither JPMorgan Chase \& Co. nor any of its affiliates makes any explicit or implied representation or warranty and none of them accept any liability in connection with this paper, including, without limitation, with respect to the completeness, accuracy, or reliability of the information contained herein and the potential legal, compliance, tax, or accounting effects thereof. This document is not intended as investment research or investment advice, or as a recommendation, offer, or solicitation for the purchase or sale of any security, financial instrument, financial product or service, or to be used in any way for evaluating the merits of participating in any transaction.

\section{Additional Information}
\label{additional-information}

Here we provide further details for selected aspects discussed in the main text. 

\subsection{Reduction run times}
\label{reduction-uj}

In this section, we provide results for the algorithmic run times of our clique-based reduction algorithm when applied to both random UJ instances and real-world networks. 
Our results are displayed in Figs.~\ref{fig:runtime-reduction-uj} and \ref{fig:runtime-reduction-networks}, respectively. For random UJ instances, we find that all reduction runs finish on sub-second time scales per instance on a laptop (taking on average $\sim 0.02\mathrm{s}$ for instances with $\sim 1000$ nodes). 
In good agreement with Eq.~(\ref{eq:reduction-runtime}) and in line with other numerical results discussed in Section \ref{numerics}, we also observe an approximately linear scaling of the run time with $\sim n$, for large instances with $n>1000$. We observe similar results for our set of ten real-world networks as described in Tab.~\ref{tab:social-networks}. 
Again, we find sub-second run times for instances as large as $n \sim 2\times 10^4$ (for the Pubmed graph), with the data fitting best to an approximate linear scaling $\sim n^{1.02(6)}$, in good agreement with Eq.~(\ref{eq:reduction-runtime}). 

\subsection{Reduction of synthetic (random) graphs}
\label{reduction-synthetic}

In this section, we provide additional results on the clique-based reduction of synthetic (random) graphs, similar to the results shown for RG graphs in the main text. Our results for random BA and ER graphs are shown in Fig.~\ref{fig:reduction-ba-er}.

\bibliography{bibliography}

\begin{thebibliography}{80}
\expandafter\ifx\csname natexlab\endcsname\relax\def\natexlab#1{#1}\fi
\expandafter\ifx\csname bibnamefont\endcsname\relax
  \def\bibnamefont#1{#1}\fi
\expandafter\ifx\csname bibfnamefont\endcsname\relax
  \def\bibfnamefont#1{#1}\fi
\expandafter\ifx\csname citenamefont\endcsname\relax
  \def\citenamefont#1{#1}\fi
\expandafter\ifx\csname url\endcsname\relax
  \def\url#1{\texttt{#1}}\fi
\expandafter\ifx\csname urlprefix\endcsname\relax\def\urlprefix{URL }\fi
\providecommand{\bibinfo}[2]{#2}
\providecommand{\eprint}[2][]{\url{#2}}

\bibitem[{\citenamefont{Papadimitriou and Steiglitz}(1998)}]{papadimitriou:98}
\bibinfo{author}{\bibfnamefont{C.~H.} \bibnamefont{Papadimitriou}}
  \bibnamefont{and}
  \bibinfo{author}{\bibfnamefont{K.}~\bibnamefont{Steiglitz}},
  \emph{\bibinfo{title}{{{Combinatorial Optimization: Algorithms and
  Complexity}}}} (\bibinfo{publisher}{Courier Corporation},
  \bibinfo{address}{North Chelmsford}, \bibinfo{year}{1998}).

\bibitem[{\citenamefont{Korte and Vygen}(2012)}]{korte:12}
\bibinfo{author}{\bibfnamefont{B.}~\bibnamefont{Korte}} \bibnamefont{and}
  \bibinfo{author}{\bibfnamefont{J.}~\bibnamefont{Vygen}},
  \emph{\bibinfo{title}{{Combinatorial Optimization}}},
  vol.~\bibinfo{volume}{2} (\bibinfo{publisher}{Springer},
  \bibinfo{address}{New York}, \bibinfo{year}{2012}).

\bibitem[{\citenamefont{Wurtz et~al.}(2022)\citenamefont{Wurtz, Lopes, Gemelke,
  Keesling, and Wang}}]{wurtz:22}
\bibinfo{author}{\bibfnamefont{J.}~\bibnamefont{Wurtz}},
  \bibinfo{author}{\bibfnamefont{P.~L.~S.} \bibnamefont{Lopes}},
  \bibinfo{author}{\bibfnamefont{N.}~\bibnamefont{Gemelke}},
  \bibinfo{author}{\bibfnamefont{A.}~\bibnamefont{Keesling}}, \bibnamefont{and}
  \bibinfo{author}{\bibfnamefont{S.}~\bibnamefont{Wang}},
  \emph{\bibinfo{title}{{Industry applications of neutral-atom quantum
  computing solving Independent Set problems}}} (\bibinfo{year}{2022}),
  \bibinfo{note}{arXiv:2205.08500}.

\bibitem[{\citenamefont{Hale}(1980)}]{hale:80}
\bibinfo{author}{\bibfnamefont{W.~K.} \bibnamefont{Hale}},
  \emph{\bibinfo{title}{{Frequency assignment: Theory and applications}}},
  \bibinfo{journal}{Proceedings of the IEEE} \textbf{\bibinfo{volume}{68}},
  \bibinfo{pages}{1497} (\bibinfo{year}{1980}).

\bibitem[{\citenamefont{Dong et~al.}(2025)\citenamefont{Dong, Goldberg, Noe,
  Parotsidis, Resende, and Spaen}}]{dong:22}
\bibinfo{author}{\bibfnamefont{Y.}~\bibnamefont{Dong}},
  \bibinfo{author}{\bibfnamefont{A.~V.} \bibnamefont{Goldberg}},
  \bibinfo{author}{\bibfnamefont{A.}~\bibnamefont{Noe}},
  \bibinfo{author}{\bibfnamefont{N.}~\bibnamefont{Parotsidis}},
  \bibinfo{author}{\bibfnamefont{M.~G.~C.} \bibnamefont{Resende}},
  \bibnamefont{and} \bibinfo{author}{\bibfnamefont{Q.}~\bibnamefont{Spaen}},
  \emph{\bibinfo{title}{A metaheuristic algorithm for large maximum weight
  independent set problems}}, \bibinfo{journal}{Networks}
  \textbf{\bibinfo{volume}{85}}, \bibinfo{pages}{91} (\bibinfo{year}{2025}),
  \eprint{https://onlinelibrary.wiley.com/doi/pdf/10.1002/net.22247},
  \urlprefix\url{https://onlinelibrary.wiley.com/doi/abs/10.1002/net.22247}.

\bibitem[{\citenamefont{Boginski et~al.}(2005)\citenamefont{Boginski, Butenko,
  and Pardalos}}]{boginski:05}
\bibinfo{author}{\bibfnamefont{V.}~\bibnamefont{Boginski}},
  \bibinfo{author}{\bibfnamefont{S.}~\bibnamefont{Butenko}}, \bibnamefont{and}
  \bibinfo{author}{\bibfnamefont{P.~M.} \bibnamefont{Pardalos}},
  \emph{\bibinfo{title}{{Statistical analysis of financial networks}}},
  \bibinfo{journal}{Computational Statistics and Data Analysis}
  \textbf{\bibinfo{volume}{48}}, \bibinfo{pages}{431} (\bibinfo{year}{2005}).

\bibitem[{\citenamefont{Kalra et~al.}({2018})\citenamefont{Kalra, Qureshi, and
  Tisi}}]{kalra:08}
\bibinfo{author}{\bibfnamefont{A.}~\bibnamefont{Kalra}},
  \bibinfo{author}{\bibfnamefont{F.}~\bibnamefont{Qureshi}}, \bibnamefont{and}
  \bibinfo{author}{\bibfnamefont{M.}~\bibnamefont{Tisi}},
  \emph{\bibinfo{title}{{Portfolio asset identification using graph algorithms
  on a Quantum Annealer}}}, \bibinfo{journal}{{SSRN}}
  (\bibinfo{year}{{2018}}),
  \urlprefix\url{{https://ssrn.com/abstract=3333537}}.

\bibitem[{\citenamefont{Herman et~al.}(2023)\citenamefont{Herman, Googin, Liu,
  Sun, Galda, Safro, Pistoia, and Alexeev}}]{herman:13}
\bibinfo{author}{\bibfnamefont{D.}~\bibnamefont{Herman}},
  \bibinfo{author}{\bibfnamefont{C.}~\bibnamefont{Googin}},
  \bibinfo{author}{\bibfnamefont{X.}~\bibnamefont{Liu}},
  \bibinfo{author}{\bibfnamefont{Y.}~\bibnamefont{Sun}},
  \bibinfo{author}{\bibfnamefont{A.}~\bibnamefont{Galda}},
  \bibinfo{author}{\bibfnamefont{I.}~\bibnamefont{Safro}},
  \bibinfo{author}{\bibfnamefont{M.}~\bibnamefont{Pistoia}}, \bibnamefont{and}
  \bibinfo{author}{\bibfnamefont{Y.}~\bibnamefont{Alexeev}},
  \emph{\bibinfo{title}{Quantum computing for finance}},
  \bibinfo{journal}{Nature Reviews Physics} \textbf{\bibinfo{volume}{5}},
  \bibinfo{pages}{450} (\bibinfo{year}{2023}),
  \urlprefix\url{https://doi.org/10.1038/s42254-023-00603-1}.

\bibitem[{\citenamefont{Abbas et~al.}(2024)\citenamefont{Abbas, Ambainis,
  Augustino, B{\"a}rtschi, Buhrman, Coffrin, Cortiana, Dunjko, Egger, Elmegreen
  et~al.}}]{abbas:23}
\bibinfo{author}{\bibfnamefont{A.}~\bibnamefont{Abbas}},
  \bibinfo{author}{\bibfnamefont{A.}~\bibnamefont{Ambainis}},
  \bibinfo{author}{\bibfnamefont{B.}~\bibnamefont{Augustino}},
  \bibinfo{author}{\bibfnamefont{A.}~\bibnamefont{B{\"a}rtschi}},
  \bibinfo{author}{\bibfnamefont{H.}~\bibnamefont{Buhrman}},
  \bibinfo{author}{\bibfnamefont{C.}~\bibnamefont{Coffrin}},
  \bibinfo{author}{\bibfnamefont{G.}~\bibnamefont{Cortiana}},
  \bibinfo{author}{\bibfnamefont{V.}~\bibnamefont{Dunjko}},
  \bibinfo{author}{\bibfnamefont{D.~J.} \bibnamefont{Egger}},
  \bibinfo{author}{\bibfnamefont{B.~G.} \bibnamefont{Elmegreen}},
  \bibnamefont{et~al.}, \emph{\bibinfo{title}{Challenges and opportunities in
  quantum optimization}}, \bibinfo{journal}{Nature Reviews Physics}
  \textbf{\bibinfo{volume}{6}}, \bibinfo{pages}{718} (\bibinfo{year}{2024}),
  \urlprefix\url{https://doi.org/10.1038/s42254-024-00770-9}.

\bibitem[{\citenamefont{Kadowaki and Nishimori}(1998)}]{kadowaki:98}
\bibinfo{author}{\bibfnamefont{T.}~\bibnamefont{Kadowaki}} \bibnamefont{and}
  \bibinfo{author}{\bibfnamefont{H.}~\bibnamefont{Nishimori}},
  \emph{\bibinfo{title}{Quantum annealing in the transverse ising model}},
  \bibinfo{journal}{Phys. Rev. E} \textbf{\bibinfo{volume}{58}},
  \bibinfo{pages}{5355} (\bibinfo{year}{1998}),
  \urlprefix\url{https://link.aps.org/doi/10.1103/PhysRevE.58.5355}.

\bibitem[{\citenamefont{Farhi et~al.}(2000)\citenamefont{Farhi, Goldstone,
  Gutmann, and Sipser}}]{farhi:00}
\bibinfo{author}{\bibfnamefont{E.}~\bibnamefont{Farhi}},
  \bibinfo{author}{\bibfnamefont{J.}~\bibnamefont{Goldstone}},
  \bibinfo{author}{\bibfnamefont{S.}~\bibnamefont{Gutmann}}, \bibnamefont{and}
  \bibinfo{author}{\bibfnamefont{M.}~\bibnamefont{Sipser}},
  \emph{\bibinfo{title}{Quantum computation by adiabatic evolution}}
  (\bibinfo{year}{2000}), \eprint{arXiv:quant-ph/0001106}.

\bibitem[{\citenamefont{Farhi et~al.}(2001)\citenamefont{Farhi, Goldstone,
  Gutmann, Lapan, Lundgren, and Preda}}]{farhi:01}
\bibinfo{author}{\bibfnamefont{E.}~\bibnamefont{Farhi}},
  \bibinfo{author}{\bibfnamefont{J.}~\bibnamefont{Goldstone}},
  \bibinfo{author}{\bibfnamefont{S.}~\bibnamefont{Gutmann}},
  \bibinfo{author}{\bibfnamefont{J.}~\bibnamefont{Lapan}},
  \bibinfo{author}{\bibfnamefont{A.}~\bibnamefont{Lundgren}}, \bibnamefont{and}
  \bibinfo{author}{\bibfnamefont{D.}~\bibnamefont{Preda}},
  \emph{\bibinfo{title}{A quantum adiabatic evolution algorithm applied to
  random instances of an np-complete problem}}, \bibinfo{journal}{Science}
  \textbf{\bibinfo{volume}{292}}, \bibinfo{pages}{472} (\bibinfo{year}{2001}),
  \urlprefix\url{https://www.science.org/doi/abs/10.1126/science.1057726}.

\bibitem[{\citenamefont{Das and Chakrabarti}(2008)}]{das:08}
\bibinfo{author}{\bibfnamefont{A.}~\bibnamefont{Das}} \bibnamefont{and}
  \bibinfo{author}{\bibfnamefont{B.~K.} \bibnamefont{Chakrabarti}},
  \emph{\bibinfo{title}{Colloquium: Quantum annealing and analog quantum
  computation}}, \bibinfo{journal}{Rev. Mod. Phys.}
  \textbf{\bibinfo{volume}{80}}, \bibinfo{pages}{1061} (\bibinfo{year}{2008}),
  \urlprefix\url{https://link.aps.org/doi/10.1103/RevModPhys.80.1061}.

\bibitem[{\citenamefont{Hauke et~al.}(2020)\citenamefont{Hauke, Katzgraber,
  Lechner, Nishimori, and Oliver}}]{hauke:20}
\bibinfo{author}{\bibfnamefont{P.}~\bibnamefont{Hauke}},
  \bibinfo{author}{\bibfnamefont{H.~G.} \bibnamefont{Katzgraber}},
  \bibinfo{author}{\bibfnamefont{W.}~\bibnamefont{Lechner}},
  \bibinfo{author}{\bibfnamefont{H.}~\bibnamefont{Nishimori}},
  \bibnamefont{and} \bibinfo{author}{\bibfnamefont{W.}~\bibnamefont{Oliver}},
  \emph{\bibinfo{title}{Perspectives of quantum annealing: methods and
  implementations}}, \bibinfo{journal}{Rep. Prog. Phys.}
  \textbf{\bibinfo{volume}{83}}, \bibinfo{pages}{054401}
  (\bibinfo{year}{2020}).

\bibitem[{\citenamefont{Farhi et~al.}(2014)\citenamefont{Farhi, Goldstone, and
  Gutmann}}]{farhi:14}
\bibinfo{author}{\bibfnamefont{E.}~\bibnamefont{Farhi}},
  \bibinfo{author}{\bibfnamefont{J.}~\bibnamefont{Goldstone}},
  \bibnamefont{and} \bibinfo{author}{\bibfnamefont{S.}~\bibnamefont{Gutmann}},
  \emph{\bibinfo{title}{A quantum approximate optimization algorithm}}
  (\bibinfo{year}{2014}), \eprint{arXiv:1411.4028}.

\bibitem[{\citenamefont{Zhou et~al.}(2020)\citenamefont{Zhou, Wang, Choi,
  Pichler, and Lukin}}]{zhou:20}
\bibinfo{author}{\bibfnamefont{L.}~\bibnamefont{Zhou}},
  \bibinfo{author}{\bibfnamefont{S.-T.} \bibnamefont{Wang}},
  \bibinfo{author}{\bibfnamefont{S.}~\bibnamefont{Choi}},
  \bibinfo{author}{\bibfnamefont{H.}~\bibnamefont{Pichler}}, \bibnamefont{and}
  \bibinfo{author}{\bibfnamefont{M.~D.} \bibnamefont{Lukin}},
  \emph{\bibinfo{title}{{Quantum Approximate Optimization Algorithm:
  Performance, mechanism, and implementation on near-term devices}}},
  \bibinfo{journal}{Phys. Rev. X} \textbf{\bibinfo{volume}{10}},
  \bibinfo{pages}{021067} (\bibinfo{year}{2020}),
  \urlprefix\url{https://link.aps.org/doi/10.1103/PhysRevX.10.021067}.

\bibitem[{\citenamefont{Lucas}(2014)}]{lucas:14}
\bibinfo{author}{\bibfnamefont{A.}~\bibnamefont{Lucas}},
  \emph{\bibinfo{title}{{Ising formulations of many NP problems}}},
  \bibinfo{journal}{Front. Physics} \textbf{\bibinfo{volume}{2}},
  \bibinfo{pages}{5} (\bibinfo{year}{2014}).

\bibitem[{\citenamefont{{Glover} et~al.}(2019)\citenamefont{{Glover},
  {Kochenberger}, and {Du}}}]{glover:18}
\bibinfo{author}{\bibfnamefont{F.}~\bibnamefont{{Glover}}},
  \bibinfo{author}{\bibfnamefont{G.}~\bibnamefont{{Kochenberger}}},
  \bibnamefont{and} \bibinfo{author}{\bibfnamefont{Y.}~\bibnamefont{{Du}}},
  \emph{\bibinfo{title}{{{Quantum Bridge Analytics I: A Tutorial on Formulating
  and Using QUBO Models}}}}, \bibinfo{journal}{4OR}
  \textbf{\bibinfo{volume}{17}}, \bibinfo{pages}{335} (\bibinfo{year}{2019}).

\bibitem[{\citenamefont{Pichler
  et~al.}(2018{\natexlab{a}})\citenamefont{Pichler, Wang, Zhou, Choi, and
  Lukin}}]{pichler:18}
\bibinfo{author}{\bibfnamefont{H.}~\bibnamefont{Pichler}},
  \bibinfo{author}{\bibfnamefont{S.-T.} \bibnamefont{Wang}},
  \bibinfo{author}{\bibfnamefont{L.}~\bibnamefont{Zhou}},
  \bibinfo{author}{\bibfnamefont{S.}~\bibnamefont{Choi}}, \bibnamefont{and}
  \bibinfo{author}{\bibfnamefont{M.~D.} \bibnamefont{Lukin}},
  \emph{\bibinfo{title}{{Quantum optimization for Maximum Independent Set using
  Rydberg atom arrays}}} (\bibinfo{year}{2018}{\natexlab{a}}),
  \bibinfo{note}{arXiv:1808.10816}.

\bibitem[{\citenamefont{Pichler
  et~al.}(2018{\natexlab{b}})\citenamefont{Pichler, Wang, Zhou, Choi, and
  Lukin}}]{pichler:18computational}
\bibinfo{author}{\bibfnamefont{H.}~\bibnamefont{Pichler}},
  \bibinfo{author}{\bibfnamefont{S.-T.} \bibnamefont{Wang}},
  \bibinfo{author}{\bibfnamefont{L.}~\bibnamefont{Zhou}},
  \bibinfo{author}{\bibfnamefont{S.}~\bibnamefont{Choi}}, \bibnamefont{and}
  \bibinfo{author}{\bibfnamefont{M.~D.} \bibnamefont{Lukin}},
  \emph{\bibinfo{title}{{Computational complexity of the Rydberg Blockade in
  two dimensions}}} (\bibinfo{year}{2018}{\natexlab{b}}),
  \bibinfo{note}{arXiv:1809.04954}.

\bibitem[{\citenamefont{Serret et~al.}(2020)\citenamefont{Serret, Marchand, and
  Ayral}}]{serret:20}
\bibinfo{author}{\bibfnamefont{M.~F.} \bibnamefont{Serret}},
  \bibinfo{author}{\bibfnamefont{B.}~\bibnamefont{Marchand}}, \bibnamefont{and}
  \bibinfo{author}{\bibfnamefont{T.}~\bibnamefont{Ayral}},
  \emph{\bibinfo{title}{{Solving optimization problems with Rydberg analog
  quantum computers: Realistic requirements for quantum advantage using noisy
  simulation and classical benchmarks}}}, \bibinfo{journal}{Phys. Rev. A}
  \textbf{\bibinfo{volume}{102}}, \bibinfo{pages}{052617}
  (\bibinfo{year}{2020}),
  \urlprefix\url{https://link.aps.org/doi/10.1103/PhysRevA.102.052617}.

\bibitem[{\citenamefont{Ebadi et~al.}(2022)\citenamefont{Ebadi, Keesling, Cain,
  Wang, Levine, Bluvstein, Semeghini, Omran, Liu, Samajdar et~al.}}]{ebadi:22}
\bibinfo{author}{\bibfnamefont{S.}~\bibnamefont{Ebadi}},
  \bibinfo{author}{\bibfnamefont{A.}~\bibnamefont{Keesling}},
  \bibinfo{author}{\bibfnamefont{M.}~\bibnamefont{Cain}},
  \bibinfo{author}{\bibfnamefont{T.~T.} \bibnamefont{Wang}},
  \bibinfo{author}{\bibfnamefont{H.}~\bibnamefont{Levine}},
  \bibinfo{author}{\bibfnamefont{D.}~\bibnamefont{Bluvstein}},
  \bibinfo{author}{\bibfnamefont{G.}~\bibnamefont{Semeghini}},
  \bibinfo{author}{\bibfnamefont{A.}~\bibnamefont{Omran}},
  \bibinfo{author}{\bibfnamefont{J.-G.} \bibnamefont{Liu}},
  \bibinfo{author}{\bibfnamefont{R.}~\bibnamefont{Samajdar}},
  \bibnamefont{et~al.}, \emph{\bibinfo{title}{{Quantum optimization of Maximum
  Independent Set using Rydberg atom arrays}}}, \bibinfo{journal}{Science}
  \textbf{\bibinfo{volume}{376}}, \bibinfo{pages}{1209} (\bibinfo{year}{2022}),
  \urlprefix\url{https://doi.org/10.1126/science.abo6587}.

\bibitem[{\citenamefont{Cain et~al.}(2023)\citenamefont{Cain, Chattopadhyay,
  Liu, Samajdar, Pichler, and Lukin}}]{cain:23}
\bibinfo{author}{\bibfnamefont{M.}~\bibnamefont{Cain}},
  \bibinfo{author}{\bibfnamefont{S.}~\bibnamefont{Chattopadhyay}},
  \bibinfo{author}{\bibfnamefont{J.-G.} \bibnamefont{Liu}},
  \bibinfo{author}{\bibfnamefont{R.}~\bibnamefont{Samajdar}},
  \bibinfo{author}{\bibfnamefont{H.}~\bibnamefont{Pichler}}, \bibnamefont{and}
  \bibinfo{author}{\bibfnamefont{M.~D.} \bibnamefont{Lukin}},
  \emph{\bibinfo{title}{{Quantum speedup for combinatorial optimization with
  flat energy landscapes}}} (\bibinfo{year}{2023}),
  \bibinfo{note}{arXiv:2306.13123}.

\bibitem[{\citenamefont{Schiffer et~al.}(2024)\citenamefont{Schiffer, Wild,
  Maskara, Cain, Lukin, and Samajdar}}]{schiffer:23}
\bibinfo{author}{\bibfnamefont{B.~F.} \bibnamefont{Schiffer}},
  \bibinfo{author}{\bibfnamefont{D.~S.} \bibnamefont{Wild}},
  \bibinfo{author}{\bibfnamefont{N.}~\bibnamefont{Maskara}},
  \bibinfo{author}{\bibfnamefont{M.}~\bibnamefont{Cain}},
  \bibinfo{author}{\bibfnamefont{M.~D.} \bibnamefont{Lukin}}, \bibnamefont{and}
  \bibinfo{author}{\bibfnamefont{R.}~\bibnamefont{Samajdar}},
  \emph{\bibinfo{title}{Circumventing superexponential runtimes for hard
  instances of quantum adiabatic optimization}}, \bibinfo{journal}{Phys. Rev.
  Res.} \textbf{\bibinfo{volume}{6}}, \bibinfo{pages}{013271}
  (\bibinfo{year}{2024}),
  \urlprefix\url{https://link.aps.org/doi/10.1103/PhysRevResearch.6.013271}.

\bibitem[{\citenamefont{Fin\ifmmode~\check{z}\else \v{z}\fi{}gar
  et~al.}(2024{\natexlab{a}})\citenamefont{Fin\ifmmode~\check{z}\else
  \v{z}\fi{}gar, Schuetz, Brubaker, Nishimori, and Katzgraber}}]{finzgar:23}
\bibinfo{author}{\bibfnamefont{J.~R.} \bibnamefont{Fin\ifmmode~\check{z}\else
  \v{z}\fi{}gar}}, \bibinfo{author}{\bibfnamefont{M.~J.~A.}
  \bibnamefont{Schuetz}}, \bibinfo{author}{\bibfnamefont{J.~K.}
  \bibnamefont{Brubaker}},
  \bibinfo{author}{\bibfnamefont{H.}~\bibnamefont{Nishimori}},
  \bibnamefont{and} \bibinfo{author}{\bibfnamefont{H.~G.}
  \bibnamefont{Katzgraber}}, \emph{\bibinfo{title}{Designing quantum annealing
  schedules using bayesian optimization}}, \bibinfo{journal}{Phys. Rev. Res.}
  \textbf{\bibinfo{volume}{6}}, \bibinfo{pages}{023063}
  (\bibinfo{year}{2024}{\natexlab{a}}),
  \urlprefix\url{https://link.aps.org/doi/10.1103/PhysRevResearch.6.023063}.

\bibitem[{\citenamefont{Fin\ifmmode~\check{z}\else \v{z}\fi{}gar
  et~al.}(2024{\natexlab{b}})\citenamefont{Fin\ifmmode~\check{z}\else
  \v{z}\fi{}gar, Kerschbaumer, Schuetz, Mendl, and Katzgraber}}]{finzgar:23b}
\bibinfo{author}{\bibfnamefont{J.~R.} \bibnamefont{Fin\ifmmode~\check{z}\else
  \v{z}\fi{}gar}},
  \bibinfo{author}{\bibfnamefont{A.}~\bibnamefont{Kerschbaumer}},
  \bibinfo{author}{\bibfnamefont{M.~J.} \bibnamefont{Schuetz}},
  \bibinfo{author}{\bibfnamefont{C.~B.} \bibnamefont{Mendl}}, \bibnamefont{and}
  \bibinfo{author}{\bibfnamefont{H.~G.} \bibnamefont{Katzgraber}},
  \emph{\bibinfo{title}{Quantum-informed recursive optimization algorithms}},
  \bibinfo{journal}{PRX Quantum} \textbf{\bibinfo{volume}{5}},
  \bibinfo{pages}{020327} (\bibinfo{year}{2024}{\natexlab{b}}),
  \urlprefix\url{https://link.aps.org/doi/10.1103/PRXQuantum.5.020327}.

\bibitem[{\citenamefont{Kim et~al.}(2024)\citenamefont{Kim, Kim, Park, Byun,
  and Ahn}}]{kim:23}
\bibinfo{author}{\bibfnamefont{K.}~\bibnamefont{Kim}},
  \bibinfo{author}{\bibfnamefont{M.}~\bibnamefont{Kim}},
  \bibinfo{author}{\bibfnamefont{J.}~\bibnamefont{Park}},
  \bibinfo{author}{\bibfnamefont{A.}~\bibnamefont{Byun}}, \bibnamefont{and}
  \bibinfo{author}{\bibfnamefont{J.}~\bibnamefont{Ahn}},
  \emph{\bibinfo{title}{Quantum computing dataset of maximum independent set
  problem on king lattice of over hundred rydberg atoms}},
  \bibinfo{journal}{Scientific Data} \textbf{\bibinfo{volume}{11}},
  \bibinfo{pages}{111} (\bibinfo{year}{2024}),
  \urlprefix\url{https://doi.org/10.1038/s41597-024-02926-9}.

\bibitem[{\citenamefont{Andrist et~al.}(2023)\citenamefont{Andrist, Schuetz,
  Minssen, Yalovetzky, Chakrabarti, Herman, Kumar, Salton, Shaydulin, Sun
  et~al.}}]{andrist:23}
\bibinfo{author}{\bibfnamefont{R.~S.} \bibnamefont{Andrist}},
  \bibinfo{author}{\bibfnamefont{M.~J.~A.} \bibnamefont{Schuetz}},
  \bibinfo{author}{\bibfnamefont{P.}~\bibnamefont{Minssen}},
  \bibinfo{author}{\bibfnamefont{R.}~\bibnamefont{Yalovetzky}},
  \bibinfo{author}{\bibfnamefont{S.}~\bibnamefont{Chakrabarti}},
  \bibinfo{author}{\bibfnamefont{D.}~\bibnamefont{Herman}},
  \bibinfo{author}{\bibfnamefont{N.}~\bibnamefont{Kumar}},
  \bibinfo{author}{\bibfnamefont{G.}~\bibnamefont{Salton}},
  \bibinfo{author}{\bibfnamefont{R.}~\bibnamefont{Shaydulin}},
  \bibinfo{author}{\bibfnamefont{Y.}~\bibnamefont{Sun}}, \bibnamefont{et~al.},
  \emph{\bibinfo{title}{Hardness of the maximum-independent-set problem on
  unit-disk graphs and prospects for quantum speedups}},
  \bibinfo{journal}{Phys. Rev. Res.} \textbf{\bibinfo{volume}{5}},
  \bibinfo{pages}{043277} (\bibinfo{year}{2023}),
  \urlprefix\url{https://link.aps.org/doi/10.1103/PhysRevResearch.5.043277}.

\bibitem[{\citenamefont{Lukin et~al.}(2001)\citenamefont{Lukin, Fleischhauer,
  Cote, Duan, Jaksch, Cirac, and Zoller}}]{lukin:01}
\bibinfo{author}{\bibfnamefont{M.~D.} \bibnamefont{Lukin}},
  \bibinfo{author}{\bibfnamefont{M.}~\bibnamefont{Fleischhauer}},
  \bibinfo{author}{\bibfnamefont{R.}~\bibnamefont{Cote}},
  \bibinfo{author}{\bibfnamefont{L.~M.} \bibnamefont{Duan}},
  \bibinfo{author}{\bibfnamefont{D.}~\bibnamefont{Jaksch}},
  \bibinfo{author}{\bibfnamefont{J.~I.} \bibnamefont{Cirac}}, \bibnamefont{and}
  \bibinfo{author}{\bibfnamefont{P.}~\bibnamefont{Zoller}},
  \emph{\bibinfo{title}{{Dipole Blockade and quantum information processing in
  mesoscopic atomic ensembles}}}, \bibinfo{journal}{Phys. Rev. Lett.}
  \textbf{\bibinfo{volume}{87}}, \bibinfo{pages}{037901}
  (\bibinfo{year}{2001}),
  \urlprefix\url{https://link.aps.org/doi/10.1103/PhysRevLett.87.037901}.

\bibitem[{\citenamefont{Levine et~al.}(2019)\citenamefont{Levine, Keesling,
  Semeghini, Omran, Wang, Ebadi, Bernien, Greiner,
  Vuleti\ifmmode~\acute{c}\else \'{c}\fi{}, Pichler et~al.}}]{levine:19}
\bibinfo{author}{\bibfnamefont{H.}~\bibnamefont{Levine}},
  \bibinfo{author}{\bibfnamefont{A.}~\bibnamefont{Keesling}},
  \bibinfo{author}{\bibfnamefont{G.}~\bibnamefont{Semeghini}},
  \bibinfo{author}{\bibfnamefont{A.}~\bibnamefont{Omran}},
  \bibinfo{author}{\bibfnamefont{T.~T.} \bibnamefont{Wang}},
  \bibinfo{author}{\bibfnamefont{S.}~\bibnamefont{Ebadi}},
  \bibinfo{author}{\bibfnamefont{H.}~\bibnamefont{Bernien}},
  \bibinfo{author}{\bibfnamefont{M.}~\bibnamefont{Greiner}},
  \bibinfo{author}{\bibfnamefont{V.}~\bibnamefont{Vuleti\ifmmode~\acute{c}\else
  \'{c}\fi{}}}, \bibinfo{author}{\bibfnamefont{H.}~\bibnamefont{Pichler}},
  \bibnamefont{et~al.}, \emph{\bibinfo{title}{{Parallel implementation of
  high-fidelity multiqubit gates with neutral atoms}}}, \bibinfo{journal}{Phys.
  Rev. Lett.} \textbf{\bibinfo{volume}{123}}, \bibinfo{pages}{170503}
  (\bibinfo{year}{2019}),
  \urlprefix\url{https://link.aps.org/doi/10.1103/PhysRevLett.123.170503}.

\bibitem[{\citenamefont{Saffman et~al.}(2010)\citenamefont{Saffman, Walker, and
  M\o{}lmer}}]{saffman:10}
\bibinfo{author}{\bibfnamefont{M.}~\bibnamefont{Saffman}},
  \bibinfo{author}{\bibfnamefont{T.~G.} \bibnamefont{Walker}},
  \bibnamefont{and}
  \bibinfo{author}{\bibfnamefont{K.}~\bibnamefont{M\o{}lmer}},
  \emph{\bibinfo{title}{{Quantum information with Rydberg atoms}}},
  \bibinfo{journal}{Rev. Mod. Phys.} \textbf{\bibinfo{volume}{82}},
  \bibinfo{pages}{2313} (\bibinfo{year}{2010}),
  \urlprefix\url{https://link.aps.org/doi/10.1103/RevModPhys.82.2313}.

\bibitem[{\citenamefont{Clark et~al.}(1990)\citenamefont{Clark, Colbourn, and
  Johnson}}]{clark:90}
\bibinfo{author}{\bibfnamefont{B.~N.} \bibnamefont{Clark}},
  \bibinfo{author}{\bibfnamefont{C.~J.} \bibnamefont{Colbourn}},
  \bibnamefont{and} \bibinfo{author}{\bibfnamefont{D.~S.}
  \bibnamefont{Johnson}}, \emph{\bibinfo{title}{{Unit disk graphs}}},
  \bibinfo{journal}{Discrete Mathematics} \textbf{\bibinfo{volume}{86}},
  \bibinfo{pages}{165} (\bibinfo{year}{1990}), ISSN \bibinfo{issn}{0012-365X},
  \urlprefix\url{https://www.sciencedirect.com/science/article/pii/0012365X9090358O}.

\bibitem[{\citenamefont{K\"onz et~al.}(2021)\citenamefont{K\"onz, Lechner,
  Katzgraber, and Troyer}}]{koenz:21}
\bibinfo{author}{\bibfnamefont{M.~S.} \bibnamefont{K\"onz}},
  \bibinfo{author}{\bibfnamefont{W.}~\bibnamefont{Lechner}},
  \bibinfo{author}{\bibfnamefont{H.~G.} \bibnamefont{Katzgraber}},
  \bibnamefont{and} \bibinfo{author}{\bibfnamefont{M.}~\bibnamefont{Troyer}},
  \emph{\bibinfo{title}{Embedding overhead scaling of optimization problems in
  quantum annealing}}, \bibinfo{journal}{PRX Quantum}
  \textbf{\bibinfo{volume}{2}}, \bibinfo{pages}{040322} (\bibinfo{year}{2021}),
  \urlprefix\url{https://link.aps.org/doi/10.1103/PRXQuantum.2.040322}.

\bibitem[{\citenamefont{Bunyk et~al.}(2014)\citenamefont{Bunyk, Hoskinson,
  Johnson, Tolkacheva, Altomare, Berkley, Harris, Hilton, Lanting, Przybysz
  et~al.}}]{bunyk:14}
\bibinfo{author}{\bibfnamefont{P.~I.} \bibnamefont{Bunyk}},
  \bibinfo{author}{\bibfnamefont{E.~M.} \bibnamefont{Hoskinson}},
  \bibinfo{author}{\bibfnamefont{M.~W.} \bibnamefont{Johnson}},
  \bibinfo{author}{\bibfnamefont{E.}~\bibnamefont{Tolkacheva}},
  \bibinfo{author}{\bibfnamefont{F.}~\bibnamefont{Altomare}},
  \bibinfo{author}{\bibfnamefont{A.~J.} \bibnamefont{Berkley}},
  \bibinfo{author}{\bibfnamefont{R.}~\bibnamefont{Harris}},
  \bibinfo{author}{\bibfnamefont{J.~P.} \bibnamefont{Hilton}},
  \bibinfo{author}{\bibfnamefont{T.}~\bibnamefont{Lanting}},
  \bibinfo{author}{\bibfnamefont{A.~J.} \bibnamefont{Przybysz}},
  \bibnamefont{et~al.}, \emph{\bibinfo{title}{Architectural considerations in
  the design of a superconducting quantum annealing processor}},
  \bibinfo{journal}{IEEE Transactions on Applied Superconductivity}
  \textbf{\bibinfo{volume}{24}}, \bibinfo{pages}{1} (\bibinfo{year}{2014}).

\bibitem[{\citenamefont{Vinci et~al.}(2015)\citenamefont{Vinci, Albash,
  Paz-Silva, Hen, and Lidar}}]{vinci:15}
\bibinfo{author}{\bibfnamefont{W.}~\bibnamefont{Vinci}},
  \bibinfo{author}{\bibfnamefont{T.}~\bibnamefont{Albash}},
  \bibinfo{author}{\bibfnamefont{G.}~\bibnamefont{Paz-Silva}},
  \bibinfo{author}{\bibfnamefont{I.}~\bibnamefont{Hen}}, \bibnamefont{and}
  \bibinfo{author}{\bibfnamefont{D.~A.} \bibnamefont{Lidar}},
  \emph{\bibinfo{title}{Quantum annealing correction with minor embedding}},
  \bibinfo{journal}{Phys. Rev. A} \textbf{\bibinfo{volume}{92}},
  \bibinfo{pages}{042310} (\bibinfo{year}{2015}),
  \urlprefix\url{https://link.aps.org/doi/10.1103/PhysRevA.92.042310}.

\bibitem[{\citenamefont{Sugie et~al.}(2021)\citenamefont{Sugie, Yoshida,
  Mertig, Takemoto, Teramoto, Nakamura, Takigawa, Minato, Yamaoka, and
  Komatsuzaki}}]{sugie:21}
\bibinfo{author}{\bibfnamefont{Y.}~\bibnamefont{Sugie}},
  \bibinfo{author}{\bibfnamefont{Y.}~\bibnamefont{Yoshida}},
  \bibinfo{author}{\bibfnamefont{N.}~\bibnamefont{Mertig}},
  \bibinfo{author}{\bibfnamefont{T.}~\bibnamefont{Takemoto}},
  \bibinfo{author}{\bibfnamefont{H.}~\bibnamefont{Teramoto}},
  \bibinfo{author}{\bibfnamefont{A.}~\bibnamefont{Nakamura}},
  \bibinfo{author}{\bibfnamefont{I.}~\bibnamefont{Takigawa}},
  \bibinfo{author}{\bibfnamefont{S.-i.} \bibnamefont{Minato}},
  \bibinfo{author}{\bibfnamefont{M.}~\bibnamefont{Yamaoka}}, \bibnamefont{and}
  \bibinfo{author}{\bibfnamefont{T.}~\bibnamefont{Komatsuzaki}},
  \emph{\bibinfo{title}{Minor-embedding heuristics for large-scale annealing
  processors with sparse hardware graphs of up to 102,400 nodes}},
  \bibinfo{journal}{Soft Computing} \textbf{\bibinfo{volume}{25}},
  \bibinfo{pages}{1731} (\bibinfo{year}{2021}),
  \urlprefix\url{https://doi.org/10.1007/s00500-020-05502-6}.

\bibitem[{\citenamefont{Byun et~al.}(2022)\citenamefont{Byun, Kim, and
  Ahn}}]{byun:22}
\bibinfo{author}{\bibfnamefont{A.}~\bibnamefont{Byun}},
  \bibinfo{author}{\bibfnamefont{M.}~\bibnamefont{Kim}}, \bibnamefont{and}
  \bibinfo{author}{\bibfnamefont{J.}~\bibnamefont{Ahn}},
  \emph{\bibinfo{title}{Finding the maximum independent sets of platonic graphs
  using rydberg atoms}}, \bibinfo{journal}{PRX Quantum}
  \textbf{\bibinfo{volume}{3}}, \bibinfo{pages}{030305} (\bibinfo{year}{2022}),
  \urlprefix\url{https://link.aps.org/doi/10.1103/PRXQuantum.3.030305}.

\bibitem[{\citenamefont{Byun et~al.}(2023)\citenamefont{Byun, Jung, Kim, Kim,
  Jeong, Jeong, and Ahn}}]{byun:23}
\bibinfo{author}{\bibfnamefont{A.}~\bibnamefont{Byun}},
  \bibinfo{author}{\bibfnamefont{J.}~\bibnamefont{Jung}},
  \bibinfo{author}{\bibfnamefont{K.}~\bibnamefont{Kim}},
  \bibinfo{author}{\bibfnamefont{M.}~\bibnamefont{Kim}},
  \bibinfo{author}{\bibfnamefont{S.}~\bibnamefont{Jeong}},
  \bibinfo{author}{\bibfnamefont{H.}~\bibnamefont{Jeong}}, \bibnamefont{and}
  \bibinfo{author}{\bibfnamefont{J.}~\bibnamefont{Ahn}},
  \emph{\bibinfo{title}{Rydberg-atom graphs for quadratic unconstrained binary
  optimization problems}} (\bibinfo{year}{2023}), \eprint{arXiv:2309.14847}.

\bibitem[{\citenamefont{Kim et~al.}(2022)\citenamefont{Kim, Kim, Hwang, Moon,
  and Ahn}}]{kim:22}
\bibinfo{author}{\bibfnamefont{M.}~\bibnamefont{Kim}},
  \bibinfo{author}{\bibfnamefont{K.}~\bibnamefont{Kim}},
  \bibinfo{author}{\bibfnamefont{J.}~\bibnamefont{Hwang}},
  \bibinfo{author}{\bibfnamefont{E.-G.} \bibnamefont{Moon}}, \bibnamefont{and}
  \bibinfo{author}{\bibfnamefont{J.}~\bibnamefont{Ahn}},
  \emph{\bibinfo{title}{{Rydberg quantum wires for Maximum Independent Set
  problems}}}, \bibinfo{journal}{Nature Physics} \textbf{\bibinfo{volume}{18}},
  \bibinfo{pages}{755} (\bibinfo{year}{2022}),
  \urlprefix\url{https://doi.org/10.1038/s41567-022-01629-5}.

\bibitem[{\citenamefont{Dlaska et~al.}(2022)\citenamefont{Dlaska, Ender, Mbeng,
  Kruckenhauser, Lechner, and van Bijnen}}]{dlaska:22}
\bibinfo{author}{\bibfnamefont{C.}~\bibnamefont{Dlaska}},
  \bibinfo{author}{\bibfnamefont{K.}~\bibnamefont{Ender}},
  \bibinfo{author}{\bibfnamefont{G.~B.} \bibnamefont{Mbeng}},
  \bibinfo{author}{\bibfnamefont{A.}~\bibnamefont{Kruckenhauser}},
  \bibinfo{author}{\bibfnamefont{W.}~\bibnamefont{Lechner}}, \bibnamefont{and}
  \bibinfo{author}{\bibfnamefont{R.}~\bibnamefont{van Bijnen}},
  \emph{\bibinfo{title}{Quantum optimization via four-body rydberg gates}},
  \bibinfo{journal}{Phys. Rev. Lett.} \textbf{\bibinfo{volume}{128}},
  \bibinfo{pages}{120503} (\bibinfo{year}{2022}),
  \urlprefix\url{https://link.aps.org/doi/10.1103/PhysRevLett.128.120503}.

\bibitem[{\citenamefont{Nguyen et~al.}(2023)\citenamefont{Nguyen, Liu, Wurtz,
  Lukin, Wang, and Pichler}}]{nguyen:23}
\bibinfo{author}{\bibfnamefont{M.-T.} \bibnamefont{Nguyen}},
  \bibinfo{author}{\bibfnamefont{J.-G.} \bibnamefont{Liu}},
  \bibinfo{author}{\bibfnamefont{J.}~\bibnamefont{Wurtz}},
  \bibinfo{author}{\bibfnamefont{M.~D.} \bibnamefont{Lukin}},
  \bibinfo{author}{\bibfnamefont{S.-T.} \bibnamefont{Wang}}, \bibnamefont{and}
  \bibinfo{author}{\bibfnamefont{H.}~\bibnamefont{Pichler}},
  \emph{\bibinfo{title}{{Quantum optimization with arbitrary connectivity using
  Rydberg atom arrays}}}, \bibinfo{journal}{PRX Quantum}
  \textbf{\bibinfo{volume}{4}}, \bibinfo{pages}{010316} (\bibinfo{year}{2023}),
  \urlprefix\url{https://link.aps.org/doi/10.1103/PRXQuantum.4.010316}.

\bibitem[{\citenamefont{de~Oliveira et~al.}(2024)\citenamefont{de~Oliveira,
  Diamond-Hitchcock, Walker, Wells-Pestell, Pelegrí, Picken, Malcolm, Daley,
  Bass, and Pritchard}}]{deoliveira:24}
\bibinfo{author}{\bibfnamefont{A.~G.} \bibnamefont{de~Oliveira}},
  \bibinfo{author}{\bibfnamefont{E.}~\bibnamefont{Diamond-Hitchcock}},
  \bibinfo{author}{\bibfnamefont{D.~M.} \bibnamefont{Walker}},
  \bibinfo{author}{\bibfnamefont{M.~T.} \bibnamefont{Wells-Pestell}},
  \bibinfo{author}{\bibfnamefont{G.}~\bibnamefont{Pelegrí}},
  \bibinfo{author}{\bibfnamefont{C.~J.} \bibnamefont{Picken}},
  \bibinfo{author}{\bibfnamefont{G.~P.~A.} \bibnamefont{Malcolm}},
  \bibinfo{author}{\bibfnamefont{A.~J.} \bibnamefont{Daley}},
  \bibinfo{author}{\bibfnamefont{J.}~\bibnamefont{Bass}}, \bibnamefont{and}
  \bibinfo{author}{\bibfnamefont{J.~D.} \bibnamefont{Pritchard}},
  \emph{\bibinfo{title}{Demonstration of weighted graph optimization on a
  rydberg atom array using local light-shifts}} (\bibinfo{year}{2024}),
  \eprint{arXiv:2404.02658}.

\bibitem[{\citenamefont{Bluvstein et~al.}(2022)\citenamefont{Bluvstein, Levine,
  Semeghini, Wang, Ebadi, Kalinowski, Keesling, Maskara, Pichler, Greiner
  et~al.}}]{bluvstein:21}
\bibinfo{author}{\bibfnamefont{D.}~\bibnamefont{Bluvstein}},
  \bibinfo{author}{\bibfnamefont{H.}~\bibnamefont{Levine}},
  \bibinfo{author}{\bibfnamefont{G.}~\bibnamefont{Semeghini}},
  \bibinfo{author}{\bibfnamefont{T.~T.} \bibnamefont{Wang}},
  \bibinfo{author}{\bibfnamefont{S.}~\bibnamefont{Ebadi}},
  \bibinfo{author}{\bibfnamefont{M.}~\bibnamefont{Kalinowski}},
  \bibinfo{author}{\bibfnamefont{A.}~\bibnamefont{Keesling}},
  \bibinfo{author}{\bibfnamefont{N.}~\bibnamefont{Maskara}},
  \bibinfo{author}{\bibfnamefont{H.}~\bibnamefont{Pichler}},
  \bibinfo{author}{\bibfnamefont{M.}~\bibnamefont{Greiner}},
  \bibnamefont{et~al.}, \emph{\bibinfo{title}{{A quantum processor based on
  coherent transport of entangled atom arrays}}}, \bibinfo{journal}{Nature}
  \textbf{\bibinfo{volume}{604}}, \bibinfo{pages}{451} (\bibinfo{year}{2022}),
  \urlprefix\url{https://doi.org/10.1038/s41586-022-04592-6}.

\bibitem[{\citenamefont{Bluvstein et~al.}(2024)\citenamefont{Bluvstein, Evered,
  Geim, Li, Zhou, Manovitz, Ebadi, Cain, Kalinowski, Hangleiter
  et~al.}}]{bluvstein:24}
\bibinfo{author}{\bibfnamefont{D.}~\bibnamefont{Bluvstein}},
  \bibinfo{author}{\bibfnamefont{S.~J.} \bibnamefont{Evered}},
  \bibinfo{author}{\bibfnamefont{A.~A.} \bibnamefont{Geim}},
  \bibinfo{author}{\bibfnamefont{S.~H.} \bibnamefont{Li}},
  \bibinfo{author}{\bibfnamefont{H.}~\bibnamefont{Zhou}},
  \bibinfo{author}{\bibfnamefont{T.}~\bibnamefont{Manovitz}},
  \bibinfo{author}{\bibfnamefont{S.}~\bibnamefont{Ebadi}},
  \bibinfo{author}{\bibfnamefont{M.}~\bibnamefont{Cain}},
  \bibinfo{author}{\bibfnamefont{M.}~\bibnamefont{Kalinowski}},
  \bibinfo{author}{\bibfnamefont{D.}~\bibnamefont{Hangleiter}},
  \bibnamefont{et~al.}, \emph{\bibinfo{title}{Logical quantum processor based
  on reconfigurable atom arrays}}, \bibinfo{journal}{Nature}
  \textbf{\bibinfo{volume}{626}}, \bibinfo{pages}{58} (\bibinfo{year}{2024}),
  \urlprefix\url{https://doi.org/10.1038/s41586-023-06927-3}.

\bibitem[{\citenamefont{Butenko et~al.}(2002)\citenamefont{Butenko, Pardalos,
  Sergienko, Shylo, and Stetsyuk}}]{butenko:02}
\bibinfo{author}{\bibfnamefont{S.}~\bibnamefont{Butenko}},
  \bibinfo{author}{\bibfnamefont{P.}~\bibnamefont{Pardalos}},
  \bibinfo{author}{\bibfnamefont{I.}~\bibnamefont{Sergienko}},
  \bibinfo{author}{\bibfnamefont{V.}~\bibnamefont{Shylo}}, \bibnamefont{and}
  \bibinfo{author}{\bibfnamefont{P.}~\bibnamefont{Stetsyuk}}, in
  \emph{\bibinfo{booktitle}{Proceedings of the 2002 ACM Symposium on Applied
  Computing}} (\bibinfo{publisher}{Association for Computing Machinery},
  \bibinfo{address}{New York, NY, USA}, \bibinfo{year}{2002}), SAC '02, pp.
  \bibinfo{pages}{542--546}, ISBN \bibinfo{isbn}{1581134452},
  \urlprefix\url{https://doi.org/10.1145/508791.508897}.

\bibitem[{\citenamefont{Butenko and Trukhanov}(2007)}]{butenko:07}
\bibinfo{author}{\bibfnamefont{S.}~\bibnamefont{Butenko}} \bibnamefont{and}
  \bibinfo{author}{\bibfnamefont{S.}~\bibnamefont{Trukhanov}},
  \emph{\bibinfo{title}{Using critical sets to solve the maximum independent
  set problem}}, \bibinfo{journal}{Operations Research Letters}
  \textbf{\bibinfo{volume}{35}}, \bibinfo{pages}{519} (\bibinfo{year}{2007}),
  ISSN \bibinfo{issn}{0167-6377},
  \urlprefix\url{https://www.sciencedirect.com/science/article/pii/S0167637706000952}.

\bibitem[{\citenamefont{Strash}(2016)}]{strash:16}
\bibinfo{author}{\bibfnamefont{D.}~\bibnamefont{Strash}}, in
  \emph{\bibinfo{booktitle}{Computing and Combinatorics}}, edited by
  \bibinfo{editor}{\bibfnamefont{T.~N.} \bibnamefont{Dinh}} \bibnamefont{and}
  \bibinfo{editor}{\bibfnamefont{M.~T.} \bibnamefont{Thai}}
  (\bibinfo{publisher}{Springer International Publishing},
  \bibinfo{address}{Cham}, \bibinfo{year}{2016}), pp.
  \bibinfo{pages}{345--356}, ISBN \bibinfo{isbn}{978-3-319-42634-1}.

\bibitem[{\citenamefont{Hespe et~al.}(2019)\citenamefont{Hespe, Schulz, and
  Strash}}]{hespe:19}
\bibinfo{author}{\bibfnamefont{D.}~\bibnamefont{Hespe}},
  \bibinfo{author}{\bibfnamefont{C.}~\bibnamefont{Schulz}}, \bibnamefont{and}
  \bibinfo{author}{\bibfnamefont{D.}~\bibnamefont{Strash}},
  \emph{\bibinfo{title}{Scalable kernelization for maximum independent sets}},
  \bibinfo{journal}{ACM J. Exp. Algorithmics} \textbf{\bibinfo{volume}{24}}
  (\bibinfo{year}{2019}), ISSN \bibinfo{issn}{1084-6654},
  \urlprefix\url{https://doi.org/10.1145/3355502}.

\bibitem[{\citenamefont{Lamm et~al.}(2019)\citenamefont{Lamm, Schulz, Strash,
  Williger, and Zhang}}]{lamm:19}
\bibinfo{author}{\bibfnamefont{S.}~\bibnamefont{Lamm}},
  \bibinfo{author}{\bibfnamefont{C.}~\bibnamefont{Schulz}},
  \bibinfo{author}{\bibfnamefont{D.}~\bibnamefont{Strash}},
  \bibinfo{author}{\bibfnamefont{R.}~\bibnamefont{Williger}}, \bibnamefont{and}
  \bibinfo{author}{\bibfnamefont{H.}~\bibnamefont{Zhang}},
  \emph{\bibinfo{title}{Exactly Solving the Maximum Weight Independent Set
  Problem on Large Real-World Graphs}} (\bibinfo{year}{2019}), pp.
  \bibinfo{pages}{144--158},
  \eprint{https://epubs.siam.org/doi/pdf/10.1137/1.9781611975499.12},
  \urlprefix\url{https://epubs.siam.org/doi/abs/10.1137/1.9781611975499.12}.

\bibitem[{\citenamefont{McCallum et~al.}(2000)\citenamefont{McCallum, Nigam,
  Rennie, and Seymore}}]{cora:00}
\bibinfo{author}{\bibfnamefont{A.~K.} \bibnamefont{McCallum}},
  \bibinfo{author}{\bibfnamefont{K.}~\bibnamefont{Nigam}},
  \bibinfo{author}{\bibfnamefont{J.}~\bibnamefont{Rennie}}, \bibnamefont{and}
  \bibinfo{author}{\bibfnamefont{K.}~\bibnamefont{Seymore}},
  \emph{\bibinfo{title}{{Automating the construction of internet portals with
  machine learning}}}, \bibinfo{journal}{Information Retrieval}
  \textbf{\bibinfo{volume}{3}}, \bibinfo{pages}{127} (\bibinfo{year}{2000}).

\bibitem[{\citenamefont{Namata et~al.}(2012)\citenamefont{Namata, London,
  Getoor, and Huang}}]{pubmed:12}
\bibinfo{author}{\bibfnamefont{G.}~\bibnamefont{Namata}},
  \bibinfo{author}{\bibfnamefont{B.}~\bibnamefont{London}},
  \bibinfo{author}{\bibfnamefont{L.}~\bibnamefont{Getoor}}, \bibnamefont{and}
  \bibinfo{author}{\bibfnamefont{B.}~\bibnamefont{Huang}}, in
  \emph{\bibinfo{booktitle}{10th International Workshop on Mining and Learning
  with Graphs}} (\bibinfo{year}{2012}), vol.~\bibinfo{volume}{8}, p.
  \bibinfo{pages}{249}.

\bibitem[{\citenamefont{Wurtz et~al.}(2023)\citenamefont{Wurtz, Bylinskii,
  Braverman, Amato-Grill, Cantu, Huber, Lukin, Liu, Weinberg, Long
  et~al.}}]{wurtz:23}
\bibinfo{author}{\bibfnamefont{J.}~\bibnamefont{Wurtz}},
  \bibinfo{author}{\bibfnamefont{A.}~\bibnamefont{Bylinskii}},
  \bibinfo{author}{\bibfnamefont{B.}~\bibnamefont{Braverman}},
  \bibinfo{author}{\bibfnamefont{J.}~\bibnamefont{Amato-Grill}},
  \bibinfo{author}{\bibfnamefont{S.~H.} \bibnamefont{Cantu}},
  \bibinfo{author}{\bibfnamefont{F.}~\bibnamefont{Huber}},
  \bibinfo{author}{\bibfnamefont{A.}~\bibnamefont{Lukin}},
  \bibinfo{author}{\bibfnamefont{F.}~\bibnamefont{Liu}},
  \bibinfo{author}{\bibfnamefont{P.}~\bibnamefont{Weinberg}},
  \bibinfo{author}{\bibfnamefont{J.}~\bibnamefont{Long}}, \bibnamefont{et~al.},
  \emph{\bibinfo{title}{Aquila: Quera's 256-qubit neutral-atom quantum
  computer}} (\bibinfo{year}{2023}), \eprint{arXiv:2306.11727}.

\bibitem[{\citenamefont{Gyger et~al.}(2024)\citenamefont{Gyger, Ammenwerth,
  Tao, Timme, Snigirev, Bloch, and Zeiher}}]{gyger:24}
\bibinfo{author}{\bibfnamefont{F.}~\bibnamefont{Gyger}},
  \bibinfo{author}{\bibfnamefont{M.}~\bibnamefont{Ammenwerth}},
  \bibinfo{author}{\bibfnamefont{R.}~\bibnamefont{Tao}},
  \bibinfo{author}{\bibfnamefont{H.}~\bibnamefont{Timme}},
  \bibinfo{author}{\bibfnamefont{S.}~\bibnamefont{Snigirev}},
  \bibinfo{author}{\bibfnamefont{I.}~\bibnamefont{Bloch}}, \bibnamefont{and}
  \bibinfo{author}{\bibfnamefont{J.}~\bibnamefont{Zeiher}},
  \emph{\bibinfo{title}{Continuous operation of large-scale atom arrays in
  optical lattices}}, \bibinfo{journal}{Phys. Rev. Res.}
  \textbf{\bibinfo{volume}{6}}, \bibinfo{pages}{033104} (\bibinfo{year}{2024}),
  \urlprefix\url{https://link.aps.org/doi/10.1103/PhysRevResearch.6.033104}.

\bibitem[{\citenamefont{Manetsch et~al.}(2024)\citenamefont{Manetsch, Nomura,
  Bataille, Leung, Lv, and Endres}}]{manetsch:24}
\bibinfo{author}{\bibfnamefont{H.~J.} \bibnamefont{Manetsch}},
  \bibinfo{author}{\bibfnamefont{G.}~\bibnamefont{Nomura}},
  \bibinfo{author}{\bibfnamefont{E.}~\bibnamefont{Bataille}},
  \bibinfo{author}{\bibfnamefont{K.~H.} \bibnamefont{Leung}},
  \bibinfo{author}{\bibfnamefont{X.}~\bibnamefont{Lv}}, \bibnamefont{and}
  \bibinfo{author}{\bibfnamefont{M.}~\bibnamefont{Endres}},
  \emph{\bibinfo{title}{A tweezer array with 6100 highly coherent atomic
  qubits}} (\bibinfo{year}{2024}), \eprint{arXiv:2403.12021}.

\bibitem[{\citenamefont{Adams et~al.}(2019)\citenamefont{Adams, Pritchard, and
  Shaffer}}]{adams:20}
\bibinfo{author}{\bibfnamefont{C.~S.} \bibnamefont{Adams}},
  \bibinfo{author}{\bibfnamefont{J.~D.} \bibnamefont{Pritchard}},
  \bibnamefont{and} \bibinfo{author}{\bibfnamefont{J.~P.}
  \bibnamefont{Shaffer}}, \emph{\bibinfo{title}{{Rydberg atom quantum
  technologies}}}, \bibinfo{journal}{Journal of Physics B: Atomic, Molecular
  and Optical Physics} \textbf{\bibinfo{volume}{53}}, \bibinfo{pages}{012002}
  (\bibinfo{year}{2019}),
  \urlprefix\url{https://dx.doi.org/10.1088/1361-6455/ab52ef}.

\bibitem[{\citenamefont{Henriet et~al.}(2020)\citenamefont{Henriet, Beguin,
  Signoles, Lahaye, Browaeys, Reymond, and Jurczak}}]{henriet:20}
\bibinfo{author}{\bibfnamefont{L.}~\bibnamefont{Henriet}},
  \bibinfo{author}{\bibfnamefont{L.}~\bibnamefont{Beguin}},
  \bibinfo{author}{\bibfnamefont{A.}~\bibnamefont{Signoles}},
  \bibinfo{author}{\bibfnamefont{T.}~\bibnamefont{Lahaye}},
  \bibinfo{author}{\bibfnamefont{A.}~\bibnamefont{Browaeys}},
  \bibinfo{author}{\bibfnamefont{G.-O.} \bibnamefont{Reymond}},
  \bibnamefont{and} \bibinfo{author}{\bibfnamefont{C.}~\bibnamefont{Jurczak}},
  \emph{\bibinfo{title}{{Quantum computing with neutral atoms}}},
  \bibinfo{journal}{{Quantum}} \textbf{\bibinfo{volume}{4}},
  \bibinfo{pages}{327} (\bibinfo{year}{2020}), ISSN \bibinfo{issn}{2521-327X},
  \urlprefix\url{https://doi.org/10.22331/q-2020-09-21-327}.

\bibitem[{\citenamefont{Butenko et~al.}(2009)\citenamefont{Butenko, Pardalos,
  Sergienko, Shylo, and Stetsyuk}}]{butenko:09}
\bibinfo{author}{\bibfnamefont{S.}~\bibnamefont{Butenko}},
  \bibinfo{author}{\bibfnamefont{P.}~\bibnamefont{Pardalos}},
  \bibinfo{author}{\bibfnamefont{I.}~\bibnamefont{Sergienko}},
  \bibinfo{author}{\bibfnamefont{V.}~\bibnamefont{Shylo}}, \bibnamefont{and}
  \bibinfo{author}{\bibfnamefont{P.}~\bibnamefont{Stetsyuk}},
  \emph{\bibinfo{title}{Estimating the size of correcting codes using extremal
  graph problems}} (\bibinfo{publisher}{Springer New York},
  \bibinfo{address}{New York, NY}, \bibinfo{year}{2009}), pp.
  \bibinfo{pages}{227--243}, ISBN \bibinfo{isbn}{978-0-387-98096-6},
  \urlprefix\url{https://doi.org/10.1007/978-0-387-98096-6_12}.

\bibitem[{\citenamefont{Chang et~al.}(2017)\citenamefont{Chang, Li, and
  Zhang}}]{chang:17}
\bibinfo{author}{\bibfnamefont{L.}~\bibnamefont{Chang}},
  \bibinfo{author}{\bibfnamefont{W.}~\bibnamefont{Li}}, \bibnamefont{and}
  \bibinfo{author}{\bibfnamefont{W.}~\bibnamefont{Zhang}}, in
  \emph{\bibinfo{booktitle}{Proceedings of the 2017 ACM International
  Conference on Management of Data}} (\bibinfo{publisher}{Association for
  Computing Machinery}, \bibinfo{address}{New York, NY, USA},
  \bibinfo{year}{2017}), SIGMOD '17, pp. \bibinfo{pages}{1181--1196}, ISBN
  \bibinfo{isbn}{9781450341974},
  \urlprefix\url{https://doi.org/10.1145/3035918.3035939}.

\bibitem[{\citenamefont{Lamm et~al.}(2017)\citenamefont{Lamm, Sanders, Schulz,
  Strash, and Werneck}}]{lamm:17}
\bibinfo{author}{\bibfnamefont{S.}~\bibnamefont{Lamm}},
  \bibinfo{author}{\bibfnamefont{P.}~\bibnamefont{Sanders}},
  \bibinfo{author}{\bibfnamefont{C.}~\bibnamefont{Schulz}},
  \bibinfo{author}{\bibfnamefont{D.}~\bibnamefont{Strash}}, \bibnamefont{and}
  \bibinfo{author}{\bibfnamefont{R.~F.} \bibnamefont{Werneck}},
  \emph{\bibinfo{title}{Finding near-optimal independent sets at scale}},
  \bibinfo{journal}{Journal of Heuristics} \textbf{\bibinfo{volume}{23}},
  \bibinfo{pages}{207} (\bibinfo{year}{2017}),
  \urlprefix\url{https://doi.org/10.1007/s10732-017-9337-x}.

\bibitem[{\citenamefont{Gro\ss{}mann et~al.}(2023)\citenamefont{Gro\ss{}mann,
  Lamm, Schulz, and Strash}}]{grossmann:23}
\bibinfo{author}{\bibfnamefont{E.}~\bibnamefont{Gro\ss{}mann}},
  \bibinfo{author}{\bibfnamefont{S.}~\bibnamefont{Lamm}},
  \bibinfo{author}{\bibfnamefont{C.}~\bibnamefont{Schulz}}, \bibnamefont{and}
  \bibinfo{author}{\bibfnamefont{D.}~\bibnamefont{Strash}}, in
  \emph{\bibinfo{booktitle}{Proceedings of the Genetic and Evolutionary
  Computation Conference}} (\bibinfo{publisher}{Association for Computing
  Machinery}, \bibinfo{address}{New York, NY, USA}, \bibinfo{year}{2023}),
  GECCO '23, pp. \bibinfo{pages}{293--302}, ISBN \bibinfo{isbn}{9798400701191},
  \urlprefix\url{https://doi.org/10.1145/3583131.3590353}.

\bibitem[{\citenamefont{Liu et~al.}(2018)\citenamefont{Liu, Safavi, Dighe, and
  Koutra}}]{liu:18}
\bibinfo{author}{\bibfnamefont{Y.}~\bibnamefont{Liu}},
  \bibinfo{author}{\bibfnamefont{T.}~\bibnamefont{Safavi}},
  \bibinfo{author}{\bibfnamefont{A.}~\bibnamefont{Dighe}}, \bibnamefont{and}
  \bibinfo{author}{\bibfnamefont{D.}~\bibnamefont{Koutra}},
  \emph{\bibinfo{title}{Graph summarization methods and applications: A
  survey}}, \bibinfo{journal}{ACM Comput. Surv.} \textbf{\bibinfo{volume}{51}}
  (\bibinfo{year}{2018}), ISSN \bibinfo{issn}{0360-0300},
  \urlprefix\url{https://doi.org/10.1145/3186727}.

\bibitem[{\citenamefont{Hashemi et~al.}(2024)\citenamefont{Hashemi, Gong, Ni,
  Fan, Prakash, and Jin}}]{hashemi:24}
\bibinfo{author}{\bibfnamefont{M.}~\bibnamefont{Hashemi}},
  \bibinfo{author}{\bibfnamefont{S.}~\bibnamefont{Gong}},
  \bibinfo{author}{\bibfnamefont{J.}~\bibnamefont{Ni}},
  \bibinfo{author}{\bibfnamefont{W.}~\bibnamefont{Fan}},
  \bibinfo{author}{\bibfnamefont{B.~A.} \bibnamefont{Prakash}},
  \bibnamefont{and} \bibinfo{author}{\bibfnamefont{W.}~\bibnamefont{Jin}},
  \emph{\bibinfo{title}{A comprehensive survey on graph reduction:
  Sparsification, coarsening, and condensation}} (\bibinfo{year}{2024}),
  \eprint{arXiv:2402.03358}.

\bibitem[{\citenamefont{Lewis and Glover}(2017)}]{lewis:17}
\bibinfo{author}{\bibfnamefont{M.}~\bibnamefont{Lewis}} \bibnamefont{and}
  \bibinfo{author}{\bibfnamefont{F.}~\bibnamefont{Glover}},
  \emph{\bibinfo{title}{Quadratic unconstrained binary optimization problem
  preprocessing: Theory and empirical analysis}}, \bibinfo{journal}{Netw.}
  \textbf{\bibinfo{volume}{70}}, \bibinfo{pages}{79} (\bibinfo{year}{2017}),
  ISSN \bibinfo{issn}{0028-3045},
  \urlprefix\url{https://doi.org/10.1002/net.21751}.

\bibitem[{\citenamefont{Glover et~al.}(2018)\citenamefont{Glover, Lewis, and
  Kochenberger}}]{glover:18b}
\bibinfo{author}{\bibfnamefont{F.}~\bibnamefont{Glover}},
  \bibinfo{author}{\bibfnamefont{M.}~\bibnamefont{Lewis}}, \bibnamefont{and}
  \bibinfo{author}{\bibfnamefont{G.}~\bibnamefont{Kochenberger}},
  \emph{\bibinfo{title}{Logical and inequality implications for reducing the
  size and difficulty of quadratic unconstrained binary optimization
  problems}}, \bibinfo{journal}{European Journal of Operational Research}
  \textbf{\bibinfo{volume}{265}}, \bibinfo{pages}{829} (\bibinfo{year}{2018}),
  ISSN \bibinfo{issn}{0377-2217},
  \urlprefix\url{https://www.sciencedirect.com/science/article/pii/S0377221717307567}.

\bibitem[{\citenamefont{Narimani et~al.}(2017)\citenamefont{Narimani, Rezaei,
  and Zaribafiyan}}]{narimani:17}
\bibinfo{author}{\bibfnamefont{A.}~\bibnamefont{Narimani}},
  \bibinfo{author}{\bibfnamefont{S.~S.~C.} \bibnamefont{Rezaei}},
  \bibnamefont{and}
  \bibinfo{author}{\bibfnamefont{A.}~\bibnamefont{Zaribafiyan}},
  \emph{\bibinfo{title}{Combinatorial optimization by decomposition on hybrid
  cpu--non-cpu solver architectures}} (\bibinfo{year}{2017}),
  \eprint{arXiv:1708.03439}.

\bibitem[{\citenamefont{Djidjev et~al.}(2020)\citenamefont{Djidjev, Pelofske,
  and Hahn}}]{djidjev:20}
\bibinfo{author}{\bibfnamefont{H.~N.} \bibnamefont{Djidjev}},
  \bibinfo{author}{\bibfnamefont{E.~A.~R.} \bibnamefont{Pelofske}},
  \bibnamefont{and} \bibinfo{author}{\bibfnamefont{G.}~\bibnamefont{Hahn}},
  \emph{\bibinfo{title}{Decomposition algorithms for solving np-hard problems
  on a quantum annealer}}, \bibinfo{journal}{Journal of Signal Processing
  Systems} \textbf{\bibinfo{volume}{93}} (\bibinfo{year}{2020}), ISSN
  \bibinfo{issn}{1939-8018},
  \urlprefix\url{https://www.osti.gov/biblio/1822729}.

\bibitem[{\citenamefont{Newman}(2010)}]{newman:10}
\bibinfo{author}{\bibfnamefont{M.~E.~J.} \bibnamefont{Newman}},
  \emph{\bibinfo{title}{Networks: an introduction}} (\bibinfo{publisher}{Oxford
  University Press}, \bibinfo{address}{Oxford; New York},
  \bibinfo{year}{2010}), ISBN \bibinfo{isbn}{9780199206650 0199206651},
  \urlprefix\url{http://www.amazon.com/Networks-An-Introduction-Mark-Newman/dp/0199206651/ref=sr_1_5?ie=UTF8&qid=1352896678&sr=8-5&keywords=complex+networks}.

\bibitem[{\citenamefont{Chaves et~al.}(2024{\natexlab{a}})\citenamefont{Chaves,
  Resende, Schuetz, Brubaker, Katzgraber, de~Arruda, and Silva}}]{chaves:24rko}
\bibinfo{author}{\bibfnamefont{A.~A.} \bibnamefont{Chaves}},
  \bibinfo{author}{\bibfnamefont{M.~G.~C.} \bibnamefont{Resende}},
  \bibinfo{author}{\bibfnamefont{M.~J.~A.} \bibnamefont{Schuetz}},
  \bibinfo{author}{\bibfnamefont{J.~K.} \bibnamefont{Brubaker}},
  \bibinfo{author}{\bibfnamefont{H.~G.} \bibnamefont{Katzgraber}},
  \bibinfo{author}{\bibfnamefont{E.~F.} \bibnamefont{de~Arruda}},
  \bibnamefont{and} \bibinfo{author}{\bibfnamefont{R.~M.~A.}
  \bibnamefont{Silva}}, \emph{\bibinfo{title}{A random-key optimizer for
  combinatorial optimization}} (\bibinfo{year}{2024}{\natexlab{a}}),
  \eprint{2411.04293}, \urlprefix\url{https://arxiv.org/abs/2411.04293}.

\bibitem[{\citenamefont{Londe et~al.}(2024{\natexlab{a}})\citenamefont{Londe,
  Pessoa, Andrade, and Resende}}]{londe:24a}
\bibinfo{author}{\bibfnamefont{M.~A.} \bibnamefont{Londe}},
  \bibinfo{author}{\bibfnamefont{L.~S.} \bibnamefont{Pessoa}},
  \bibinfo{author}{\bibfnamefont{C.~E.} \bibnamefont{Andrade}},
  \bibnamefont{and} \bibinfo{author}{\bibfnamefont{M.~G.}
  \bibnamefont{Resende}}, \emph{\bibinfo{title}{Biased random-key genetic
  algorithms: A review}}, \bibinfo{journal}{European Journal of Operational
  Research}  (\bibinfo{year}{2024}{\natexlab{a}}), ISSN
  \bibinfo{issn}{0377-2217},
  \urlprefix\url{https://www.sciencedirect.com/science/article/pii/S0377221724002303}.

\bibitem[{\citenamefont{Londe et~al.}(2024{\natexlab{b}})\citenamefont{Londe,
  Pessoa, Andrade, and Resende}}]{londe:24b}
\bibinfo{author}{\bibfnamefont{M.~A.} \bibnamefont{Londe}},
  \bibinfo{author}{\bibfnamefont{L.~S.} \bibnamefont{Pessoa}},
  \bibinfo{author}{\bibfnamefont{C.~E.} \bibnamefont{Andrade}},
  \bibnamefont{and} \bibinfo{author}{\bibfnamefont{M.~G.~C.}
  \bibnamefont{Resende}}, \emph{\bibinfo{title}{Early years of biased
  random-key genetic algorithms: A systematic review}}
  (\bibinfo{year}{2024}{\natexlab{b}}), \eprint{2405.01765},
  \urlprefix\url{https://arxiv.org/abs/2405.01765}.

\bibitem[{\citenamefont{Schuetz et~al.}(2022)\citenamefont{Schuetz, Brubaker,
  Montagu, van Dijk, Klepsch, Ross, Luckow, Resende, and
  Katzgraber}}]{schuetz:22}
\bibinfo{author}{\bibfnamefont{M.~J.} \bibnamefont{Schuetz}},
  \bibinfo{author}{\bibfnamefont{J.~K.} \bibnamefont{Brubaker}},
  \bibinfo{author}{\bibfnamefont{H.}~\bibnamefont{Montagu}},
  \bibinfo{author}{\bibfnamefont{Y.}~\bibnamefont{van Dijk}},
  \bibinfo{author}{\bibfnamefont{J.}~\bibnamefont{Klepsch}},
  \bibinfo{author}{\bibfnamefont{P.}~\bibnamefont{Ross}},
  \bibinfo{author}{\bibfnamefont{A.}~\bibnamefont{Luckow}},
  \bibinfo{author}{\bibfnamefont{M.~G.} \bibnamefont{Resende}},
  \bibnamefont{and} \bibinfo{author}{\bibfnamefont{H.~G.}
  \bibnamefont{Katzgraber}}, \emph{\bibinfo{title}{Optimization of
  robot-trajectory planning with nature-inspired and hybrid quantum
  algorithms}}, \bibinfo{journal}{Phys. Rev. Appl.}
  \textbf{\bibinfo{volume}{18}}, \bibinfo{pages}{054045}
  (\bibinfo{year}{2022}),
  \urlprefix\url{https://link.aps.org/doi/10.1103/PhysRevApplied.18.054045}.

\bibitem[{\citenamefont{Chaves et~al.}(2024{\natexlab{b}})\citenamefont{Chaves,
  Resende, and Silva}}]{chaves:24}
\bibinfo{author}{\bibfnamefont{A.~A.} \bibnamefont{Chaves}},
  \bibinfo{author}{\bibfnamefont{M.~G.~C.} \bibnamefont{Resende}},
  \bibnamefont{and} \bibinfo{author}{\bibfnamefont{R.~M.~A.}
  \bibnamefont{Silva}}, in \emph{\bibinfo{booktitle}{Metaheuristics: 15th
  International Conference, MIC 2024, Lorient, France, June 4-7, 2024,
  Proceedings, Part I}} (\bibinfo{publisher}{Springer-Verlag},
  \bibinfo{address}{Berlin, Heidelberg}, \bibinfo{year}{2024}{\natexlab{b}}),
  pp. \bibinfo{pages}{15--20}, ISBN \bibinfo{isbn}{978-3-031-62911-2},
  \urlprefix\url{https://doi.org/10.1007/978-3-031-62912-9_3}.

\bibitem[{\citenamefont{Kirkpatrick et~al.}(1983)\citenamefont{Kirkpatrick,
  Gelatt, and Vecchi}}]{kirkpatrick:83}
\bibinfo{author}{\bibfnamefont{S.}~\bibnamefont{Kirkpatrick}},
  \bibinfo{author}{\bibfnamefont{C.~D.} \bibnamefont{Gelatt}},
  \bibnamefont{and} \bibinfo{author}{\bibfnamefont{M.~P.}
  \bibnamefont{Vecchi}}, \emph{\bibinfo{title}{{Optimization by Simulated
  Annealing}}}, \bibinfo{journal}{Science} \textbf{\bibinfo{volume}{220}},
  \bibinfo{pages}{671} (\bibinfo{year}{1983}),
  \urlprefix\url{https://www.science.org/doi/abs/10.1126/science.220.4598.671}.

\bibitem[{\citenamefont{Perseguers}(2024)}]{perseguers:24}
\bibinfo{author}{\bibfnamefont{S.}~\bibnamefont{Perseguers}},
  \emph{\bibinfo{title}{Hardness-dependent adiabatic schedules for analog
  quantum computing}} (\bibinfo{year}{2024}), \eprint{2410.08995},
  \urlprefix\url{https://arxiv.org/abs/2410.08995}.

\bibitem[{\citenamefont{Leyton-Brown et~al.}(2014)\citenamefont{Leyton-Brown,
  Hoos, Hutter, and Xu}}]{leyton-brown:14}
\bibinfo{author}{\bibfnamefont{K.}~\bibnamefont{Leyton-Brown}},
  \bibinfo{author}{\bibfnamefont{H.~H.} \bibnamefont{Hoos}},
  \bibinfo{author}{\bibfnamefont{F.}~\bibnamefont{Hutter}}, \bibnamefont{and}
  \bibinfo{author}{\bibfnamefont{L.}~\bibnamefont{Xu}},
  \emph{\bibinfo{title}{Understanding the empirical hardness of np-complete
  problems}}, \bibinfo{journal}{Commun. ACM} \textbf{\bibinfo{volume}{57}},
  \bibinfo{pages}{98} (\bibinfo{year}{2014}), ISSN \bibinfo{issn}{0001-0782},
  \urlprefix\url{https://doi.org/10.1145/2594413.2594424}.

\bibitem[{\citenamefont{Song et~al.}(2021)\citenamefont{Song, Kim, Hwang, Lee,
  and Ahn}}]{song:21}
\bibinfo{author}{\bibfnamefont{Y.}~\bibnamefont{Song}},
  \bibinfo{author}{\bibfnamefont{M.}~\bibnamefont{Kim}},
  \bibinfo{author}{\bibfnamefont{H.}~\bibnamefont{Hwang}},
  \bibinfo{author}{\bibfnamefont{W.}~\bibnamefont{Lee}}, \bibnamefont{and}
  \bibinfo{author}{\bibfnamefont{J.}~\bibnamefont{Ahn}},
  \emph{\bibinfo{title}{Quantum simulation of cayley-tree ising hamiltonians
  with three-dimensional rydberg atoms}}, \bibinfo{journal}{Phys. Rev. Res.}
  \textbf{\bibinfo{volume}{3}}, \bibinfo{pages}{013286} (\bibinfo{year}{2021}),
  \urlprefix\url{https://link.aps.org/doi/10.1103/PhysRevResearch.3.013286}.

\bibitem[{\citenamefont{Rossi and Ahmed}(2015)}]{rossi:15}
\bibinfo{author}{\bibfnamefont{R.~A.} \bibnamefont{Rossi}} \bibnamefont{and}
  \bibinfo{author}{\bibfnamefont{N.~K.} \bibnamefont{Ahmed}}, in
  \emph{\bibinfo{booktitle}{AAAI}} (\bibinfo{year}{2015}),
  \urlprefix\url{http://networkrepository.com}.

\bibitem[{\citenamefont{Barthelemy}(2011)}]{barthelemy:11}
\bibinfo{author}{\bibfnamefont{M.}~\bibnamefont{Barthelemy}},
  \emph{\bibinfo{title}{Spatial networks}}, \bibinfo{journal}{Physics Reports}
  \textbf{\bibinfo{volume}{499}}, \bibinfo{pages}{1} (\bibinfo{year}{2011}),
  ISSN \bibinfo{issn}{0370-1573},
  \urlprefix\url{https://www.sciencedirect.com/science/article/pii/S037015731000308X}.

\bibitem[{\citenamefont{Barab\'asi and Albert}(1999)}]{barabasi-albert:99}
\bibinfo{author}{\bibfnamefont{A.-L.} \bibnamefont{Barab\'asi}}
  \bibnamefont{and} \bibinfo{author}{\bibfnamefont{R.}~\bibnamefont{Albert}},
  \emph{\bibinfo{title}{Emergence of scaling in random networks}},
  \bibinfo{journal}{Science} \textbf{\bibinfo{volume}{286}},
  \bibinfo{pages}{509} (\bibinfo{year}{1999}),
  \eprint{https://www.science.org/doi/pdf/10.1126/science.286.5439.509},
  \urlprefix\url{https://www.science.org/doi/abs/10.1126/science.286.5439.509}.

\bibitem[{\citenamefont{Wurtz et~al.}(2024)\citenamefont{Wurtz, Sack, and
  Wang}}]{wurtz:24}
\bibinfo{author}{\bibfnamefont{J.}~\bibnamefont{Wurtz}},
  \bibinfo{author}{\bibfnamefont{S.~H.} \bibnamefont{Sack}}, \bibnamefont{and}
  \bibinfo{author}{\bibfnamefont{S.-T.} \bibnamefont{Wang}},
  \emph{\bibinfo{title}{Solving non-native combinatorial optimization problems
  using hybrid quantum-classical algorithms}}, \bibinfo{journal}{IEEE
  Transactions on Quantum Engineering} pp. \bibinfo{pages}{1--15}
  (\bibinfo{year}{2024}).

\end{thebibliography}

\end{document}